\documentclass[pra,aps,nobibnotes,nofootinbib, amssymb,floatfix,a4paper,longbibliography,10pt]{revtex4-2}
\usepackage{adjustbox,amsmath,amsfonts,amsthm,bbm,cancel,enumerate,enumitem,graphicx,mathtools,tikz,pgfplots,pifont,soul,thmtools}
\usepackage[normalem]{ulem}
\pgfplotsset{compat=1.18} 

\usepackage{hyperref}
\hypersetup{colorlinks=true,linkcolor=blue,citecolor=blue,urlcolor=blue}
\renewcommand{\bibinfo}[2]{%
  \ifstrequal{#1}{title}
    {\textit{#2}}
    {#2}%
}
\let\OO\O

\def\C{ {\cal C} }
\def\D{ {\cal D} }

\def\H{ {\cal H} }

\def\O{ {\cal O} }

\def\Rc{ {\cal R} }

\newcommand{\bra}[1]{\langle {#1} |}
\newcommand{\ket}[1]{| {#1} \rangle}
 
\newcommand{\ketbra}[2]{\ensuremath{\left|#1\right\rangle\!\!\left\langle#2\right|}}
\newcommand{\braket}[2]{\ensuremath{\!\left\langle#1|#2\right\rangle}\!}

\newcommand{\dev}{\mathrm{dev}}
\newcommand{\vol}{\mathrm{vol}}

\renewcommand{\v}[1]{\ensuremath{#1}}

\renewcommand{\i}{\textbf{i}}
\newcommand{\ot}{\otimes}
\newcommand{\norm}[1]{\left\lVert#1\right\rVert}
\newcommand{\comp}{\mathrm{Complexity}}

\DeclareMathOperator{\polylog}{polylog}

\begin{document}

\title{Physical regimes of quartic quantum advantage in elastic media simulation}
\author{David Jennings}
\author{Kamil Korzekwa\textsuperscript{*}}
\author{Matteo Lostaglio}
\author{Tam\'as Vaszary\textsuperscript{*}}
\affiliation{PsiQuantum, 700 Hansen Way, Palo Alto, CA 94304, USA}
\date\today

\renewcommand{\thefootnote}{\fnsymbol{footnote}} 
\footnotetext[1]{Corresponding authors' email addresses: kkorzekwa@psiquantum.com and tamas.vaszary@maths.ox.ac.uk}
\renewcommand{\thefootnote}{\arabic{footnote}}

\begin{abstract}
    Quantum algorithms for elastic media simulation have been conjectured to offer up to a quartic quantum advantage over classical methods; however, the devil is always in the details. Here, we determine key physical conditions under which this quartic separation could be achieved in the relevant parameters. For continuum elastic dynamics at fixed volume with spatial convergence order $m$, we find that the end-to-end speedup is of order $(m+4)/(m+1)$ in the simulation time, and $4/(m+1)$ in the accuracy, with exponential memory advantage. For elastic lattice systems, we define microscopically resolved bulk observables (MRBOs) as those that can be efficiently extracted from a quantum state, but cannot be resolved by a known continuum model with fewer degrees of freedom than the lattice. We then argue that the estimation of MRBOs is a candidate problem for quartic quantum advantage in problem size. We propose MRBOs for pseudorandom heterogeneous materials as an explicit end-to-end application where this occurs, while admitting efficient `quantum cheese' encodings of the material properties. We thus demonstrate how the interplay between regularity, physical scale, observable sensitivity and quantum encodability constrains the relation between classical and quantum complexity.
\end{abstract}

\maketitle
\tableofcontents


\section{Introduction}
\label{sec:intro}

Elastic dynamics in heterogeneous media play a central role across a wide range of scientific and engineering applications. In geophysics, seismic waves probe the structure of the Earth's interior and are used in imaging, hazard assessment, and subsurface characterization~\cite{VirieuxOperto2009}. In structural mechanics and materials science, elastic responses underlie nondestructive testing, vibration analysis, fracture diagnostics, and the design of engineered media for wave guiding or damping~\cite{SuYeLu2006}. Similar phenomena arise in acoustics, aerospace engineering, and the study of metamaterials~\cite{HusseinLeamyRuzzene2014,CummerChristensenAlu2016}. In all of these settings, spatial inhomogeneity in the mass density and elastic moduli is essential: interfaces, defects, layering, inclusions, and continuously varying material parameters strongly affect scattering, mode conversion, localization, and attenuation~\cite{Hashin1983}. Accurate simulation of elastic dynamics in heterogeneous media is therefore both practically important and computationally demanding.

Classically, elastic dynamics are simulated using either continuum models~\cite{Landau_book,Love_book} or microscopic lattice models~\cite{born1996dynamical}. When continuum elasticity is applicable, its governing equations are typically discretized using finite-difference, finite-element, finite-volume, or spectral methods~\cite{Virieux1986,Komatitsch1999,ZhangWang2022}. The spatial mesh must be sufficiently fine to resolve the shortest relevant wavelengths, rapid variations in material parameters, interfaces, defects, and other small-scale geometric features. Moreover, time-domain simulations must satisfy stability constraints that relate the time step to the mesh spacing. Consequently, simulations combining large three-dimensional domains with fine spatial structure, high target accuracy, or long evolution times can incur substantial computational costs~\cite{tsuboi2016trillion}. 

The continuum description can break down when the relevant wavelengths or the characteristic length scales of material variations become comparable to the spacing between the material's microscopic constituents, or when atomistic structure and individual interactions must be resolved explicitly. The material must then be represented through its microscopic degrees of freedom, using discrete approaches such as lattice dynamics, spring-network models, or atomistic molecular dynamics~\cite{TogoTanaka2015,OstojaStarzewski2002,ThompsonEtAl2022}. The most direct example of such continuum breakdown is a crystalline solid excited at wavelengths comparable to the interatomic or unit-cell spacing, where the full lattice dispersion becomes relevant rather than its long-wavelength elastic limit~\cite{born1996dynamical,TogoTanaka2015}. Closely related breakdowns occur in amorphous solids and glasses, where heterogeneous local elastic response produces excess or localized vibrational modes and the Debye continuum description fails at sufficiently short wavelengths~\cite{mizuno2016relation}; in granular and jammed materials, where vibrations can resolve the disordered contact network~\cite{silbert2005vibrations}; and in phononic crystals and elastic metamaterials when a homogenized, frequency-independent continuum description cannot capture the nontrivial dispersion, band gaps, and local resonances caused by differences between the materials in each unit cell~\cite{HusseinLeamyRuzzene2014,CummerChristensenAlu2016}. Similar considerations apply to disordered alloys, nanoporous materials, superlattices, and nanocomposites whenever microscopic scattering, localization, or vibrational modes control the quantity of interest. In such cases, increasing the resolution of a continuum discretization does not recover the missing physics: the microscopic structure itself must be included in the model. Such simulations can likewise be costly because they may involve very large numbers of interacting constituents and must resolve their fastest timescales~\cite{JiaEtAl2020}. 

\begin{figure}[t!]
    \centering
    \includegraphics[width=0.5\textwidth]{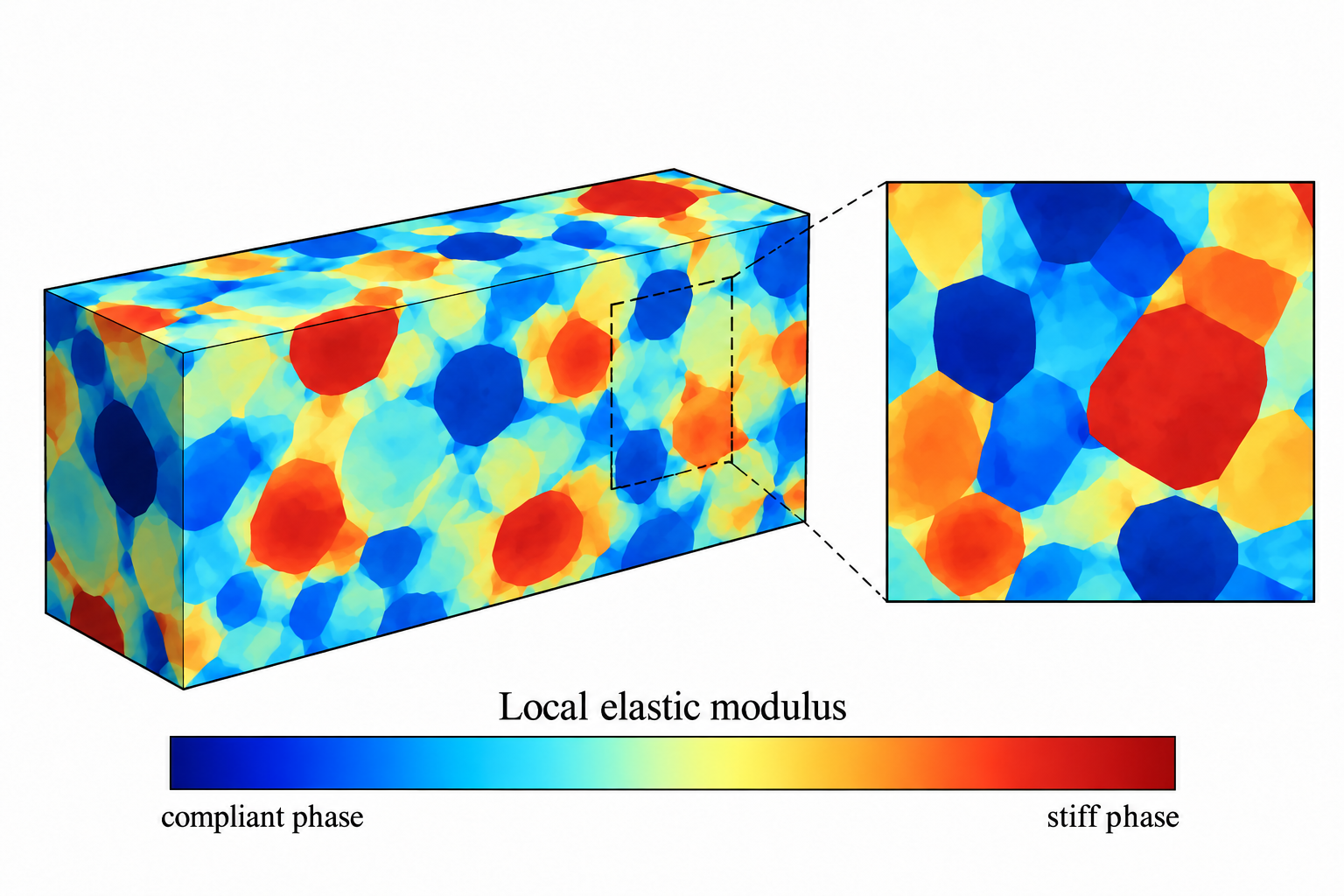}
    \caption{\textbf{Quantum simulation of heterogeneous material.} Shown is a highly heterogeneous elastic system with spatially varying elastic modulus. How do we construct quantum algorithms to simulate such systems? How do regularity and scale impact any potential quantum speedup? }
    \label{fig:heterogenous-material}
\end{figure}

These challenges motivate the search for alternative computational paradigms, including quantum algorithms for dynamical systems. Since its original formulation as a theoretical proposal~\cite{feynman1982simulating}, quantum computation has evolved rapidly into a field encompassing both active experimental development~\cite{preskill2018quantum} and continuing algorithmic advances~\cite{dalzell2025quantum}. Quantum devices are now capable of executing nontrivial coherent computations~\cite{arute2019quantum}, while fault-tolerant architectures remain a major long-term objective~\cite{acharya2023suppressing,psiquantum2025manufacturable}. At the same time, quantum-algorithmic theory has developed a range of methods offering provable asymptotic advantages for structured problems in scientific computing, including Hamiltonian simulation, amplitude estimation, linear-system solving, and linear differential equations~\cite{berry2007efficient,low2019hamiltonian,brassard2002quantum,harrow2009quantum,berry2014high,lin2022lecturenotesquantumalgorithms,li2026quantum}. These developments naturally raise the question of whether the dynamics of heterogeneous elastic systems can be reformulated so as to exploit such quantum primitives (see Fig.~\ref{fig:heterogenous-material}).

More broadly, there has been growing interest in quantum algorithms for classical dynamical systems, such as fluid dynamics~\cite{li2025potential,penuel2024detailed,jennings2026end}, plasma physics~\cite{vaszary2026solving,christensen2026improved,berntson2026end}, and general nonlinear systems embedded into larger linear dynamics~\cite{liu2021efficient,wu2024quantum,jennings2026quantum}. While the details differ across problem classes, the central idea is to encode the state of a discretized classical system into amplitudes of a quantum state and then use coherent quantum evolution to simulate the dynamics. In favorable settings, this can lead to improved complexity for evolving the system and estimating selected observables. Note, however, that such improvements are typically conditional: they rely on efficient state preparation, succinct access to spatially varying coefficients, and an output model in which one seeks a limited set of observables rather than the full classical field. Nevertheless, these methods provide a plausible route to quantum advantage for important families of classical simulation tasks.

Within this broader program, several recent works have developed quantum algorithms specifically for wave propagation and elastic dynamics. At the continuum level, Ref.~\cite{costa2019quantum} developed a Hamiltonian-simulation algorithm for the homogeneous scalar wave equation, including nontrivial boundary geometries, while Ref.~\cite{nakanishi2026hamiltonian} constructed explicit Trotter circuits for three-dimensional elasticity in homogeneous isotropic media.  More generally,  Ref.~\cite{bosch2025quantum} formulated a quantum-algorithmic framework for lossless wave equations with spatially varying anisotropic material coefficients, including elasticity, although efficient access to these coefficients is assumed rather than explicitly constructed and consideration of regularity and spatial discretization errors are not included. A separate line of work starts from discrete oscillator systems rather than a continuum elastic equation. Ref.~\cite{babbush2023exponential} developed a quantum algorithm for general networks of coupled harmonic oscillators, allowing nonuniform masses and spring constants provided that they can be queried efficiently. Subsequent work considered translationally invariant harmonic lattices~\cite{li2025exponential} and specialized the oscillator algorithm to a periodic elastic-network model of graphene with identical masses and couplings~\cite{kolotouros2026quantum}. In Ref.~\cite{sakamoto2026quantum} it was shown that short-time dynamics generated by any geometrically local interactions can be efficiently dequantized, ruling out an exponential quantum advantage in that regime, while leaving room for polynomial time advantages and exponential space savings at longer times. More recently, Ref.~\cite{schade2026quantum} extended oscillator simulations to non-Markovian dissipation, external sources, and time-dependent material parameters. They also establish worst-case BQP-completeness for dissipative oscillator networks and locality-based classical simulation bounds for spatially local observables. Together, these results establish the relevance of Hamiltonian simulation to both continuum waves and microscopic lattices, but do not provide an end-to-end analysis that connects heterogeneous continuum elasticity to the corresponding microscopic-lattice regime while accounting for physical accuracy, material regularity, and observable-extraction costs.

\begin{figure}[t!]
    \centering
    \includegraphics[width=0.85\textwidth]{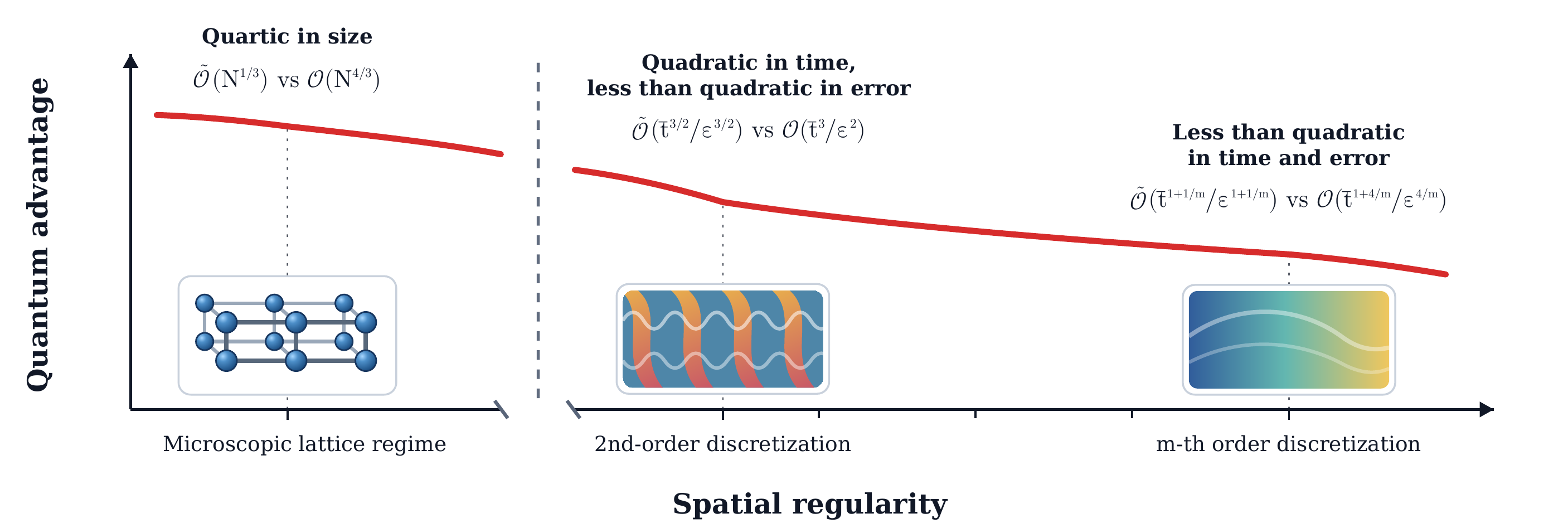}
    \caption{\textbf{Quantum advantage versus regularity.} We analyze how the runtime complexity of estimating an energy fraction observable to some error $\epsilon$ is affected by regularity of the dynamics. Within the continuum regime at fixed volume, the regularity of the material parameters and solution limits the applicable spatial discretization order $m$. For example, when only a second-order discretization is possible, the quantum complexity scales as $\tilde{\O}(\bar t^{3/2}/\epsilon^{3/2})$, compared with the classical scaling $\O(\bar t^{3}/\epsilon^{2})$, corresponding to a quadratic speedup in simulation time $\bar{t}$ and a weaker improvement in error $\epsilon$. As higher-order discretizations become applicable, the potential advantage decreases (note, however, that here we speak of an effective regularity, where $m$ can also be lower than~2, yielding a better than quadratic advantage). To the left of the dashed line, continuum elasticity breaks down and the $N$ lattice degrees of freedom now represent the actual physical microscopic system rather than a numerical discretization. In this regime, the respective quantum and classical scalings with system size are $\tilde{\O}(N^{1/3})$ and $\O(N^{4/3})$, yielding a quartic speedup. The red curve is a schematic guide to the eye, while the illustrations represent microscopic lattice, rapidly varying continuum, and smooth continuum models. 
    \label{fig:advantage-regimes}}
\end{figure}

We address this gap by developing a quantum algorithm for elastic dynamics in heterogeneous media that treats the continuum and microscopic-lattice regimes within a common framework. We first formulate the problem within continuum elasticity with spatially varying density and Lam\'e parameters and construct an explicit local lattice discretization in terms of coupled oscillators. The resulting second-order equations of motion are embedded into a first-order Schrödinger evolution generated by a sparse Hermitian Hamiltonian, enabling their simulation using block-encoding and quantum Hamiltonian-simulation techniques. We explicitly construct the required block-encoding from value-oracle access to the local material parameters. Our end-to-end analysis accounts for the errors and costs associated with spatial discretization, state preparation, Hamiltonian simulation, and observable extraction. It shows that the potential quantum advantage in the continuum setting at fixed volume depends strongly on the regularity of the material parameters and the solution, which limits the applicable discretization order and determines how the number of lattice sites $N$ scales with the target accuracy $\epsilon$. High regularity permits efficient high-order classical discretizations and therefore limits the potential advantage, whereas lower regularity yields a more favorable quantum-to-classical complexity comparison. This trade-off between the potential quantum advantage and problem regularity is illustrated schematically in Fig.~\ref{fig:advantage-regimes}. For example, when regularity permits only a second-order spatial discretization, the algorithm provides at most a quadratic speedup with respect to simulation time $\bar{t}$, while the dependence on the target error $\epsilon$ improves only from inverse-quadratic to inverse-three-halves scaling. As higher discretization orders become applicable, the potential advantage becomes progressively smaller, while lower regularity problems admit a larger quantum advantage. This is in accord with recent work that gives an end-to-end quantum algorithm for the three-dimensional neutron-diffusion eigenvalue problem with piecewise-constant material coefficients, showing how low solution regularity can increase uniform-mesh costs~\cite{childs2026quantum}.

The more favorable complexity comparison obtained at lower regularity points naturally toward the limiting case in which material variations occur on microscopic length scales. When these length scales become comparable to the spacing between the material's constituents, the separation of scales underlying continuum elasticity can be lost, and an atomistic or lattice model becomes the appropriate description. In this microscopic lattice regime, the number of degrees of freedom $N$ grows with the physical size of the system, rather than through numerical refinement of a continuum approximation within a fixed physical volume. The comparison is therefore between the quantum and classical costs of simulating the same physical lattice system. In this setting, we define [microscopically resolved bulk observables (MRBOs) that cannot be resolved by a known continuum model, but can be efficiently extracted without a size-dependent overhead from the quantum state representing the lattice system. We also give an explicit efficient encoding for a class of complex pseudorandom heterogeneous geometries that provide a concrete realization of the material-access model rather than leaving it solely as an abstract oracle assumption. Under these conditions, we show that one can expect a quartic quantum speedup in the scaling with system size, making the estimation of MRBOs in the microscopic lattice regime the most promising regime for quantum advantage identified in this work (see Fig.~\ref{fig:advantage-regimes}).

The remainder of the paper is organized as follows. In Sec.~\ref{sec:continuum_setup}, we formulate the continuum elasticity problem and describe the complexity of its classical simulation. We also explain how to map the continuum dynamics to a local system of coupled oscillators. In Sec.~\ref{sec:continuum_algo}, we develop the corresponding quantum algorithm by embedding the oscillators' equation of motion into a Schr\"odinger evolution and using the Hamiltonian simulation. For that, we also construct the required block-encodings, initial-state preparation, and observable extraction subroutines, and finish with the analysis of the resulting end-to-end complexity and a discussion of the quantum speedup regime. In Sec.~\ref{sec:lattice_setup}, we turn to microscopic lattice systems beyond continuum elasticity and discuss criteria for continuum breakdown. We also introduce the notion of MRBOs as quantities that require a full lattice simulation to be estimated accurately while retaining a signal independent of $N$, and establish sufficient criteria for their efficient quantum readout. In Sec.~\ref{sec:lattice_algo}, we then show how heterogeneous elastic media can be generated and accessed efficiently on a quantum computer and examine an explicit pseudorandom lattice model. We also analyze the end-to-end quantum and classical complexities in this setting and delineate the regime in which a quartic quantum speedup in system size can be sustained. We conclude in Sec.~\ref{sec:discussion} with a discussion of the results, limitations, and possible extensions.


\section{Continuum elasticity}
\label{sec:continuum_setup}


\subsection{Model}
\label{sec:continuum_model}

The macroscopic state of a three-dimensional isotropic material in classical elasticity theory is described by the displacement vector field $\bar{{u}}(\bar{{x}},\bar{t})$, where $\bar{{x}}$ denotes position and $\bar{t}$ denotes time. Here, and throughout, we place an overbar on selected quantities expressed in physical units to distinguish them from the corresponding dimensionless quantities used later in our quantum algorithm, for which we use the same symbols without an overbar for convenience. The displacement field evolves according to the linear elastodynamic equation for a heterogeneous isotropic elastic medium~\cite{Love_book,Landau_book}:
\begin{equation}
    \label{eq:wave_eq}
    \bar{\rho} \frac{\partial^2\bar{{u}}}{\partial \bar{t}^2}=({\lambda}+{\mu})\nabla (\nabla \cdot \bar{{u}}) +(\nabla\cdot\bar{{u}})\nabla{\lambda}+\left[(\nabla\bar{{u}})^\mathsf{T}+\nabla\bar{{u}}\right]\nabla{\mu}+ {\mu}  \nabla^2 \bar{{u}},
\end{equation}
where $\bar{\rho}(\bar{{x}})$ is the density of the material, while ${\lambda}(\bar{{x}})$ and ${\mu}(\bar{{x}})$ are the \emph{elastic moduli}, also known as the \emph{Lam\'e parameters}. These encode the elastic properties of the material: ${\mu}$ measures the material's resistance to changes in shape at constant volume, whereas ${\lambda+2\mu/3}$ (known as the bulk modulus) measures the material's resistance to expansion or compression (i.e., changes in volume). The physical ranges of the Lam\'e parameters are ${\mu}> 0$ and ${\lambda}\geq -(2/3){\mu}$. In the special case when $\bar{\rho}$, ${\lambda}$ and ${\mu}$ are constant in~$\bar{{x}}$, Eq.~\eqref{eq:wave_eq} simplifies and gives the \emph{elastic wave equation}, which is homogeneous and reads
\begin{align}
    \label{eq:wave_eq_hom}
    \bar{\rho} \frac{\partial^2\bar{{u}}}{\partial \bar{t}^2}=({\lambda}+{\mu})\nabla (\nabla \cdot \bar{{u}}) + {\mu}  \nabla^2 \bar{{u}}.
\end{align}
Then, the \emph{compressional}- or \emph{longitudinal wave speed} is ${v}_l=\sqrt{({\lambda}+2{\mu})/\bar{\rho}}$, whereas the \emph{shear}- or \emph{transverse wave speed} is ${v}_t=\sqrt{{\mu}/\bar{\rho}}$. Furthermore, in the above homogeneous case, we have the scalar ${\gamma}=\nabla\cdot\bar{{u}}$ satisfying the wave equation in its standard form. To see this, we take the divergence of Eq.~\eqref{eq:wave_eq_hom} and get
\begin{align}
    \label{eq:gamma_evol}
    \frac{\partial^2{\gamma}}{\partial \bar{t}^2}=c^2\nabla^2{\gamma},
\end{align}
where $c^2=({\lambda}+2{\mu})/\bar{\rho} $. This means that, if ${\gamma}=\partial{{\gamma}}/\partial \bar{t}=0$ at $\bar{t}=0$, then ${\gamma}=0$ for all $\bar{t}>0$, and so Eq.~\eqref{eq:wave_eq_hom} simplifies to the standard three-dimensional wave equation with wave speed $\sqrt{{\mu}/\bar{\rho}}$, for which a quantum algorithm was developed in Ref.~\cite{costa2019quantum}.

Classical simulations of Eq.~\eqref{eq:wave_eq} are typically performed using spatial finite-difference, finite-volume, or finite-element discretizations combined with explicit time integration~\cite{Virieux1986,Komatitsch1999,ZhangWang2022}. Rather than provide an exhaustive classical analysis, we consider a standard finite-difference method and discuss how spatial regularity in this setting constrains the potential quantum speedup. For concreteness, we consider a time-domain simulation in which $N$ degrees of freedom are propagated forward in time. The number of degrees of freedom is in turn related to the target spatial discretization error threshold $\epsilon_{\rm SD}$ in some functional form $N= N(\epsilon_{\rm SD})$, and if time-step size is $\Delta t$ then the classical complexity for the time-stepping method is
\begin{equation}
    \comp_C
    =\O\!\left(
     \frac{\bar{t}}{\Delta t}N(\epsilon_{\rm SD})
    \right).
\end{equation}
We shall now consider a situation where the dynamics has spatial regularity of some degree that can be exploited by a spatial discretization method. We focus on a fixed three-dimensional physical domain, and a spatial discretization with characteristic mesh spacing $a$ containing $N=\Theta(a^{-3})$ degrees of freedom. Assuming sufficiently smooth material parameters and solutions, a stable $m$-th-order spatial discretization has global error
\begin{align}
    \epsilon_{\rm SD}=\O(c_m(\bar{t})a^m),
\end{align}
 while an $m'$-th-order explicit time integrator has global error
\begin{align}
    \epsilon_{\rm TD}=\O(c_{m'}(\bar{t})\Delta t^{\,m'}).
\end{align}
Here $c_m(\bar{t})$ and $c_{m'}(\bar{t})$ are independent of the spatial and temporal discretization scales, $a$ and $\Delta t$, respectively. Additionally, assuming uniform-in-time bounds on the relevant derivatives of the exact solution and uniform stability of the numerical evolution, the discretization errors accumulate at most linearly in simulation time $\bar{t}$, such that $c_m(\bar{t})=\O(\bar{t})$ and $c_{m'}(\bar{t})=\O(\bar{t})$. Moreover, explicit stability requires the CFL condition $ \Delta t=\O(a) = \O(N^{-\frac{1}{3}})$. We assume $\Delta t = \Theta(N^{-\frac{1}{3}})$, and since each time step requires $\O(N)$ operations for a local order-$m$ discretization (with constant $m$) and the total number of time-steps is $\bar{t}/\Delta t$, the resulting classical runtime complexity of the method is 
\begin{equation}
    \comp_C
    =\O\!\left(
    \bar{t}\,
    N^{\frac{4}{3}}
    \right).
\end{equation}
We now assume the global spatial and temporal discretization errors obey $\epsilon_{\rm SD}= \O(\epsilon)$ and $\epsilon_{\rm TD}= \O(\epsilon)$, which are satisfied if $\bar{t} a^m = \O(\epsilon)$ and $\bar{t} \Delta t^{m'}= \O(\epsilon)$, with $a,\Delta t \rightarrow 0$. The scaling of the number of lattice points on the cubic grid with $a$ implies the first condition is
\begin{equation}
    \bar{t} N^{-\frac{m}{3}} = \O(\epsilon),
\end{equation}
while the assumed scaling of $\Delta t$ implies the second condition is
\begin{equation}
    \bar{t} N^{-\frac{m'}{3}} = \O(\epsilon),
\end{equation}
as $N\rightarrow \infty$. These together imply that the conditions hold if
\begin{equation}
    \label{eq:temp_eq}
    N = \Omega\left (\left(\frac{\bar{t}}{\epsilon} \right)^{\max \{\frac{3}{m},\frac{3}{m'} \}} \right ).
\end{equation}
Taking the optimal $N$ scaling implies
\begin{equation}
    \comp_C
    =\O\!\left(
    \bar{t}\,
    \left (\frac{\bar{t}}{\epsilon} \right)^{\max \{\frac{4}{m},\frac{4}{m'} \}} 
    \right).
    \label{eq:comp_C}
\end{equation}
This provides the classical runtime complexity of the finite-difference algorithm.


\subsection{Discrete oscillator mapping}
\label{sec:discrete_mapping}

The continuum elasticity theory introduced above can be understood as the long-wavelength limit of a more fundamental lattice dynamics. Appendix~\ref{app:elastic-theory} develops this lattice-based formulation in detail, starting from microscopic displacement degrees of freedom, strain-displacement and constitutive maps, and showing how continuum strain, stress, and elastic equations of motion emerge in the long-wavelength limit. As explained there, this correspondence is not unique: a family of distinct lattice models can reproduce the same continuum elastic theory while differing in their behavior at wavelengths comparable to the lattice spacing. Here, we introduce one particular model of coupled oscillators that is especially well suited to quantum simulation, and whose long-wavelength dynamics reproduces the continuum elasticity. Its construction and correspondence with the continuum theory are derived in detail in Appendix~\ref{app:micro}. 

Consider a Cartesian three-dimensional periodic cubic lattice with unit vectors $\{\v{e}_1,\v{e}_2,\v{e}_3\}$, where each site is characterized by a position index vector $\v{n}=(n_1,n_2,n_3)$ belonging to the set of all $N$ site vectors $\v{n}\in\Lambda := \{1,\dots,N^{1/3}\}^3$, and we assume $N^{1/3}$ is an integer. An oscillator at site $\v{n}$ has a mass ${\rho}_{\v{n}}$, and oscillators are coupled locally to each other via two types of cell interactions: deviatoric (penalizing changes in local shape) and volumetric (penalizing changes in local volume). We consider the evolution of the system of oscillators, where the displacement ${u}_{\v{n},i}$ in direction $i$ of the oscillator at site $\v{n}$ evolves according to
\begin{align}
    \label{eq:discrete_eom}
    {\rho}_{\v{n}}\frac{d^2 {u}_{\v{n},i}}{d {t}^2}
    =
    & \frac{1}{2}\sum_{j=1}^3
    \left[
    {\kappa}^{\dev}_{\v{n}+\v{e}_j}
    {\xi}^\dev_{{\v{n}+\v{e}_j},ij}
    -
    {\kappa}^{\dev}_{\v{n}-\v{e}_j}
    {\xi}^\dev_{{\v{n}-\v{e}_j},ij}
    \right] 
    +
    \frac{1}{2}    \left[
    {\kappa}^{\vol}_{\v{n}+\v{e}_i}
    {\xi}^\vol_{\v{n}+\v{e}_i}
    -
    {\kappa}^{\vol}_{\v{n}-\v{e}_i}
    {\xi}^\vol_{\v{n}-\v{e}_i}
    \right],
\end{align}
where the discrete strains are given by
\begin{subequations}
\begin{align}
    {\xi}^\dev_{\v{n},ij}&= \frac{1}{4}\left[{u}_{\v{n}+\v{e}_j,i}-{u}_{\v{n}-\v{e}_j,i}+{u}_{\v{n}+\v{e}_i,j}-{u}_{\v{n}-\v{e}_i,j}\right]
    -\frac{\delta_{ij}}{6} \sum_{k=1}^3\left[{u}_{\v{n}+\v{e}_k,k}-{u}_{\v{n}-\v{e}_k,k}\right],
    \label{eq:strain_dev}\\
    {\xi}^\vol_{\v{n}}&= \frac{1}{2} \sum_{k=1}^3\left[{u}_{\v{n}+\v{e}_k,k}-{u}_{\v{n}-\v{e}_k,k}\right].
    \label{eq:strain_vol}    
\end{align}
\end{subequations}
This dimensionless lattice model is then related to the continuum elasticity model discussed in the previous section in the following way. The physical position $\bar{\v{x}}$ is related to a site position $\v{n}$ via $\bar{\v{x}}=a\v{n}$, where $a$ is constant lattice spacing. The dimensionless quantities are related to the physical ones by:
\begin{equation}
    \label{eq:non_dim}
    u_{\v{n},i}, = \frac{\bar{u}_i(a\v{n})}{u_0},\qquad    {\kappa}^\dev_{\v{n}}= 
    \frac{2{\mu}(a\v{n})}{\lambda_0},\qquad
    {\kappa}^\vol_{\v{n}} = \frac{{\lambda}(a\v{n})+\frac{2}{3}{\mu}(a\v{n})}{\lambda_0},\qquad
    {\rho}_{\v{n}} = \frac{\bar{\rho}(a \v{n})}{\rho_0},\qquad t = \frac{\bar{t}}{a\sqrt{\frac{\rho_0}{\lambda_0}}},
\end{equation}
where $u_0$, $\lambda_0$, and $\rho_0$ are reference scales introduced for nondimensionalization and chosen such that $|u_{\v{n},i}|$, ${\kappa}^\dev_{\v{n}}$, ${\kappa}^\vol_{\v{n}}$, and $1/\rho_{\v{n}}$ are less than or equal to 1 for all $\v{n}$.

The above discrete mapping introduces a discretization error $\epsilon_{\rm DM}$ independently of how the discrete equations of motion from Eq.~\eqref{eq:discrete_eom} are subsequently solved. As shown in Appendix~\ref{app:micro}, comparing the dimensionless discrete equations to non-dimensionalized continuum elasticity equations yields a consistency error of order $\O(a^4)$. Transforming back to physical variables, this yields an error of order $\O(a^2)$. Assuming sufficiently smooth material parameters and displacement fields, together with uniform stability of the discrete evolution, the resulting solution error accumulates at most linearly with physical simulation time $\bar{t}$. Consequently, the solution of the discrete model, when mapped back to physical variables, approximates the solution of the continuum elasticity equation with error
\begin{equation}
    \label{eq:discrete_mapping_error}
    \epsilon_{\rm DM}
    =
    \O(\bar{t} a^2)
    =
    \O(\bar{t} N^{-2/3}),
\end{equation}
where in the last equality we used $a=\Theta(N^{-1/3})$ for a fixed three-dimensional physical volume. Hence, to achieve the discretization error $\epsilon_{\rm DM}$ it suffices to choose
\begin{equation}
    N
    =
    \Theta\!\left[
        \left(
            \frac{\bar{t}}{\epsilon_{\rm DM}}
        \right)^{3/2}
    \right].
\end{equation}
This error is intrinsic to the discrete mapping and is present even if the discrete equations are solved exactly. Any additional error arising from an approximate numerical or quantum solution of the discrete dynamics is independent of this mapping error and can be combined with it using the
triangle inequality.

Although the discrete model above is second-order accurate, a straightforward generalization (sketched in Appendix~\ref{app:kordermodel}) extends it to $m$-th order. In such a case, the above expressions would be replaced by
\begin{equation}
    \label{eps_DM_k}
    \epsilon^{(m)}_{\rm DM}
    =
    \O(\bar{t} a^m)
    =
    \O(\bar{t} N^{-m/3}),\qquad N
    =
    \Theta\!\left[
        \left(
            \frac{\bar{t}}{\epsilon_{\rm DM}^{(m)}}
        \right)^{3/m}
    \right].
\end{equation}


\section{Quantum algorithm for continuum elasticity}
\label{sec:continuum_algo}

In this section, we present a quantum algorithm for simulating the continuum elastic dynamics, which is achieved in a series of steps. First, in Sec.~\ref{sec:ham_emb}, we embed the discretized model of continuum elasticity from Eq.~\eqref{eq:discrete_eom} into a Schr\"{o}dinger evolution generated by a Hamiltonian $H$, and employ a Hamiltonian simulation technique based on Quantum Singular Value Transformation (QSVT)~\cite{gilyen2019qsvt, lin2022lecturenotesquantumalgorithms, martyn2021grand}. This simulation technique requires the construction of a unitary block-encoding of the Hamiltonian $H$, which we describe in Sec.~\ref{sec:BE}, and an efficient preparation of a quantum state encoding the initial conditions of the problem, which we describe in Sec.~\ref{sec:ini_state}. Having established how to prepare a quantum state encoding the continuum elastic system at a final time, we then discuss in Sec.~\ref{sec:extraction} how physically relevant observables can be extracted from it. Finally, in Sec.~\ref{sec:continuum_complexity}, we present the end-to-end computational complexity of our quantum algorithm and discuss regimes of quantum speedup.


\subsection{Hamiltonian embedding}
\label{sec:ham_emb}

We start by defining the Hilbert spaces relevant for quantum encoding of the problem at hand. First, we have the vertex space 
\begin{equation}
    \H_V=\text{span}\{|\v{n}\rangle= |n_1, n_2, n_3\rangle:\: \v{n}\in\Lambda\},    
\end{equation}
so that $\text{dim}(\H_V)=N$. Next, the two types of interactions are each assigned their spaces:
\begin{align}
    \H_I^\dev & =\text{span}\left\{|e^\dev_{\v{n},ij}\rangle:\: \v{n}\in\Lambda;\: i,j\in \{1,2,3\} \right\}, \\
    \H_I^\vol \ &=\text{span}\left\{|e^\vol_{\v{n}}\rangle:\: \v{n}\in\Lambda\right\}.
\end{align} 
The dimensions of these spaces are $\text{dim}(\H^\dev_I)=9N$ and $\text{dim}(\H^\vol_I)=N$, respectively. The two interaction spaces are combined into a single larger space 
\begin{align}
    \H_I = \H_I^\dev \oplus \H_I^\vol,
\end{align} so that vectors corresponding to the two interaction types are orthogonal and can be treated within a common Hilbert space with $\text{dim}(\H_I)=10N$. Moreover, since each vertex carries the same set of interaction labels, we can decompose the interaction space as
\begin{equation}
    \H_I \simeq \H_V \ot \H_S,
\end{equation}
and analogously for $\H_I^\dev$ and $\H_I^\vol$, where
\begin{align}
    \H_S = \mathrm{span} \left\{ \{\ket{\mathrm{dev}, i,j}: \; i,j \in \{1,2,3\}\} \cup \{ \ket{\mathrm{vol}}\}\right\}  .
\end{align}
Finally, we shall have a register for the Cartesian direction index as well, for which we define the corresponding Hilbert space as 
\begin{align}
    \H_3=\text{span}\{|i\rangle:\: i\in\{1,2,3\}\}.
\end{align}

We construct the displacement vector $u(t)\in \H_V \otimes \H_3$ and the mass matrix $M: \H_V \otimes \H_3 \to \H_V \otimes \H_3$ as
\begin{equation}
    \label{eq:u_and_M_2}
    u(t) =\sum_{\v{n}\in\Lambda} \sum_{i=1}^3 {u}_{\v{n},i}(t) |\v{n},i\rangle ,\qquad
    M= \sum_{\v{n}\in\Lambda} {\rho}_{\v{n}}   \ketbra{\v{n}}{\v{n}}\otimes I_3,
\end{equation}
where $I_3$ is the identity operator on $\H_3$. The discrete strains defined in Eqs.~\eqref{eq:strain_dev}-\eqref{eq:strain_vol} determine a strain-displacement map 
\begin{equation}
    D=\begin{pmatrix}
        D^\dev \\
        D^\vol
    \end{pmatrix}:\H_V\otimes\H_3\longrightarrow\H_I
\end{equation}
where
\begin{subequations}
\label{eq:strain_displacement_components}
\begin{align}
    \!\!\!D^\dev
    &\!=\!
    \sum_{\v{n}\in\Lambda}
    \sum_{i,j=1}^{3}
    |\v{e}_{\v{n},ij}^\dev\rangle
    \Bigg[
        \frac{1}{4}
        \Big(
            \langle\v{n}\!+\!\v{e}_j,i|
            -
            \langle\v{n}\!-\!\v{e}_j,i|
            +
            \langle\v{n}\!+\!\v{e}_i,j|
            -
            \langle\v{n}\!-\!\v{e}_i,j|
        \Big)
        -
        \frac{\delta_{ij}}{6}
        \sum_{k=1}^{3}
        \Big(
            \langle\v{n}\!+\!\v{e}_k,k|
            -
            \langle\v{n}\!-\!\v{e}_k,k|
        \Big)
    \Bigg],
    \label{eq:D_dev}
    \\
    \!\!\!D^\vol
    &\!=\!
    \frac{1}{2}
    \sum_{\v{n}\in\Lambda}
    |\v{e}_{\v{n}}^\vol\rangle
    \sum_{k=1}^{3}
    \Big(
        \langle\v{n}\!+\!\v{e}_k,k|
        -
        \langle\v{n}\!-\!\v{e}_k,k|
    \Big).
    \label{eq:D_vol}
\end{align}
\end{subequations}
By construction, it satisfies
\begin{align}
    D u
    &=
    \sum_{\v{n}\in\Lambda}
    \sum_{i,j=1}^{3}
    \xi_{\v{n},ij}^\dev
    |\v{e}_{\v{n},ij}^\dev\rangle
    +
    \sum_{\v{n}\in\Lambda}
    \xi_{\v{n}}^\vol
    |\v{e}_{\v{n}}^\vol\rangle.
    \label{eq:discrete_strain_displacement}
\end{align}
The spatially varying stiffness coefficients, ${\kappa}^\dev_{\v{n}}$ and ${\kappa}^\vol_{\v{n}}$, can be encoded in the constitutive operator
\begin{equation}
    C=C^\dev\oplus C^\vol:\H_I\longrightarrow\H_I,
\end{equation}
where
\begin{align}
    C^\dev
    &=
    \sum_{\v{n}\in\Lambda}
    \sum_{i,j=1}^{3}
    \kappa^\dev_{\v{n}}
    \ketbra{\v{e}_{\v{n},ij}^\dev}
            {\v{e}_{\v{n},ij}^\dev},
    \qquad
    C^\vol
    =
    \sum_{\v{n}\in\Lambda}
    \kappa^\vol_{\v{n}}
    \ketbra{\v{e}_{\v{n}}^\vol}
            {\v{e}_{\v{n}}^\vol}.
\end{align}
Then, introducing the stiffness operator,
\begin{equation}
    \label{eq:stiffness_operator}
    K:=D^\dagger C D,
\end{equation}
the lattice equation of motion from Eq.~\eqref{eq:discrete_eom} takes the form
\begin{equation}
    \label{eq:eom_for_u_K}
    M\Ddot{u}(t)=-Ku(t).
\end{equation}

To obtain a factorization $K=B^\dagger B$ suitable for the Hamiltonian embedding, we define
\begin{equation}
    \label{eq:B_factorization}
    B:=C^{1/2} D :
    \H_V\otimes\H_3\longrightarrow \H_I.
\end{equation}
Because both the interaction space and the constitutive operator decompose into deviatoric and volumetric sectors, we have
\begin{equation}
    B = \begin{pmatrix}
        B^\dev \\
        B^\vol
    \end{pmatrix},
    \qquad
    B^\dev:=(C^\dev)^{1/2}D^\dev,
    \qquad
    B^\vol:=(C^\vol)^{1/2}D^\vol.
\label{eq:Brectangular}
\end{equation}
It follows that Eq.~\eqref{eq:eom_for_u_K} can equivalently be written as
\begin{equation}
    \label{eq:eom_for_u}
    M\Ddot{u}(t)=-B^\dagger B u(t).
\end{equation}
Finally, we perform the mass rescaling
\begin{equation}
    \label{eq:scaling_def}
    y(t)=M^{1/2}u(t),
    \qquad
    \tilde B=BM^{-1/2},
\end{equation}
and analogously for the terms $B^\dev$ and $B^\vol$. This enables us to rewrite the equation of motion as
\begin{equation}
    \label{eq:scaled_matrix_ode}
    \Ddot{y}(t)
    =
    -\tilde B^\dagger \tilde B y(t).
\end{equation}
    
Following the strategy of Refs.~\cite{costa2019quantum,babbush2023exponential}, we then define the Hamiltonian 
\begin{align}
    \label{eq:ham}
    H=\begin{pmatrix}
        0&\tilde{B}^\dagger\,\\
        \tilde{B}&0
    \end{pmatrix},
\end{align}
and note that a normalized quantum state
\begin{align}
    \label{eq:psi_t}
    |\psi(t)\rangle\propto
    \begin{pmatrix}
        \dot{y}(t)\\
      -\i\tilde{B} y(t)
    \end{pmatrix},
\end{align}
evolves under the Schr\"{o}dinger equation according to this Hamiltonian:
\begin{align}
    \label{eq:Schrodinger}
    \frac{d}{dt}|\psi(t)\rangle = -\i H |\psi(t)\rangle\qquad \Longrightarrow\qquad |\psi(t)\rangle = \exp(-\i Ht)|\psi(0)\rangle.
\end{align}
This evolution takes place in the Hilbert space 
\begin{equation}
    \label{eq:hilbert}
    \H=(\H_V\otimes\H_3)\oplus \H_I \cong \H_V\ot (\H_3 \oplus \H_S)  ,
\end{equation}
which has dimension $\text{dim}(\H)=13N$. Note that the kinetic energy $T(t)$ and potential energy $U(t)$ of the discrete oscillator system are given by
\begin{subequations}
\begin{align}
    \label{eq:kinetic}
    T(t)&= \frac{1}{2} \dot{u}^\mathsf{T}(t) M \dot{u}(t) = \frac{1}{2}\norm{\dot{y}(t)}^2,\\
    U(t)&= \frac{1}{2} u^\mathsf{T}(t) B^\dagger B u(t)=\frac{1}{2}\norm{-\i\tilde{B} y(t)}^2,
    \label{eq:potential}
\end{align}
\end{subequations}
with the total energy $E=T(t)+U(t)$ being conserved as long as $u(t)$ satisfies the equation of motion from Eq.~\eqref{eq:eom_for_u}. Hence, the normalization factor in Eq.~\eqref{eq:psi_t} is in fact $1/\sqrt{2E}$. 

Now, assuming access to a $(\alpha_H,n_H,0)$-block-encoding $U_H$ of the Hamiltonian $H$, the number of queries to $U_H$ needed to block-encode the evolution operator $\exp(-\i Ht)$ up to $l_2$-type error $\epsilon_{\text{HS}}$ is~\cite{Hamiltonian_simulation_2}:
\begin{align}
    Q_{\text{HS}}= \O\left(\alpha_H t + \log\left(\frac{1}{\epsilon_{\text{HS}}}\right)\right).
\end{align}
Hence, expressing the complexity in \emph{physical time} $\bar{t}$ (recall Eq.~\eqref{eq:non_dim}), and employing the fact that the block-encoding prefactor is a constant, as we show in the next section, we get
\begin{equation}
    \label{eq:Q_HS}
    Q_{\text{HS}} = \O\left(\bar{t}\:N^{1/3}
    + \log\left(\frac{1}{\epsilon_{\text{HS}}}\right)\right).
\end{equation}


\subsection{Block-encoding}
\label{sec:BE}

Our aim here is to unitarily block-encode the Hamiltonian $H$ from Eq.~\eqref{eq:ham}, which we will achieve in a few steps. First, we embed the rectangular matrix $B$ into a square matrix (for simplicity, we keep the same notation):
\begin{equation}
    B = \begin{pmatrix}
        B^\dev & 0\\ 
        B^\vol & 0
    \end{pmatrix},
\end{equation}
where the four blocks are labeled by a register $r$, e.g., $\bra{1}_r B \ket{0}_r =  B^\vol$.
Then, access to an $(\alpha_{\tilde{B}},n_{\tilde{B}},0)$-block-encoding $U_{\tilde{B}}$ of $\tilde{B}= B M^{-1/2}$ (where $M^{-1/2}$ is here understood as padded with zeroes) provides a block-encoding of $H$ (padded with zeroes) via
\begin{equation}
    U_H = \ketbra{0}{1}\ot U_{\tilde{B}}^\dagger+\ketbra{1}{0} \ot U_{\tilde{B}},
\end{equation}
with the same block-encoding prefactor and number of ancillary qubits, involving one call to $U_{\tilde{B}}$ and one call to its inverse. 

One can then construct a unitary block-encoding $U_{\tilde{B}}$ of $\tilde{B}$ from unitary block-encodings $U_{\tilde{B}^{\dev}}$ of $\tilde{B}^{\dev} = B^{\dev} M^{-1/2}$ and  $U_{\tilde{B}^{\vol}}$ of $\tilde{B}^{\vol} = B^{\vol} M^{-1/2}$. To do so, introduce two qubits $r$ and $f$ and define: 
\begin{align}
    R \ket{0}_r &= \frac{\alpha_{\tilde{B}^\dev}}{\alpha_{\tilde{B}}} \ket{0}_r + \frac{\alpha_{\tilde{B}^\vol}}{\alpha_{\tilde{B}}}\ket{1}_r, \\ 
    \mathrm{SELECT} &= \ketbra{0}{0}_r \otimes U_{\tilde{B}^{\dev}} + \ketbra{1}{1}_r \otimes U_{\tilde{B}^{\vol}}, \\
    U_{\tilde{B}} &= (\mathrm{SELECT} \otimes I_f) (R \otimes I \otimes I_f) \mathrm{CNOT}_{r \rightarrow f}.
\end{align}
The block-encoding parameters are
\begin{equation}
 \alpha_{\tilde{B}} = \sqrt{\alpha_{\tilde{B}^\dev}^2+ \alpha_{\tilde{B}^\vol}^2}
    \qquad
    n_{\tilde{B}} = \max\{n_{\tilde{B}^\dev},n_{\tilde{B}^\vol}\} +1,
\end{equation}
where the $+1$ refers to the flag qubit $f$. 

In order to block-encode ${\tilde{B}^{\dev}}$ and ${\tilde{B}^{\vol}}$, we first recall that these matrices can be factorized in the following way:
\begin{equation}
    \tilde{B}^{\mathrm{dev}}=\left(C^{\dev}\right)^{1/2}D^{\dev}
M^{-1/2},
\qquad
\tilde{B}^{\vol}=\left(C^{\vol}\right)^{1/2}D^{\vol}M^{-1/2},    
\end{equation}
This way all dependence on the material parameters is contained in the diagonal matrices $M^{-1/2}$, $(C^{\dev})^{1/2}$, and $(C^{\mathrm{vol}})^{1/2}$, whereas $D^{\dev}$ and $D^{\vol}$ encode only the spatial discretization.

We first consider the material-dependent diagonal matrices $M^{-1/2}$, $(C^{\dev})^{1/2}$, and $(C^{\vol})^{1/2}$. We assume coherent access to the dimensionless material parameters through a reversible fixed-precision value oracle
\begin{align}
    O_{\mathrm{mat}}:
    \ket{\v n}\ket{0}\ket{0}\ket{0}
    \longmapsto
    \ket{\v n}
    \ket{\rho_{\v n}}
    \ket{\kappa_{\v n}^{\dev}}
    \ket{\kappa_{\v n}^{\vol}},
    \label{eq:material-value-oracle}
\end{align}
implementable using $\polylog(N)$ gates. Such an oracle may arise, for example, when the material parameters are specified by efficiently evaluable analytic functions, piecewise descriptions, or succinct procedural generators, and we will discuss it in more detail in Sec.~\ref{sec:geometry}. We assume the $N$-independent bounds
\begin{equation}
    1\leq \rho_{\v n}\leq\rho_{\max}=\O(1),
    \qquad
    0\leq\kappa_{\v n}^{\dev}\leq1,
    \qquad
    0\leq\kappa_{\v n}^{\vol}\leq1,
    \label{eq:material-parameter-bounds}
\end{equation}
where the lower bound on $\rho_{\v n}$ and the upper bounds on $\kappa_{\v{n}}^{\dev}$ and $\kappa_{\v{n}}^{\vol}$ follow from our nondimensionalization. Using standard reversible fixed-point arithmetic, the oracle $O_{\mathrm{mat}}$ can be used to coherently compute
\begin{equation}
    \rho_{\v n}^{-1/2},\qquad
    \rho_{\v n}^{1/2},\qquad
    \bigl(\kappa_{\v n}^{\dev}\bigr)^{1/2},\qquad
    \bigl(\kappa_{\v n}^{\vol}\bigr)^{1/2}
    \label{eq:derived-material-values}
\end{equation}
to any prescribed precision, with gate complexity polynomial in the number of precision bits. The bounds in Eq.~\eqref{eq:material-parameter-bounds} ensure that the inverse operations do not introduce an $N$-dependent precision overhead. Recall that to block-encode a diagonal matrix with entries $a_{\v n}$ satisfying $|a_{\v n}|\leq1$, one can first compute $a_{\v n}$ into a work register, then perform a controlled rotation whose successful amplitude is $a_{\v n}$, and finally uncompute the work register. Applying this standard construction with $a_{\v n}$ given by quantities appearing in Eq.~\eqref{eq:derived-material-values} gives block-encodings of $M^{-1/2}$, $(C^{\dev})^{1/2}$, and $(C^{\vol})^{1/2}$, respectively. In the ideal value-oracle model, these are $(1,1,0)$-block-encodings with one signal ancilla; finite-precision arithmetic instead gives approximate block-encodings with controllable error. 

We next consider the discretization matrices $D^{\dev}$ and $D^{\vol}$, which are independent of the material parameters and can be block-encoded using the standard sparse-access construction~\cite{lin2022lecturenotesquantumalgorithms}. From their definitions above, and recalling Eqs.~\eqref{eq:D_dev}-\eqref{eq:D_vol}, their row and column sparsities are given by: 
\begin{equation}
    s_c(D^{\dev})=14,\qquad s_r(D^{\dev})=6,
    \qquad
    s_c(D^{\vol})=2,\qquad s_r(D^{\vol})=6,
\end{equation}
while their largest matrix elements satisfy
\begin{equation}
    \|D^{\dev}\|_{\max}=\frac{1}{3},
    \qquad
    \|D^{\vol}\|_{\max}=\frac{1}{2}.
\end{equation}
The standard sparse-access block-encoding of a matrix $A$ with row sparsity $s_r$, column sparsity $s_c$, and maximal entry $\|A\|_{\max}$ yields block-encoding prefactor $\alpha_A=\sqrt{s_r s_c}\,\|A\|_{\max}$. Therefore,
\begin{equation}
    \alpha_{D^{\dev}}=\frac{2\sqrt{21}}{3},
    \qquad
    \alpha_{D^{\vol}}=\sqrt{3}.
\end{equation}
Since both matrices have constant sparsity and their nonzero entries and locations are determined entirely by the fixed lattice geometry, the corresponding sparse-access oracles can be implemented using reversible index arithmetic with gate complexity $\polylog(N)$. Denoting the corresponding numbers of block-encoding ancillas by $n_{D^{\dev}}$ and $n_{D^{\vol}}$, both scale as $\O(\log N)$.

Putting all the above ingredients together, we can construct $(\alpha_H,n_H,0)$-block-encoding of $H$ with
\begin{equation}
    \label{eq:BE_params}
    \alpha_H =\sqrt{\frac{28}{3} + 3} = \sqrt{\frac{37}{3}}=\O(1),\qquad n_H=\O(\log N).
\end{equation}
Moreover, to assess the gate complexity of preparing $U_H$, note that a single application of $U_H$ involves one call to $U_{\tilde{B}}$ and one to $U_{\tilde{B}}^\dagger$. Each of these, in turn, contains one controlled call to each of $U_{\tilde{B}^{\dev}}$ and $U_{\tilde{B}^{\vol}}$ (or their inverses). Consequently, one application of $U_H$ requires four calls to the block-encoding of $M^{-1/2}$, two calls to that of $(C^{\dev})^{1/2}$, two calls to that of $(C^{\vol})^{1/2}$, and two calls each to the block-encodings of $D^{\dev}$ and $D^{\vol}$, where calls to a unitary and its inverse are counted equally. As described above, block-encodings of $D^{\dev}$ and $D^{\vol}$ require $\polylog(N)$ gates. Therefore, as long as the oracles encoding material parameters can be prepared with $\polylog(N)$ gates, the block-encoding of $H$ can be efficiently implemented.


\subsection{Initial state preparation}
\label{sec:ini_state}

We aim to prepare the normalized initial state
\begin{equation}
    \ket{\psi(0)}
    =
    \frac{1}{\sqrt{2E}}
    \begin{pmatrix}
        \dot y(0)\\
        -\i\tilde B y(0)
    \end{pmatrix}
    =
    \frac{1}{\sqrt{2E}}
    \begin{pmatrix}
        \sqrt{M}\dot u(0)\\
        -\i B u(0)
    \end{pmatrix}.
    \label{eq:target-initial-state}
\end{equation}
A prior construction (see Lemma~10 in Ref.~\cite{babbush2023exponential}) incurs an overhead $\sqrt{E_{\max}/E}$, where $E_{\max}$ is a worst-case block-encoding parameter. Following this protocol in our setting would give us a complexity $\Theta(\|u(0)\|/\sqrt{E}) = \O(\sqrt{N}/\sqrt{Na^2}) = \O(N^{1/3})$ (see Appendix~\ref{app:priorstateprep} for details). To avoid this overhead, we consider an alternative state preparation that exploits the assumed structures: smooth initial data and efficiently implementable high-precision value access. The corresponding overhead is reduced to a constant in this case. 

In addition to the material-value oracle $O_{\mathrm{mat}}$ from Eq.~\eqref{eq:material-value-oracle}, we assume coherent fixed-precision value-oracle access to the initial displacement and velocity:
\begin{align}
    O_u^{\mathrm{val}}:
    \ket{\v n,i}\ket{0}
    &\longmapsto
    \ket{\v n,i}\ket{u_{\v n,i}(0)},
    \label{eq:value-oracle-u}\\
    O_{\dot u}^{\mathrm{val}}:
    \ket{\v n,i}\ket{0}
    &\longmapsto
    \ket{\v n,i}\ket{\dot u_{\v n,i}(0)},
    \label{eq:value-oracle-udot}
\end{align}
implementable using $\polylog(N)$ gates, as is the case, e.g., when the initial fields are specified by efficiently computable analytic functions, piecewise-polynomial descriptions, or other succinct classical circuits. Here we assume for simplicity exact access, but we could also extend to include data loading precision, which would add logarithmic factors in the inverse error.   As described in Sec.~\ref{sec:BE}, $O_{\mathrm{mat}}$, together with reversible fixed-point arithmetic, provides the values $\sqrt{\rho_{\v n}}$, $\sqrt{\kappa_{\v n}^{\dev}}$, and $\sqrt{\kappa_{\v n}^{\vol}}$ needed to evaluate the components of Eq.~\eqref{eq:target-initial-state}. 

Introduce the disjoint index set 
\begin{equation} 
    \mathcal J = \mathcal J_T \sqcup \mathcal J_{\mathrm{dev}} \sqcup \mathcal J_{\mathrm{vol}}, 
    \label{eq:initial-index-set} 
\end{equation} 
where 
\begin{align} 
    \mathcal J_T &:= \{(\v n,i): \v n\in\Lambda,\ i\in\{1,2,3\}\},\\ 
    \mathcal J_{\mathrm{dev}} &:= \{(\v n,i,j): \v n\in\Lambda,\ i,j\in\{1,2,3\}\},\\ \mathcal J_{\mathrm{vol}} &:= \{\v n: \v n\in\Lambda\}. 
\end{align} 
Its cardinality is 
\begin{equation} 
    |\mathcal J|=3N+9N+N=13N. 
\end{equation} 
For each $\ell\in\mathcal J$, define the corresponding component of the unnormalized target vector by 
\begin{equation} 
    f_\ell := 
    \begin{cases} 
        \sqrt{\rho_{\v n}}\, \dot u_{\v n,i}(0), & \ell \in \mathcal{J}_T, \\[1mm] 
        -\i\sqrt{\kappa_{\v n}^{\mathrm{dev}}}\, \xi_{\v n,ij}^{\mathrm{dev}}[u(0)], & \ell \in \mathcal{J}_{\mathrm{dev}}, \\[1mm] 
        -\i\sqrt{\kappa_{\v n}^{\mathrm{vol}}}\, \xi_{\v n}^{\mathrm{vol}}[u(0)], & \ell \in \mathcal{J}_{\mathrm{vol}}. 
    \end{cases} 
    \label{eq:initial-components} 
\end{equation} 
With this notation,
\begin{equation} 
    \ket{\psi(0)} = \frac{a}{\sqrt{2E}} \sum_{\ell\in\mathcal J} q_\ell\ket{\ell} ,\qquad q_\ell:=\frac{f_\ell}{a}.
    \label{eq:initial-state-components} 
\end{equation} 

Now note that, for smooth continuous initial data,  each $q_\ell$ is of order $\O(1)$. Indeed, Eq.~\eqref{eq:non_dim} and the strain expansions from Appendix~\ref{app:micro} (see Eqs.~\eqref{eq:vol_expansion} and \eqref{eq:dev_expansion}) imply
\begin{align}
    \frac{\dot u_{\v n,i}(0)}{a}
    &=
    \frac{1}{u_0}\sqrt{\frac{\rho_0}{\lambda_0}}\,
    \frac{\partial\bar u_i(a\v n,0)}{\partial\bar t},
    \nonumber\\
    \frac{\xi_{\v n}^{\vol}[u(0)]}{a}
    &=
    \frac{1}{u_0}\operatorname{tr}\varepsilon(a\v n,0)
    +\O(a^2),
    \qquad
    \frac{\xi_{\v n,ij}^{\dev}[u(0)]}{a}
    =
    \frac{1}{u_0}\varepsilon_{ij}^{\dev}(a\v n,0)
    +\O(a^2).
    \label{eq:scaled-initial-data}
\end{align}
Thus, if the physical initial velocity and strain fields are uniformly bounded, there exists a known constant $C_{\mathrm{ini}}>0$, independent of $a$ and $N$, such that
\begin{equation}
    |q_\ell|\leq C_{\mathrm{ini}}
    \qquad\text{for all }\ell\in\mathcal J.
    \label{eq:initial-entry-bound}
\end{equation}
Each $q_\ell$ can be evaluated coherently with $\O(1)$ calls to the initial-data and material-value oracles and $\O(\log N)$ gates for reversible index arithmetic. Moreover,
\begin{equation}
    \frac{2E}{a^2}
    =
    \sum_{\ell\in\mathcal J}|q_\ell|^2,
\end{equation}
and, for a fixed physical initial condition carrying a nonzero amount of kinetic or elastic energy over a region of nonzero physical volume, the sampled values satisfy $\sum_\ell|q_\ell|^2=\Theta(N)$. Consequently,
\begin{equation}
    \frac{2E}{a^2}=\Theta(N).
    \label{eq:initial-energy-scaling}
\end{equation}

Next, we prepare a uniform superposition over the $13N$ valid component labels, 
\begin{equation} 
    \ket{0} \longmapsto \frac{1}{\sqrt {13N}} \sum_{\ell\in\mathcal J} \ket{\ell}. 
    \label{eq:uniform-component-superposition} 
\end{equation} 
Using the initial-data and material-value oracles described above, we coherently calculate $q_\ell$ in a work register:
\begin{equation}
    \frac{1}{\sqrt {13N}}
    \sum_{\ell\in\mathcal J}\ket{\ell}\ket{0}
    \longmapsto
    \frac{1}{\sqrt {13N}}
    \sum_{\ell\in\mathcal J}\ket{\ell}\ket{q_\ell}.
    \label{eq:coherent-q-evaluation}
\end{equation}
We then append a success qubit and perform a controlled rotation whose successful amplitude is $q_\ell/C_{\mathrm{ini}}$ (or alternatively use quantum rejection sampling/inequality testing methods). Including the sign of the kinetic entries and the phase $-\i$ of the potential entries, this implements 
\begin{equation}   
    \ket{q_\ell}\ket{0} \longmapsto \ket{q_\ell} \left[ \sqrt{1-\frac{|q_\ell|^2}{C_{\mathrm{ini}}^2}} \ket{0} + \frac{q_\ell}{C_{\mathrm{ini}}} \ket{1} \right]. 
    \label{eq:initial-controlled-rotation} 
\end{equation} 
Finally, we uncompute the value and work registers. Denote by $\mathcal A_{\mathrm{ini}}$ the unitary comprising the uniform superposition preparation, the coherent evaluation of $q_\ell$, the controlled rotation, and the subsequent uncomputation. Suppressing the work registers returned to $\ket{0}$, its action on the all-zero input is
\begin{equation}
    \mathcal A_{\mathrm{ini}}\ket{0}
    =
    \sqrt{1-p_{\mathrm{ini}}}\,
    \ket{\Phi^\perp}\ket{0}
    +
    \sqrt{p_{\mathrm{ini}}}\,
    \ket{\psi(0)}\ket{1},
    \label{eq:initial-preparation-unitary}
\end{equation}
where the last register is the success qubit, $\ket{\Phi^\perp}$ denotes the normalized state in the unsuccessful branch, and
\begin{align} 
    p_{\mathrm{ini}} &= \frac{1}{13N C_{\mathrm{ini}}^2} \sum_{\ell\in\mathcal J}|q_\ell|^2 =
     \frac{1} {13N \, C_{\mathrm{ini}}^2}\cdot\frac{2E}{a^2} = \Theta(1). 
    \label{eq:initial-success-probability} 
\end{align} 

The target state can therefore be prepared to Euclidean error $\epsilon_{\mathrm{SP}}$ using fixed-point amplitude amplification with
\begin{equation}
    Q_{\mathrm{SP}}
    =
    \O\left(
        \frac{1}{\sqrt{p_{\mathrm{ini}}}}
        \log\frac{1}{\epsilon_{\mathrm{SP}}}
    \right)
    =
    \O\left(\log\frac{1}{\epsilon_{\mathrm{SP}}}\right)
    \label{eq:Q_SP}
\end{equation}
applications of $\mathcal A_{\mathrm{ini}}$ and its inverse. Under the assumed efficient implementations of the value oracles and reversible arithmetic, the gate complexity differs from this query complexity by at most a $\polylog(N/\epsilon_{\mathrm{SP}})$ factor. 


\subsection{Observable extraction}
\label{sec:extraction}

Recalling Eqs.~\eqref{eq:kinetic}-\eqref{eq:potential}, it is clear that energy observables can be extracted from the two components of the state $|\psi(t)\rangle$: projective measurements on the first component living in $\H_V\otimes\H_3$ yield kinetic energy observables, while projective measurements on $\H_I\cong \H_V\ot\H_S$ yield potential energy observables. More precisely, let $\Rc\subseteq\Lambda$ be a chosen spatial region, and let $\Pi_{\Rc}$ project onto the vertex states associated with that region:
\begin{equation}
    \label{eq:vertex_projector}
    \Pi_{\Rc}:= \sum_{\v{n}\in\Rc}\ketbra{\v{n}}{\v{n}}. 
\end{equation}
We can then introduce the following projectors on
$\H\cong (\H_V\ot\H_3)\oplus(\H_V\ot\H_S)$:
\begin{equation}
    \label{eq:energy_projectors}
    P_{\Rc}^{T}:=
    (\Pi_{\Rc}\ot I_3)\oplus 0_{\H_I},
    \qquad
    P_{\Rc}^{U}
    :=
    0_{\H_V\ot\H_3}
    \oplus(\Pi_{\Rc}\ot I_S).
\end{equation}
Consequently,
\begin{equation}
    \label{eq:fractions}
    \langle\psi(t)|P_{\Rc}^{T}|\psi(t)\rangle
    =
    \frac{T_{\Rc}(t)}{E},
    \qquad
    \langle\psi(t)|P_{\Rc}^{U}|\psi(t)\rangle
    =
    \frac{U_{\Rc}(t)}{E},
\end{equation}
where $T_\Rc(t)$ and $U_\Rc(t)$ are kinetic and potential energies in spatial region $\Rc$. Thus, the probability weight in the vertex block gives the fraction of the total energy stored as kinetic energy in $\mathcal R$, while the probability weight in the corresponding interaction subspace gives the fraction stored as potential energy there. As we prove in Appendix~\ref{app:info}, the knowledge of these energy fractions for the discrete oscillator model provides an estimate of the corresponding energy fractions for the continuous elastic system up to a discrete mapping error $\epsilon_{\rm DM}$ from Eq.~\eqref{eq:discrete_mapping_error}.

In order to estimate the considered energy fractions, note that we can then decompose the final state $\ket{\psi(t)}$ as
\begin{align}
    \ket{\psi(t)} &=  P^T_{\Rc} \ket{\psi(t)} + (I-P^T_{\Rc})\ket{\psi(t)},\qquad \|P^T_{\Rc} \ket{\psi(t)}\|=\sqrt{\frac{T_{\Rc}(t)}{E}},\\
    \ket{\psi(t)} &=  P^U_{\Rc} \ket{\psi(t)} + (I-P^U_{\Rc})\ket{\psi(t)},\qquad \|P^U_{\Rc} \ket{\psi(t)}\|=\sqrt{\frac{U_{\Rc}(t)}{E}}.
\end{align}
Now, one can perform amplitude estimation on $|\psi(t)\rangle$~\cite{gilyen2019qsvt}, to obtain $\epsilon_{\rm OE}$-estimates of $T_{\Rc}(t)/E$ and $U_{\Rc}(t)/E$ with constant success probability. More precisely, assuming that the estimated fractions are $\Theta(1)$, one needs
\begin{align}   
    \label{eq:Q_OE}
    Q_{\rm OE} =\mathcal{O}\!\left(\frac{1}{\epsilon_\text{OE}}\right),
\end{align}
calls to the state-preparation unitary for $|\psi(t)\rangle$ and its inverse. 


\subsection{End-to-end complexity and speedup regime}
\label{sec:continuum_complexity}

Let us now combine these ingredients to obtain the asymptotic scaling of the total complexity cost for preparing the initial state, evolving it, and extracting the information about energy fraction in a given region\footnote{Note that, instead of estimating the energy fraction with high accuracy, one can also consider an alternative decision problem, where it is promised that the energy fraction is above $\tau+\Delta$ or below $\tau-\Delta$ with $\Delta=\Theta(1)$, and the task is to determine which case holds. Then, producing an estimate to additive accuracy $\epsilon$ smaller than $\Delta$ suffices, allowing for the choice $\epsilon=\Theta(1)$, and so the complexity depends asymptotically only on the simulation time.}. We work under the assumption that the material and initial-data oracles have efficient implementations and require $\polylog(N)$ gates. Then, the total number of elementary quantum gates needed to run our algorithm scales as 
\begin{equation}
  \comp_Q= \tilde{\O}((Q_{\rm SP} +  Q_{\rm HS}) \times Q_{\rm OE}),
\end{equation}
where $\tilde{\O}$ suppresses factors that are polylogarithmic in $N$ and $1/\epsilon$. Employing the derived expressions for the query complexities, Eqs.~\eqref{eq:Q_SP},~\eqref{eq:Q_HS},~\eqref{eq:Q_OE}, and suppressing the polylogarithmic factors, we arrive at
\begin{equation}
    \label{eq:comp_q_with_N}
    \comp_Q = \tilde{\O}\left(\frac{\bar{t}\:N^{1/3}}{\epsilon_\text{OE}}
         \right).
\end{equation}
Now, recall that we focus on simulations at fixed volume, where the number of spatial discretization points $N$ controls the accuracy with which our discrete oscillator model recovers the continuous elastic system. In particular, as discussed in Sec.~\ref{sec:extraction}, the extracted energy fraction of the discrete model recovers the actual energy fraction of the elastic system up to error $\epsilon_{\rm DM}$ from Eq.~\eqref{eq:discrete_mapping_error}. We can thus rewrite the above expression as
\begin{equation}
    \comp_Q= \tilde{\O}\left(\frac{\bar{t}^{3/2}}{\epsilon_\text{OE} \epsilon_{\rm DM}^{1/2}}
         \right).
\end{equation}
Finally, since the total error is upper bounded by the sum of all errors, we can choose them all to be equal to $\O(\epsilon)$ to arrive at
\begin{align}
    \label{eq:final_comp}
    \comp_Q&= \tilde{\O}\left(\left({\bar{t}}/{ \epsilon}\right)^{3/2}\right).
\end{align}

Let us also comment on the generalized model with an $m$-th-order spatial discretization. A standard $m$-th-order finite-difference stencil couples each lattice site to $\O(m)$ neighbors along the coordinate directions. Since the spatial dimension and the number of displacement components are fixed, the resulting operators have row and column sparsities $\O(m)$. Their nonzero entries and locations remain efficiently computable under the same material-access model, and the standard sparse-access construction therefore yields block-encodings with normalization and implementation cost growing at most linearly with $m$. Apart from this $\O(m)$ prefactor, the change to Eq.~\eqref{eq:final_comp} arises from the improved scaling of discrete mapping error $\epsilon_{\rm DM}^{(m)}$ in Eq.~\eqref{eps_DM_k}. More precisely, starting with Eq.~\eqref{eq:comp_q_with_N}, using the expression for $N$ from Eq.~\eqref{eps_DM_k}, assuming constant $m$, and making all errors scale equally, we arrive at the following expression for complexity:
\begin{equation}
    \label{eq:final_comp_k}
    \comp_Q= \tilde{\O}\left(\left({\bar{t}}/{ \epsilon}
         \right)^{1+1/m}\right),
\end{equation}
with the previous scenario recovered by setting $m=2$.

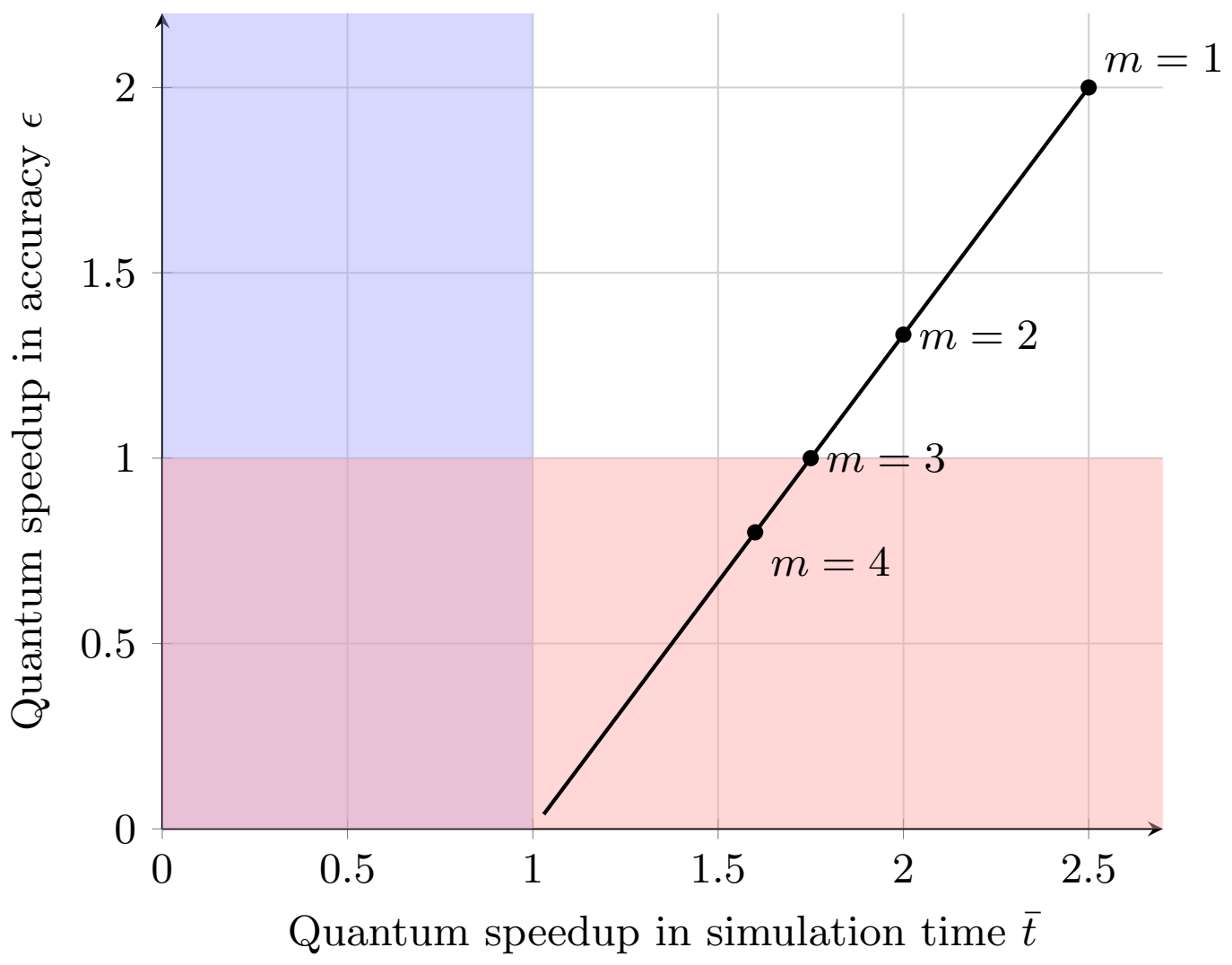
\begin{figure}[t]
    \centering
    \begin{tikzpicture}
        \def\xmax{2.7}
        \def\ymax{2.2}

        \begin{axis}[
            width=0.5\textwidth,
            axis equal image,
            xmin=0, xmax=\xmax,
            ymin=0, ymax=\ymax,
            axis lines=left,
            xlabel={Quantum speedup in simulation time $\bar{t}$},
            ylabel={Quantum speedup in accuracy $\epsilon$},
            xtick distance=0.5,
            ytick distance=0.5,
            grid=major,
            grid style={gray!35},
            clip=false,
        ]
            \addplot[
                draw=none,
                fill=blue!35,
                fill opacity=0.45,
                forget plot,
            ] coordinates {(0,0) (1,0) (1,\ymax) (0,\ymax)}
              \closedcycle;

            \addplot[
                draw=none,
                fill=red!35,
                fill opacity=0.45,
                forget plot,
            ] coordinates {(0,0) (\xmax,0) (\xmax,1) (0,1)}
              \closedcycle;

            \addplot[
                domain=1:100,
                samples=200,
                thick,
                black,
            ] (
                {(4+x)/(1+x)},
                {4/(1+x)}
            );

            \addplot[
                only marks,
                mark=*,
                mark size=1.5pt,
                black,
            ] coordinates {
                (2.5,2)
                (2,1.333333)
                (1.75,1)
                (1.6,0.8)
            };

            \node[above right] at (axis cs:2.5,2) {$m=1$};
            \node[right] at (axis cs:2,1.333333) {$m=2$};
            \node[right] at (axis cs:1.75,1) {$m=3$};
            \node[below right] at (axis cs:1.6,0.8) {$m=4$};
        \end{axis}
    \end{tikzpicture}
    \caption{\textbf{Quantum speedup regime.} Quantum speedup in simulation time $\bar{t}$ and accuracy $\epsilon$ for the continuum model as a function of the regularity order $m$, for $m\in\{1,2,3,4\}$. A speedup of order below $1$ indicates that the classical algorithm outperforms the quantum algorithm. The blue area indicates no speedup in $\bar{t}$, and the red area no speedup in $\epsilon$.}
    \label{fig:regularity-speedup}
\end{figure}

The potential quantum advantage for continuum elastic system simulation depends crucially on the spatial-discretization order $m$. Recall that we compare against a classical finite-difference method with forward in time propagation. Taking the time-integration order $n$ to be at least as large as $m$, the classical complexity from Eq.~\eqref{eq:comp_C} is
\begin{equation}
    \label{eq:classical-k-regimes}
    \comp_C
    =
    \O\!\left[
        \bar t
        \left(\frac{\bar t}{\epsilon}\right)^{4/m}
    \right].
\end{equation}
Comparing this with Eq.~\eqref{eq:final_comp_k}, the speedup is at most of order
\begin{align}
    \frac{m+4}{m+1} \quad \textrm{in simulation time $\bar{t}$}, \qquad    \frac{4}{m+1} \quad \textrm{in accuracy $\epsilon$},
\end{align} 
which we illustrate in Fig.~\ref{fig:regularity-speedup}. The admissible values of $m$ are determined by the regularity of the material parameters and of the resulting displacement field. Smooth coefficient profiles and compatible smooth initial data can support high-order discretizations. Conversely, sharp interfaces, rapidly varying material parameters, nonsmooth initial data, or insufficiently accurate interface treatments may restrict the global convergence order even when a formally high-order interior stencil is used. In the following, $m$ therefore denotes the convergence order actually attained by the discretization.

Both speedups decrease in $m$, so the regime most favorable to the quantum algorithm is low-order spatial convergence. For $m=2$, the improvement is quadratic in simulation time (from $\bar{t}^3$ to $\bar{t}^{3/2}$) and of order $4/3$ in precision (from $\epsilon^{-2}$ to $\epsilon^{-3/2}$). Potentially, the $m=1$ regime may also be relevant, e.g., when interface or boundary effects give an effective slower convergence. In this case, the simulation time speedup would be of order $2.5$ (between quadratic and cubic) and the accuracy speedup would be quadratic. For $m=4$, the scaling in accuracy becomes worse than classical. The origin of this crossover is the additional $\epsilon^{-1}$ cost associated with the amplitude estimation subroutine used by the quantum algorithm to extract a classical observable from a quantum state. Thus, the expected separation is controlled directly by the highest
\emph{realized} spatial convergence order.

Note that the quantum advantage discussed above is less optimistic than the quartic speedup previously identified for quantum wave-equation simulation~\cite{costa2019quantum} or more complex, physically realistic systems such as in Refs.~\cite{bosch2025quantum, kolotouros2026quantum,schade2026quantum}. In those works, the quartic speedup is in terms of the number $N$ of degrees of freedom simulated, with classical complexity scaling as $\O(N^{4/3})$\footnote{For the three-dimensional local lattice with $N$ microscopic degrees of freedom at fixed microscopic spacing, its linear dimension scales as $\Theta\left(N^{1/3}\right)$. Thus, if the wave propagation speed is uniformly bounded in $N$, classical time-stepping methods resolving propagation across the system require $\Theta\left(N^{1/3}\right)$ time steps. Since a standard explicit  classical integrator generically requires $\O(N)$ operations per time step, the resulting classical computational complexity is $\O\left(N^{4/3}\right)$ up to factors associated with the physical evolution time and the chosen temporal accuracy.} and quantum coherent evolution as $\tilde{\O}(N^{1/3})$. The same scalings appear here for us, and so this is the relevant speedup when we wish to consider scaling up the physical size of the system. However, given a fixed size for our system, for continuum simulations $N$ is typically not an independent problem parameter, since it must be chosen to achieve the desired physical accuracy $\epsilon$ (similar to analysis in Ref.~\cite{linden2022quantum}), and this in turn is impacted by properties such as regularity. A quartic speedup in simulation time $\bar t$ and error $\epsilon$ arises in the $m\rightarrow 0$ limit, and so in the following, we turn to the regime in which the continuum model breaks down and the physics to be resolved requires a lattice description. In this regime, $N$ becomes an independent problem parameter, corresponding to the physical size of the system, and the quartic speedup becomes possible.


\section{Microscopic elasticity beyond continuum} 
\label{sec:lattice_setup} 

\subsection{Continuum breakdown}
\label{sec:breakdown}

So far, we have assumed that the continuum elasticity equation provides an appropriate physical model of the material, with the lattice entering only as a numerical discretization of the continuum dynamics. We now consider situations in which this assumption breaks down and the microscopic degrees of freedom of the material must be resolved explicitly. To make this concrete, consider a regular cubic lattice system of volume $V$, lattice spacing $a$, and $N$ lattice points. We focus on a lattice that is highly heterogeneous, with couplings that rapidly vary. Now, any continuum model for this lattice system involves spatially continuous states in the volume $V$. Nyquist sampling says that if the continuum states have spatial frequencies within a bandwidth of Fourier scale $K_c$, then any discretization must have $\Delta x = \Theta(1/K_c)$ in order to describe the state to a fixed error threshold. For such a band-limited system the number of degrees of freedom is $N_c \sim V K_c^3$. 

Recall from before that, for a fixed volume $V$, the quantum speedup rests critically on the number of degrees of freedom the classical algorithm must simulate to achieve the desired accuracy for the problem at hand. This is a function of (a) the particular underlying lattice properties, (b) the set of initial conditions considered, and (c) the set of observables of interest. Firstly, it is clear that $N_c =O(N)$, since in the worst case we do a direct simulation of the lattice. However, there exist problems for which $N_c \ll N$, and in this case it is computationally more efficient to work with the continuum model. This is a regime similar to the one discussed earlier, where increasing regularity allows for decreasing the number of degrees of freedom in the continuum model. However, if instead the problem requires $N_c= \Theta(N)$, then the classical algorithm has complexity $\Theta(N^{4/3})$, and there is no advantage in using the continuum model over a direct lattice simulation. This limiting regime has the largest separation between classical and quantum complexities for a fixed volume $V$. 

Employing the Nyquist sampling, we can now deduce a condition on $K_c$ such that we are in this optimal regime. Specifically, we must have
\begin{equation}
    K^3_c = \Theta \left (\frac{N}{V} \right ).
\end{equation} 
However, since $V = \Theta(Na^3)$ this implies that 
\begin{equation}
    K_c = \Theta(1/a).
\end{equation}
Therefore, we obtain the largest computational separation between the quantum and classical algorithms when the estimation problem \emph{requires} $K_c = \Theta(1/a)$. The presence of lattice-scale Fourier components in the material parameters, initial conditions, or observables does not by itself establish this requirement. Their effects on the requested outputs may be still captured by a lower-bandwidth continuum model, possibly supplemented with effective constitutive parameters or a microscopic reconstruction. This is highly problem-specific and we do not delve into this here. 

Instead, we shall address the quantum side of this regime. Specifically, in such situations we face potential data-loading challenges, and it is not obvious that a highly heterogeneous lattice model can even be efficiently encoded, simulated on a quantum computer, and the observable of interest efficiently extracted. To explore this problem, we begin in Sec.~\ref{sec:lattice_model} by considering general lattice models of elasticity, before in Sec.~\ref{sec:msbo} describing observables that require lattice scale simulations to be estimated. In the following Sec.~\ref{sec:lattice_algo}, we will then construct an explicit toy model of a pseudorandom binary lattice that is highly heterogeneous, and provide an example of a lattice-scale observable that admits an efficient quantum simulation. For such discrete systems the speedup can then be viewed in terms of scaling up the system size, and therefore is potentially quartic. Such lattice settings therefore provide candidate regimes where a quantum computer could give sizable speedups over classical methods.


\subsection{Model}
\label{sec:lattice_model}

Within the harmonic approximation, elastic systems are naturally described as networks of coupled oscillators. For an atomic crystal, for example, the oscillator coordinates are the displacements of individual atoms from their equilibrium positions and the couplings are determined by the interatomic force constants. More generally, a unit cell may contain several atoms or other internal degrees of freedom, and the interactions may be anisotropic or extend beyond nearest neighbors. The resulting equations of motion take the general form
\begin{equation}
     M\ddot u = -  K u,
    \label{eq:microscopic_eom}
\end{equation}
where $ M$ is the mass matrix and $ K$ is the stiffness matrix. Such harmonic lattice models are standard in the description of phonons and lattice vibrations and have recently also been targeted by quantum algorithms~\cite{babbush2023exponential,li2025exponential}. In what follows we shall assume a regular cubic lattice $\Lambda$ in 3-d with lattice spacing $a$. A detailed exposition of the following can be found in Appendix~\ref{app:elastic-theory}. For each vertex $n\in \Lambda$ we have a 3-dimensional displacement $u_{n} = (u_{n,i})$, where the index $i$ runs over $i\in\{1,2,3\}$ for the three lattice dimensions. We combine these data into a single displacement variable
\begin{equation}
    u(t)
    =
    \sum_{ n\in\Lambda}
    \sum_{i=1}^{3}
    u_{ n,i}(t)
    | n\rangle\otimes|i\rangle
    \in
    \H_V\otimes\H_3.
\end{equation}
From the displacement $u$ we have the strain variables $\xi := D u$, where $D$ is the strain-displacement map transforming the individual displacements at lattice points into the interaction space $\H_I$, describing the different bond couplings on the lattice. The strengths of these couplings are encoded in the constitutive map $C: \H_I \rightarrow \H_I$, given by
\begin{equation}
    C = \sum_{n \in \Lambda} \ketbra{n}{n}\otimes C_n,
    \label{eq:constitutive}
\end{equation}
where $\H_I \cong \H_V \otimes \H_S$ denotes the decomposition of the interaction space into a lattice labeling component and a local strain space $\H_S$ describing the couplings local to a vertex. The term $C_n$ acts on the strain space $\H_S$ and encodes the material properties for the strain channels at the vertex $n$. Then, it can be shown that
\begin{equation}
    \label{eq:K_decomposition}
     K = D^\dagger C D,
\end{equation}
which separates out the force into kinematical degrees of freedom in $D$ from material degrees of freedom in $C$. Within this setting we shall consider performing discrete Fourier transforms on $\H_V$, which allow us to analyze inhomogeneities over the lattice in Fourier space.


\subsection{Microscopically resolved bulk observable estimation}

\label{sec:msbo}

In order to assess an end-to-end quantum advantage, it is crucial that relevant classical observables can be extracted from the final quantum state without introducing an overhead dependent on system size. This issue is particularly important in the microscopic setting: resolving the full lattice is computationally meaningful only when other coarse-grained methods to estimate the observable of interest are not known, since otherwise a coarser continuum model may suffice. We therefore focus now on observables that both require a microscopic description and remain compatible with efficient quantum readout. We call such a quantity a \emph{microscopically resolved bulk observable} (MRBO). The term ``bulk'' refers to the size of the measured signal, whereas ``microscopic'' refers to the spatial length scales on which the signal depends. 

An MRBO estimation task suitable for the proposed quantum algorithm should satisfy the following criteria that allow macroscopic observables whose predictions depend on microscopic dynamics.

\begin{enumerate}
    \item \emph{Efficient bulk readout.} 
        Recall that the simulation Hilbert space is given by
        \begin{equation}
            \ket{\psi(t)}\in\H,\qquad \H=(\H_V\otimes\H_3)\oplus\H_I,
        \end{equation}
        where for local interactions $\H_I$ can be split as $\H_I\cong\H_V\otimes\H_S$ with $\H_S$ carrying information about local strain channels. Then, let $\mathcal A$ be a Hermitian observable acting on $\H$. After rescaling, we may take $\|\mathcal A\|\leq1$ and write the desired physical scalar as
        \begin{equation}
            A(t)
            =
            \alpha_{\mathcal A}\,
            \langle\psi(t)|\mathcal A|\psi(t)\rangle,
            \label{eq:general-observable-normalization}
        \end{equation}
        where the scale $\alpha_{\mathcal A}$ is assumed to be known. Estimating $A(t)$ to additive error $\epsilon_A$ requires estimating $\langle\psi(t)|\mathcal A|\psi(t)\rangle$  to additive error $\epsilon_A/\alpha_{\mathcal A}$. Consequently, if $\alpha_{\mathcal A}$ grows with system size, we generically incur a corresponding precision overhead. For example, absolute energy observables typically have $\alpha_{\mathcal A}=\Theta(N)$ when the total energy is extensive in $N$, as is normally the case for lattice models. By contrast, when $\alpha_{\mathcal A}=\mathcal O(1)$, fixed additive accuracy does not introduce a precision overhead dependent on system size. The required physical accuracy can be achieved without a readout precision overhead growing with $N$. In particular, additive physical error $\epsilon_A$ requires normalized error $\epsilon_A/\alpha_{\mathcal A}$. As we have seen, for energy fractions at fixed additive accuracy, this introduces no size-dependent precision requirement since $\alpha_{\mathcal A} =1$. At fixed relative accuracy, a sufficient signal condition is
        \begin{equation}
            \langle\psi(t)|\mathcal A|\psi(t)\rangle=\Theta(1).
            \label{eq:bulk-signal-criterion}
        \end{equation}
    
    \item \emph{Need for microscopic simulation.} 
            At the requested time $t$ and target accuracy $\epsilon$, the observable $\mathcal{A}$ of interest does not currently admit a predictive analytical or coarse-grained description that avoids resolving the microscopic lattice dynamics. The best established methods for calculating it in the regime of interest therefore involve a microscopic simulation of $\Theta(N)$ degrees of freedom. A stronger \emph{microscopic sensitivity} criterion would additionally require the observable to be sensitive to the particular microscopic configuration. We do not impose this stronger condition: an observable may be self-averaging, taking nearly the same value across typical random realizations, while still satisfying our criterion if predicting that value requires simulating a representative microscopic realization.
        
    \item \emph{Efficient quantum access.}
        Preparation of the encoded initial state, coherent access to the lattice parameters, and implementation of the observable measurement must have costs
        \begin{equation}
            G_{\mathrm{input}},\;G_{\mathrm{obs}}
            =\operatorname{polylog}(N),
            \label{eq:efficient-access-criterion}
        \end{equation}
        under the stated input model. Initial-state preparation includes the material-dependent energy weighting used in the encoding.
\end{enumerate}

Concerning Criterion 1, as explained in Sec.~\ref{sec:extraction}, energy fractions provide a particularly simple example. Choosing $\mathcal A=P^T_{\mathcal R}$ or $P^U_{\mathcal R}$ from Eq.~\eqref{eq:energy_projectors} (i.e., projectors on selected vertices $\Rc\subseteq \Lambda$ of the kinetic and potential energy sectors) and setting $\alpha_{\mathcal A}=1$, gives
\begin{equation}
A(t) = \langle\psi(t)|P^T_{\mathcal R}|\psi(t)\rangle
=\frac{T_{\mathcal R}(t)}{E},
\qquad
A(t) =\langle\psi(t)|P^U_{\mathcal R}|\psi(t)\rangle
=\frac{U_{\mathcal R}(t)}{E},
\end{equation}
where $E>0$ is the conserved total energy. Thus, these expectations give the kinetic or potential energy contributions associated with the selected vertices $\mathcal R\subseteq\Lambda$, expressed as fractions of the total energy. Recovering the corresponding absolute energies instead requires $\alpha_{\mathcal A}=E$. For fixed additive accuracy in an energy fraction, the required normalized precision is independent of $N$, regardless of the fraction's magnitude. For fixed relative accuracy, a sufficient condition for avoiding a system-size-dependent precision overhead is that the selected fraction is $\Theta(1)$. This holds when the selected kinetic or potential energy contributions constitute a constant fraction of the total energy; extensive support alone does not guarantee it.

Concerning Criterion 2, it can be sharpened in many inequivalent ways, corresponding to different notions of `coarse-graining'. For definiteness, consider a one-dimensional nearest-neighbour lattice with spacing $a$, equal
site masses $m$, and positive bond stiffnesses $\kappa_n$. The displacement
of site $n$ is $u_n(t)$, and
\begin{equation}
 U[u]=\frac12\sum_n\kappa_n(u_{n+1}-u_n)^2,
 \qquad
 m\ddot u_n=\kappa_n(u_{n+1}-u_n)
              -\kappa_{n-1}(u_n-u_{n-1}).
 \label{eq:lattice}
\end{equation}
Let the stiffnesses vary independently from bond to bond, with
\begin{equation}
 \kappa_n=
 \begin{cases}
  \kappa^{(1)},&\text{with probability }p,\\
  \kappa^{(0)},&\text{with probability }1-p,
 \end{cases}
 \qquad
 \kappa^{(0)},\kappa^{(1)}>0,
 \quad 0<p<1,
\end{equation}
so that the material varies on scale $a$.

The lattice is the physical model for which one can construct continuum approximations intended to describe displacements varying on a scale $\lambda\gg a$ by standard homogenization techniques (see Appendix~\ref{app:1d_lattice} for details). To average the microscopic structure, choose representative intervals of length $\ell=N_\ell a$ satisfying
\begin{equation}
     a\ll\ell\ll\lambda.
    \label{eq:separation}
\end{equation}
Each interval contains many bonds, while the continuum strain varies little across it. One determines the elastic coefficient by minimizing the lattice energy over internal displacements at fixed average strain, and let this `microscopically relaxed' model define the continuum energy. As the number of independent bonds in a representative interval grows, the fraction of bonds with stiffness $\kappa^{(1)}$ convergences to $p$ and the material is represented by a constitutive matrix with effective Hook constant given by the harmonic average of the constants of the two materials:
\begin{align}
     \kappa_{\mathrm H}
    :=\left(
        \frac{1-p}{\kappa^{(0)}}+\frac{p}{\kappa^{(1)}}
    \right)^{-1}.
\end{align}
    An alternative coarse-graining procedure would instead involve inserting a cutoff in Fourier modes, effectively freezing all modes above $O(1/\lambda)$. For the above example, this alternative procedure leads to continuum equations with effective arithmetic mean stiffness:
  \begin{align}
       \kappa_{\mathrm A}
    =(1-p)\kappa^{(0)}+p\kappa^{(1)}.
  \end{align}


\section{Quantum algorithm for microscopic elasticity beyond continuum}
\label{sec:lattice_algo}

The quantum algorithm for microscopic lattice elasticity follows the same overall structure as the continuum algorithm developed in Sec.~\ref{sec:continuum_algo}. Thus, we start in Sec.~\ref{sec:steps} with a summary of necessary modifications, and then proceed to the aspects that are particular to elasticity beyond the continuum. Namely, in Sec.~\ref{sec:geometry}, we explain how microscopically varying heterogeneous material properties can be encoded and accessed efficiently without explicitly storing their values at every lattice site; and in Sec.~\ref{sec:binary-random-msbo}, we explain how one might construct a heterogeneous lattice model possessing a microscopically resolved bulk observable of the type introduced in Sec.~\ref{sec:msbo}. Finally, in Sec.~\ref{sec:lattice_complexity}, we provide the resulting end-to-end quantum complexity and identify the corresponding regime of quantum speedup.


\subsection{Algorithmic steps}
\label{sec:steps}

While the microscopic lattice model in Eq.~\eqref{eq:microscopic_eom} need not have the specific deviatoric and volumetric interactions introduced in Sec.~\ref{sec:discrete_mapping}, the quantum construction developed for these interactions in Sec.~\ref{sec:continuum_algo} extends naturally to a much broader class of harmonic lattices. Physical interatomic interactions are typically local or finite ranged, so that the strain-displacement map $D$ and the constitutive matrix $C$ have bounded or otherwise low sparsity, since each microscopic degree of freedom generates interactions with only a limited number of neighbors. Moreover, mechanical stability of the equilibrium configuration requires the harmonic potential energy to be nonnegative, and hence the stiffness matrix $K=D^\dagger C D$ is positive semidefinite. It can therefore be factorized as $K=B^\dagger B$, with the form of $B=\sqrt{C}D$ determined by the microscopic interactions through $D$, and by material properties through $C$. Provided that both $D$ and $\sqrt{C}$ remain sparse and efficiently accessible, the mass rescaling and Hamiltonian embedding derived in Sec.~\ref{sec:ham_emb} carry over unchanged, with only the microscopic interaction matrix $B$ taking a different form. Thus, the details of the block-encoding and Hamiltonian simulation described in Secs.~\ref{sec:ham_emb}-\ref{sec:BE} must be adapted to the particular microscopic interaction model (see, e.g., Ref.~\cite{kolotouros2026quantum}), but the essential algorithmic structure is unchanged. Moreover, since the initial state preparation from Sec.~\ref{sec:ini_state}, as well as the observable extraction from Sec.~\ref{sec:extraction}, were designed for the discrete model approximating the continuum, they can be reused in an unchanged manner in the current microscopic lattice setup. What then remains to be addressed, and to which we now turn, is how to efficiently prepare the material oracle for $C$ for interesting heterogeneous systems, and what kind of systems could benefit from the microscopic lattice modeling, i.e., which systems have microscopically sensitive bulk observables.


\subsection{Efficient geometry encoding}
\label{sec:geometry}

Spatially heterogeneous or random fields are widely used in classical computational materials science to construct synthetic models of heterogeneous media (recall Fig.~\ref{fig:heterogenous-material}). A common approach is to begin with a random field and introduce spatial correlations through a prescribed local transformation, such as convolution with a smoothing kernel, before thresholding the resulting field to define distinct material phases. For example, thresholded Gaussian random fields have been used to generate statistically controlled models of porous media and heterogeneous microstructures, while related random-field constructions have been employed to model closed-cell foams, porous ceramics, and three-dimensional mineral and geological structures~\cite{hyman2014stochastic,hyman2015statistical}. Such models are useful when the objective is not to reproduce one specific experimentally measured sample, but rather to generate an ensemble of statistically representative microstructures with prescribed quantities such as volume fraction, correlation length, characteristic feature size, or phase morphology. The generated material field can subsequently be supplied to a conventional finite-difference, finite-volume, or finite-element solver for elasticity, diffusion, or transport. In this sense, procedural generation of heterogeneous media is already well established classically; the principal difficulty for a quantum algorithm is to provide coherent access to such a field without first explicitly storing its value at every lattice site.

Reference~\cite{barthe2026make} addresses precisely this input-access problem by constructing a compact reversible oracle for a broad class of pseudorandom local microstructures. The starting point is a binary field of pseudorandomly distributed ``centres''. Given a lattice coordinate $n$ encoded in binary, an efficiently reversible pseudorandom permutation $P_m$ is applied and the site is marked whenever $c_{m,J}(n) =1$, where
\begin{equation}
    c_{m,J}(n) =  {1}\!\left[P_{m}(n)<J\right].
\end{equation}
Since $P_m$ is a permutation, exactly $J$ sites are marked, while their spatial locations appear pseudorandom up to the quality of the chosen permutation. More structured media are then obtained by applying a fixed local rule to this centre field. In particular, for a kernel $\Xi=\{(\eta_r,\alpha_r)\}_{r=1}^{R}$, the construction evaluates
\begin{equation}
    f_{\Xi,m}(n)
    =
    \sum_{r=1}^{R}
    \alpha_r c_{m,J}(n-\eta_r),
\end{equation}
and we may subsequently threshold $f_{\Xi,m}$ to obtain a binary material geometry. Crucially, when the kernel contains a fixed number of nonzero entries, the value of the geometry at a requested site can be computed by a reversible circuit whose size is polynomial in the number of coordinate qubits $\log_2 N$, even though the complete lattice contains $N$ sites. Thus, rather than storing an explicit $\O(N)$-sized realization of the material, the quantum algorithm stores a compact generative rule and evaluates the local geometry coherently when required. 

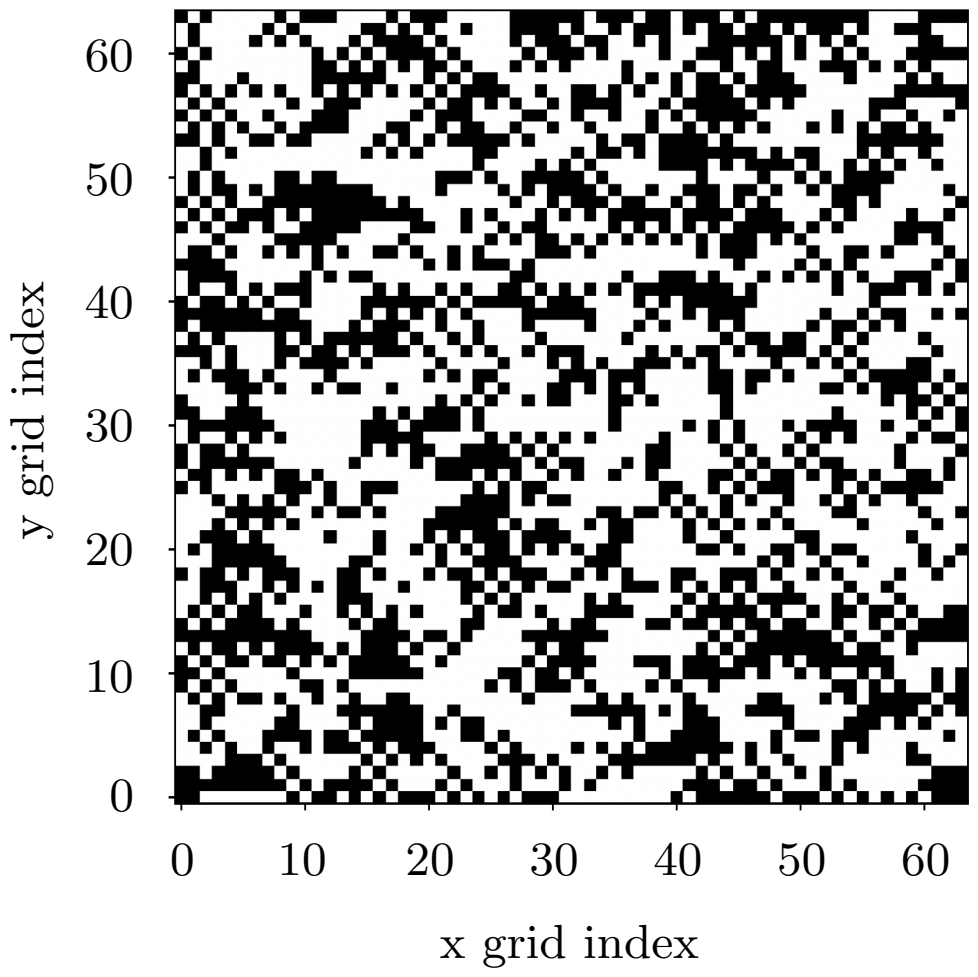
\begin{figure}
    \centering
    \begin{tikzpicture}[font=\rmfamily]
        \node[anchor=south west, inner sep=0] (pattern) at (0,0) {\includegraphics[width=0.3\linewidth]{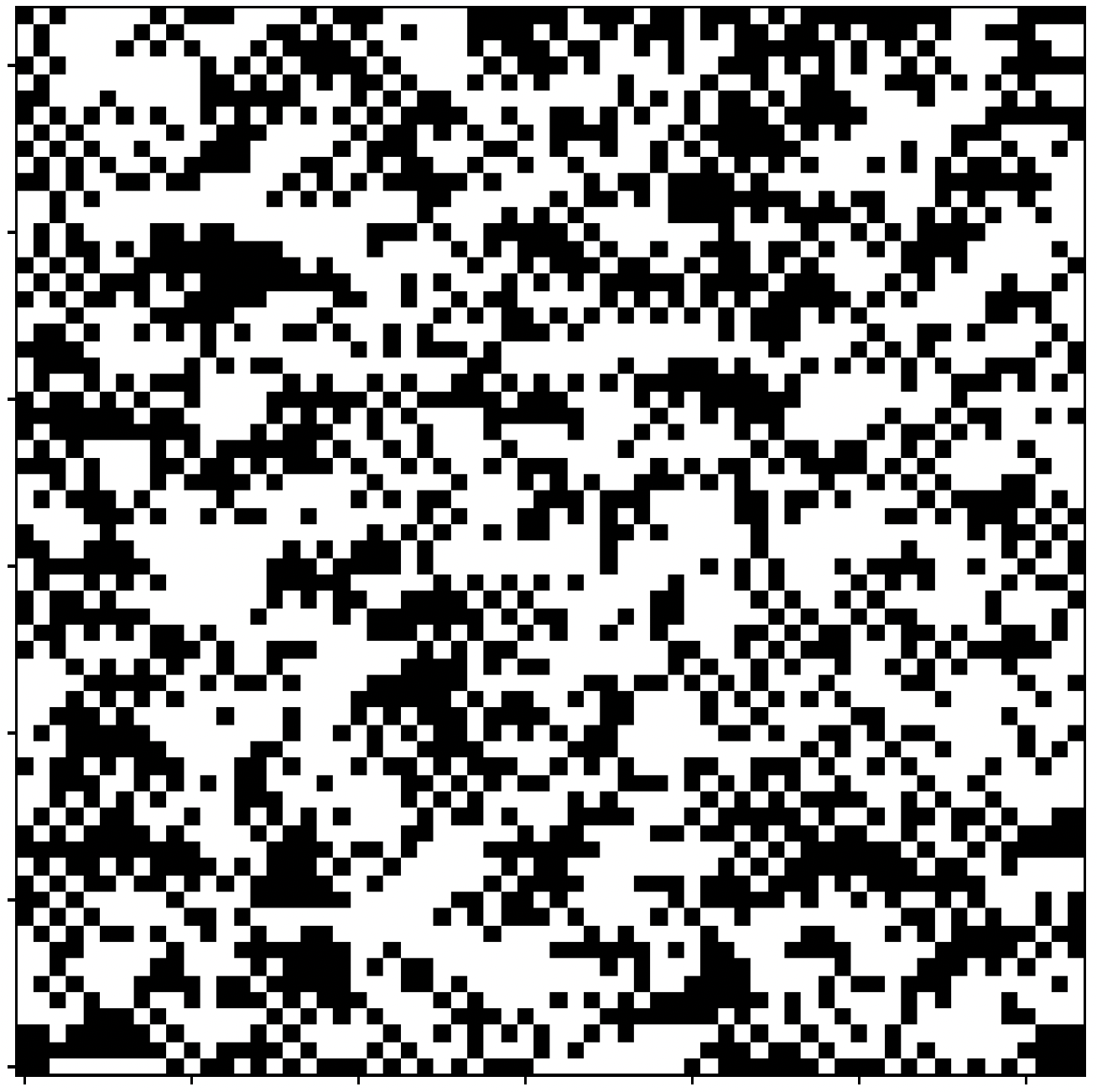}};
        \begin{scope}[x={(pattern.south east)}, y={(pattern.north west)}]
            \node[font=\fontsize{9}{17}\selectfont]
              at (0.5,-0.16) {x grid index};
        
            \node[font=\fontsize{9}{17}\selectfont, rotate=90]
              at (-0.16,0.5) {y grid index};
        
            \foreach \x/\label in {0.025/0,0.172/10,0.329/20,0.481/30,0.633/40,0.785/50,0.937/60}
              \node[font=\fontsize{9}{12}\selectfont, below=2pt]
                at (\x,0) {\label};
        
            \foreach \y/\label in {0.025/0,0.172/10,0.329/20,0.481/30,0.633/40,0.785/50,0.937/60}
              \node[font=\fontsize{9}{12}\selectfont, left=2pt]
            at (0,\y) {\label};
        \end{scope}
    \end{tikzpicture}
    \caption{\textbf{Simple example of a random material.} The material structure was generated using the quantum cheese method of Ref.~\cite{barthe2026make}. It has been carved in a $64\times 64$ lattice by setting as material 2 (black) the points $\{(-1, 0), (0, -1), (1,0)\}$ around $1024$ random centers, with all remaining grid points set to material 1 (white).}
    \label{fig:cheese}
\end{figure}

The same construction can be used to generate spatially varying material parameters rather than a binary solid--void geometry. Suppose that the material consists of two isotropic elastic phases, so that the local constitutive matrices $C_n$ at site $n$ from Eq.~\eqref{eq:constitutive} are given by either $C^{(0)}$ or $C^{(1)}$. Starting from the pseudorandom centre field, one first constructs a local scalar field $f_{\Xi,m}({n})$ by applying a fixed convolution kernel as described above, and then assigns the material parameters according to a threshold $\tau$:
\begin{equation}
    C_n
    =
    \begin{cases}
        C^{(0)}, & f_{\Xi,{m}}({n}) < \tau,\\[3pt]
        C^{(1)}, & f_{\Xi,{m}}({n}) \geq \tau.
    \end{cases}
\end{equation}
The resulting constitutive matrix $C$ therefore varies pseudorandomly across the lattice, while the choice of the local kernel controls spatial correlations: a sharply localized kernel produces rapidly varying disorder, whereas a broader smoothing kernel produces extended regions in which neighboring lattice sites are likely to share the same material parameters. This way, a compact pseudorandom generative rule can describe a two-phase heterogeneous elastic medium with controllable spatial structure, without explicitly storing the local constitutive matrices $C_n$ at every lattice site, see Fig.~\ref{fig:cheese} for an example.

Under the assumption that the convolution kernel contains a fixed number $R=\O(1)$ of nonzero entries, that its coefficients are bounded so that the convolution accumulator requires $s=\O(1)$ qubits, and that the entries of local constitutive matrices are represented with a fixed $b$-bit precision, the resulting material oracle remains efficient in the lattice size. More precisely, the pseudorandom-centre oracle of Ref.~\cite{barthe2026make} uses $\O(\log N)$ ancillary qubits and, for an $m$-round pseudorandom permutation, has gate complexity $\O(m\log N+\log^2 N)$. The convolution step increases this cost only by a factor proportional to $R$, so that the complete thresholded two-phase material oracle has complexity
\begin{equation}
    G_{\mathrm{mat}}
    =
    \O\!\left(R(m \log N+\log^2 N+\log N+s)\right).
\end{equation}
Taking $m=\Theta(\log N)$, as proposed in Ref.~\cite{barthe2026make}, and fixed $R$, $s$, and $b$, gives
\begin{equation}
    G_{\mathrm{mat}}=\O(\log^2 N),
\end{equation}
while the total number of coordinate, work, accumulator, and material-output qubits is $\O(\log N)$. Thus, a pseudorandom two-phase elastic medium of size $N$ can be queried coherently using only logarithmically many qubits and a number of gates polynomial in $\log N$, provided that the local kernel and material precision do not grow with the system size. For explicit resource estimates, see Ref.~\cite{barthe2026make}.


\subsection{Constructing microscopically resolved bulk observables}
\label{sec:binary-random-msbo}

We now demonstrate how one could construct a lattice system with microscopically resolved bulk observables introduced in Sec.~\ref{sec:msbo}. At the highest level, this amounts to a situation where microscopic features couple non-trivially to macroscopic observables. For simplicity, we consider a periodic cubic lattice $\Lambda=\mathbb Z_{L}^{\times 3}$ with $N=L^3$ sites and lattice spacing $a$. Denote by $\mathcal K$ the reciprocal lattice with wavevectors $k$ satisfying
\begin{equation}
    k_i=\frac{2\pi j}{L a},
        \quad
        j=-\lfloor L/2\rfloor,\ldots,\lceil L/2\rceil-1,
\end{equation}
and let $F_V$ be discrete Fourier transform of the spatial lattice index in the vertex space:
\begin{equation}
    F_V
    =
    \frac{1}{\sqrt N}
    \sum_{k\in\mathcal K}\sum_{n\in\Lambda}
    e^{-\i k\cdot x_n}
    \ketbra{k}{n},
\end{equation}
where $x_n=an$. We assume a uniform mass density, so that the mass matrix may be absorbed into the choice of units. The system evolves according to the oscillator equation of motion from Eq.~\eqref{eq:microscopic_eom} with a stiffness matrix decomposed according to Eq.~\eqref{eq:K_decomposition}, i.e., $K=D^\dagger C D$ with $D$ being translation-invariant and $C$ being local and positive semidefinite. Due to the translation-invariance of the strain-displacement matrix $D$, it is diagonal in the Fourier basis:
\begin{equation}
    \widehat{D}:= (F_V\otimes I_S) D (F_V^\dagger \ot I_3) = \sum_k \ketbra{k}{k} \otimes \widehat{D}(k),
\end{equation}
with $\widehat{D}(k)$ mapping $\H_3$ into $\H_S$ for all $k$. 

For the constitutive matrix $C$, which encodes spatially varying material properties, we assume that it takes one of two values:
\begin{equation}
    C
    =
    \sum_{ n\in\Lambda}
    \ketbra{n}{n}
    \otimes C_{ n},
    \qquad
    C_{ n}
    =
    (1-\eta_{ n})C^{(0)}
    +
    \eta_{ n}C^{(1)},
    \label{eq:binary-constitutive-field}
\end{equation}
where $\eta_{ n}\in\{0,1\}$ is a binary random. For the explicit calculation below, we also assume that the variables $\eta_{ n}$ are independent and identically distributed, with $\mathbb P(\eta_{ n}=1) =p$ and $p\in(0,1)$, although similar results are expected for correlated structures with different scales present (in particular: for the pseudorandom quantum cheese structure from the previous section). Moreover, for simplicity, we focus on $C^{(0)}$ and $C^{(1)}$ that have a common eigenvector $v \in \mathcal{H}_S$ with 
\begin{equation}
    C^{(r)} v = c^{(r)} v,\qquad c^{(0)}\neq c^{(1)}.
\end{equation}

As for the initial condition, we characterize it by a long-wavelength wavevector $k_0 \neq 0$ such that
\begin{equation}
    \| k_0 a\|_\infty
    \leq
    \gamma_0,
    \qquad
    0<\gamma_0\ll1.
    \label{eq:compatible-long-wave-polarization}
\end{equation}
We then take the following macroscopic initial perturbation:
\begin{equation}
    {u}_{ n}(0)= e^{\i k_0\cdot x_{ n}} {u}_0,
    \qquad
    \dot u_{ n}(0)
    =
    0,
    \label{eq:long-wavelength-initial-displacement}
\end{equation}
where $x_{ n} = a n$. We assume that $v$ is in the range of $\widehat{D}(k_0)$ and that ${u}_0 \in \mathcal{H}_3$ is chosen so that $\widehat{D}( k_0){u}_0= v$. Note that we use complex plane-wave notation for convenience; a real physical initial displacement may be obtained by taking the real part. Eq.~\eqref{eq:long-wavelength-initial-displacement} gives
\begin{equation}
    \left(D {u}(0)\right)_{ n}=e^{\i k_0\cdot x_{ n}} v,
    \label{eq:long-wavelength-initial-strain}
\end{equation}
and so the initial strain is entirely macroscopic: it varies over a length scale much larger than the lattice spacing.

Next, we define the energy-weighted strain:
\begin{equation}
    {s}(t)
    :=
    Bu(t),\qquad B=C^{1/2}D,
    \label{eq:elastic-energy-amplitude}
\end{equation}
which satisfies $ U(t)=\frac{1}{2} \|{s}(t)\|_2^2$. Thus, $s(t)$ is precisely the interaction-space component used to encode the elastic potential energy in the quantum state. Under the initial condition from Eq.~\eqref{eq:long-wavelength-initial-strain}, we have
\begin{equation}
    s_{ n}(0)=
    \sqrt{c_{ n}}\,
    e^{\i k_0\cdot x_{ n}} v,
    \qquad
    c_{ n}
    =
    (1-\eta_{ n})c^{(0)}+\eta_{ n}c^{(1)}.
    \label{eq:binary-energy-amplitude}
\end{equation}
Although the unweighted strain is long-wavelength, multiplication by the random factor $\sqrt{c_{ n}}$ generates lattice-scale structure in the elastic energy amplitude. To isolate this microscopic structure, we choose a dimensionless wavevector cutoff
\begin{equation}
    \gamma_0<\gamma_{\mathrm{mic}}<\pi,
    \qquad
    \gamma_{\mathrm{mic}}
    =
    \Theta(1),
    \label{eq:microscopic-cutoff}
\end{equation}
which sets the scale at which microscopic features become important. We then define the Fourier-space projector on the microscopic sector:
\begin{equation}
    \widehat{\Pi}_{\mathrm{mic}}
    :=
    \sum_{ k:\,
    \| k a\|_\infty\geq\gamma_{\mathrm{mic}}}
    \ketbra{k}{k},
    \qquad
    \Pi_{\mathrm{mic}}
    :=
    F_V^\dagger
    \widehat{\Pi}_{\mathrm{mic}}
    F_V.
    \label{eq:microscopic-fourier-projector}
\end{equation}
The wavevectors selected by $\widehat{\Pi}_{\mathrm{mic}}$ have physical magnitudes of order $a^{-1}$ and therefore correspond to wavelengths of order $a$, whereas the ones retained by $I-\widehat{\Pi}_{\mathrm{mic}}$ have wavelengths much larger than $a$ and are ordinarily described by continuum elasticity. This distinction is already visible for a uniform nearest-neighbor lattice, whose scalar dispersion relation is
\begin{equation}
    \omega_{\mathrm{lat}}^2( k)
    =
    \frac{4c^2}{a^2}
    \sum_{j=1}^{3}\sin^2\!\left(\frac{k_j a}{2}\right)
    =c^2\| k\|^2+\O(c^2a^2\| k\|^4),
    \label{eq:lattice-dispersion}
\end{equation}
so that the continuum dispersion is accurate for $\| k\|a\ll1$, but differs by an order-one amount when $\| k\|a=\Theta(1)$.

We shall use this microscopic elastic energy as our candidate MRBO. More precisely, the associated quadratic observable on the full quantum Hilbert space is given by
\begin{equation}
    P_{\mathrm{mic}}^U
    :=
    0_{\H_V\otimes\H_3}
    \oplus
    (\Pi_{\mathrm{mic}}\ot I_S),
    \label{eq:microscopic-stress-energy-observable}
\end{equation}
which is the analogue of $P_\Rc^U$ from Eq.~\eqref{eq:energy_projectors}, with the potential energy accumulated in the spatial region $\Rc$ replaced by the potential energy stored in the microscopic degrees of freedom. Indeed, its expectation value is
\begin{equation}
    U_{\mathrm{mic}}(t)
    :=
    \langle\psi(t)|
    P_{\mathrm{mic}}^U
    |\psi(t)\rangle
    =
    \frac{
        \left\|
            (\Pi_{\mathrm{mic}}\ot I_S) s(t)
        \right\|_2^2
    }{
        2E
    },
    \label{eq:microscopic-stress-energy-fraction}
\end{equation}
which is the fraction of the total elastic energy stored in microscopic Fourier components of the energy-weighted strain. 

The efficient quantum access criterion for the above problem is satisfied. To see this, note that the long-wavelength plane-wave initial state can be prepared using a quantum Fourier transform and a constant-size polarization register for $u_0$. The only non-trivial part for encoding the dynamics lies in preparing the constitutive matrix $C$, which can be efficiently constructed using the pseudorandom material oracle from Sec.~\ref{sec:geometry}. The observable projector can be implemented by applying $F_V$, reversibly testing whether $\| k a\|_\infty\geq\gamma_{\mathrm{mic}}$, and then undoing $F_V$. These operations all have gate complexity polylogarithmic in $N$, so that Eq.~\eqref{eq:efficient-access-criterion} is satisfied and the system can be efficiently encoded and simulated on a quantum computer.

\begin{figure}
    \centering
    \includegraphics[height=0.25\linewidth]{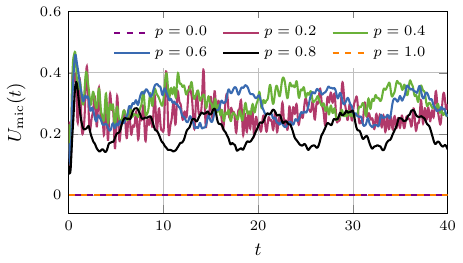}
    \hspace{1cm}
    \raisebox{0.04\linewidth}{%
        \includegraphics[height=0.2\linewidth]{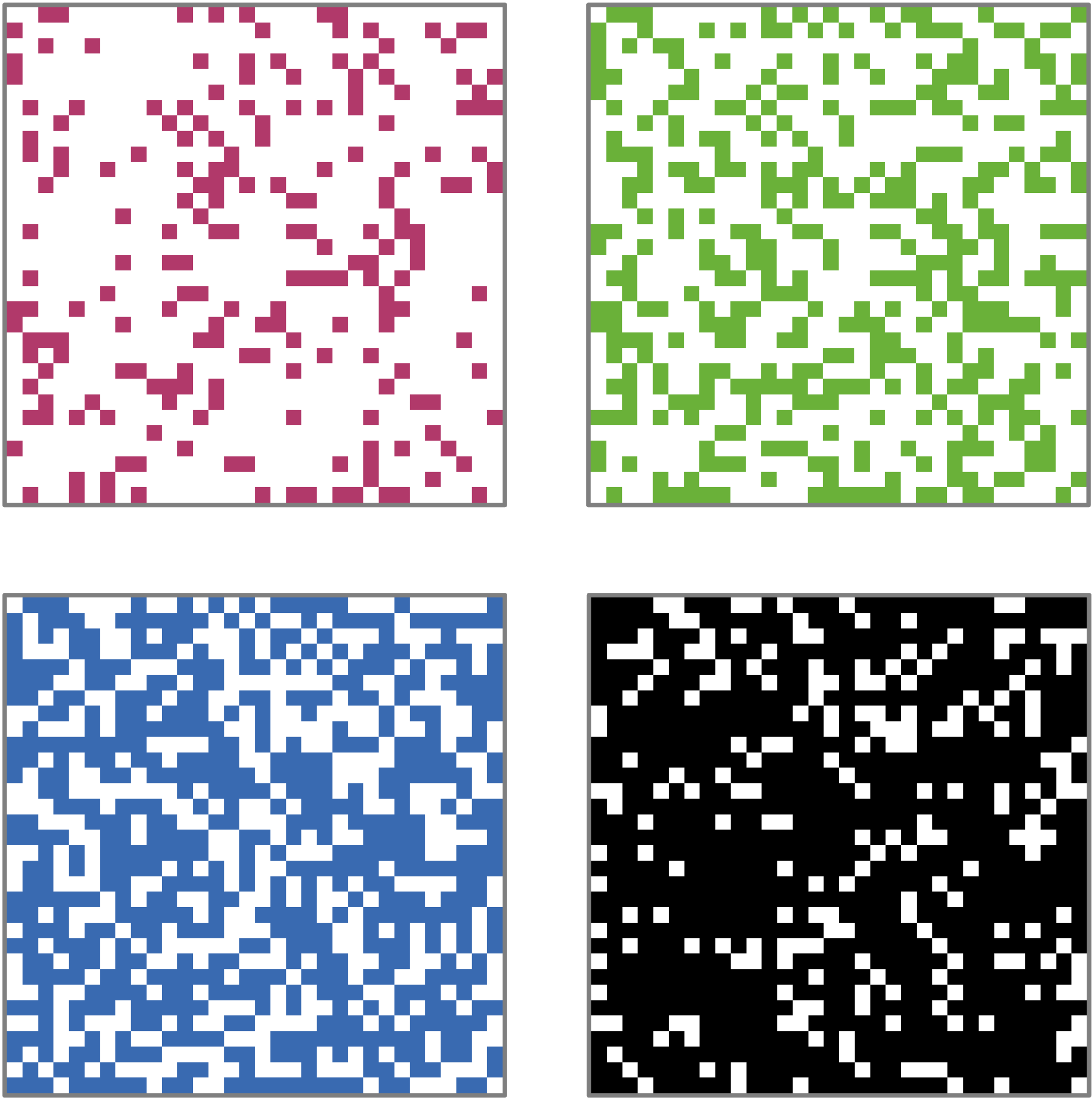}
    }
    \par\vspace{1cm}
    \includegraphics[height=0.25\linewidth]{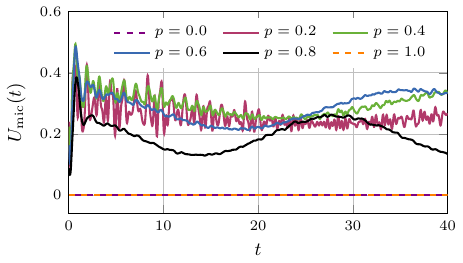}
    \hspace{1cm}
    \raisebox{0.04\linewidth}{%
        \includegraphics[height=0.2\linewidth]{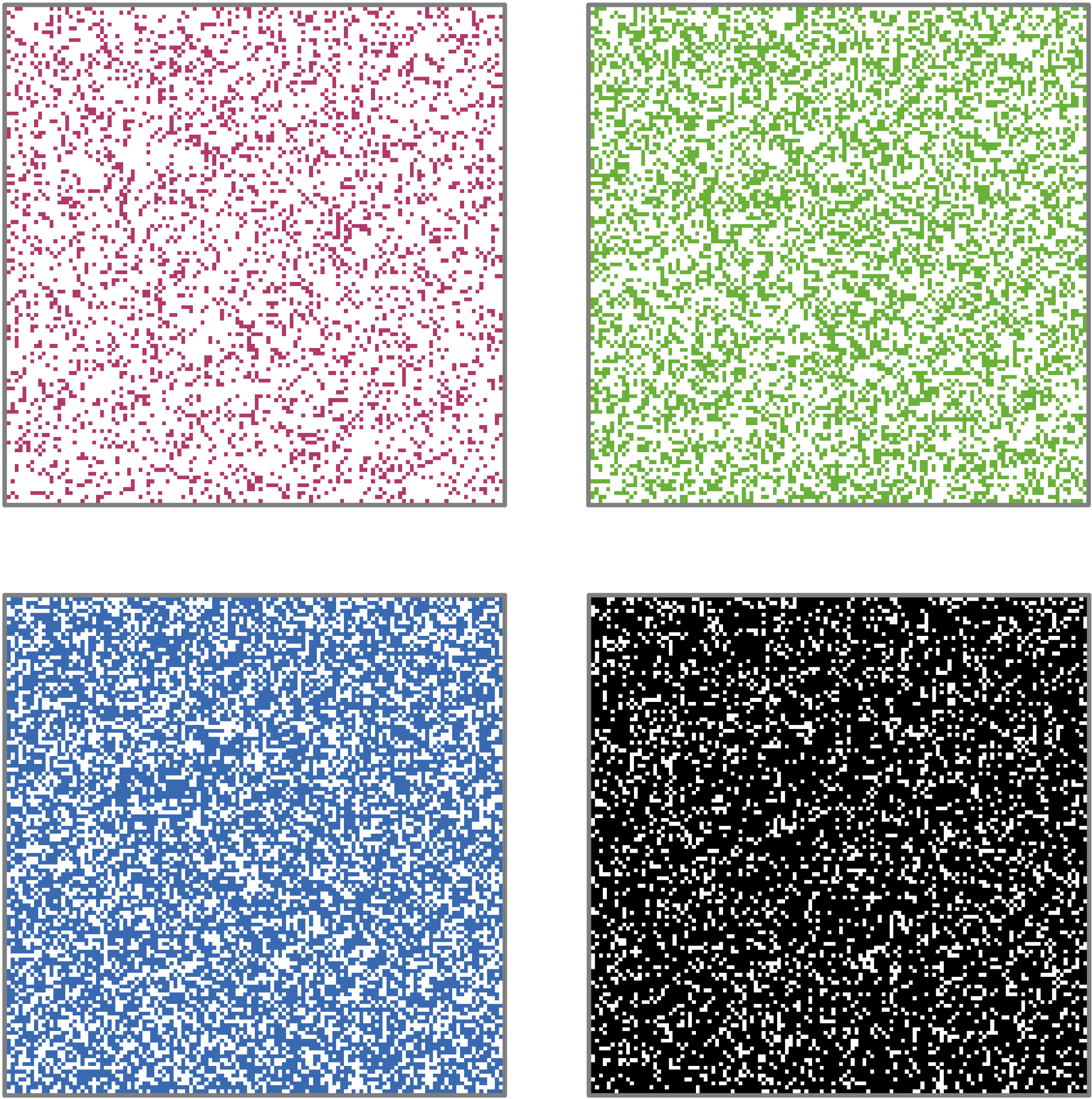}
    }
    \caption{\textbf{Microscopically resolved bulk observables for 2-d heterogeneous lattices.} Simulation of the microscopic potential energy $U_{\rm{mic}}(t)$ over time $t$ for $L\times L$ square lattices (top: $L=32$; bottom: $L=128$) with unit lattice constant $a=1$ (positions of vertices being integers $(n_x,n_y)$) and material properties at each vertex sampled identically and independently from two possibilities. The 2D square lattice uses a discrete symmetric strain tensor $\epsilon_{ij}=(\partial_i u_j+\partial_j u_i)/2$, with derivatives represented by periodic finite differences. The three strain channels at every vertex are $\xi=[(\epsilon_{xx}+\epsilon_{yy})/\sqrt{2},\,(\epsilon_{xx}-\epsilon_{yy})/\sqrt{2},\,\sqrt{2}\epsilon_{xy}]$, comprising the volumetric strain and the two components of deviatoric strain in two dimensions. Local constitutive matrices at each vertex are given by $C^{(1)}=\operatorname{diag}(20,18,18)$ with probability $p$ or $C^{(0)}=\operatorname{diag}(4,2,2)$ otherwise (actual realizations of these random material lattices are shown on the right, with colors corresponding to the values of $p$ shown on the left). The initial velocity is zero, the initial displacement is $u_{(n_x,n_y)}(0) =(0, \cos (2 \pi n_x/L) )$, and the microscopic cutoff is $\gamma_{\mathrm{mic}} =0.5 > 2\pi/L$, which ensures that the homogeneous lattice has no microscopic energy. }
    \label{fig:2d-lattice-sim}
\end{figure}

In the described setting, the leading-order homogeneous continuum model predicts zero microscopic-mode energy at all times. Its constant coefficients preserve spatial Fourier modes: an initial plane-wave displacement with zero initial velocity generates displacement and strain only at the original wavevector $k_0$. Consequently, given that $\gamma_0<\gamma_{\rm mic}$, the energy carried by modes above $\gamma_{\rm mic}$ remains zero at all times. But the numerical simulations for a two-dimensional square lattice clearly show that $U_{\rm mic}(t)$ is nonzero and of order $\Theta(1)$ at all times $t$ for $p\in(0,1)$ (see Fig.~\ref{fig:2d-lattice-sim}). Hence, $U_{\rm mic}(t)$ can be considered an MRBO if compared against a continuum homogenized model.

Interestingly, we show that one can construct better coarse-grained models than the fully homogenized one described above, that can predict the value of $U_{\mathrm{mic}}(0)$ without resorting to full lattice simulations handling $\Theta(N)$ degrees of freedom. In Appendix~\ref{app:MSBO-details}, we show that, with probability approaching one for increasing lattice sizes $N$, one has
\begin{equation}
    \left\|
        (\Pi_{\mathrm{mic}}\ot I_S) s(0)
    \right\|_2^2
    =
    N\|v \|^2
    \phi_{\mathrm{mic}}
    p(1-p)
    \left(
        \sqrt{c^{(1)}}-\sqrt{c^{(0)}}
    \right)^2
    +
    o(N),
    \label{eq:microscopic-energy-numerator2}
\end{equation}
where $\phi_{\mathrm{mic}} = \Theta(1)$ is the fraction of all lattice wavevectors $k$ contained in the microscopic sector. In addition, it is readily checked that the norm of $s(0)$ satisfies
\begin{equation}
    \|s(0)\|_2^2
    =
    N\|v \|^2
    \left[
        (1-p)c^{(0)}+pc^{(1)}
    \right]
    +
    o(N).
    \label{eq:total-energy-denominator}
\end{equation}
Since $\dot u(0)=0$, all of the initial energy is potential energy and $
    2E
    =
    \|s(0)\|_2^2$ and consequently the microscopic energy at $t=0$ admits a simple description in terms of the parameters $p, c^{(0)}, c^{(1)}, \phi_{\mathrm{mic}}$:
\begin{equation}
    U_{\mathrm{mic}}(0)
    =
    \phi_{\mathrm{mic}}
    \frac{
        p(1-p)
        \left(
            \sqrt{c^{(1)}}-\sqrt{c^{(0)}}
        \right)^2
    }{
        (1-p)c^{(0)}+pc^{(1)}
    }
    +
    o(1).
    \label{eq:binary-msbo-signal}
\end{equation}
What this showcases is that augmenting a continuum model with the above-mentioned parameters and Eq.~\eqref{eq:binary-msbo-signal} allows one to describe the microscopic energy at $t=0$ without the need to handle $\Theta(N)$ degrees of freedom. Moreover, numerical simulations for $t>0$ strongly suggest that $U_{\rm mic}(t)$ does not strongly depend on the microscopic structure, as different random realizations of the lattice yield similar results. Although deriving such a coarse-grained predictor for $U_{\rm mic}(t)$ may be nontrivial, its existence highlights the challenge of combining efficient bulk readout with microscopic sensitivity.

What we considered here was just a toy model of an MRBO with respect to the simplest continuum model. More sophisticated coarse-graining techniques exist~\cite{ye2024asymptotic}, and continuum models can always be augmented by statistical data coming from microscopic calculations, like in the example above. However, there are known elastic phenomena that are currently hard to explain with continuum models, and those are the best candidates for MRBOs, and in turn for the highest potential quantum advantage. These include short-wavelength phonon scattering by embedded nanoparticles, where lattice dispersion and finite phonon bandwidth cause deviations from heterogeneous continuum predictions~\cite{Kakodkar2015}; strongly frequency- and polarization-dependent phonon transmission across atomic interfaces, particularly near lattice cutoff frequencies where continuum acoustic-mismatch predictions fail~\cite{Schelling2002}; and nonaffine elastic responses and microscopic vibrational structure in amorphous solids below the length scale at which classical elasticity becomes accurate~\cite{Tanguy2002}. Within the harmonic regime, our quantum algorithm could probe energy signatures associated with these phenomena through measurements of spatial energy retention, reflected and transmitted wave-packet energies, and energy redistribution across wavevector regions. These provide candidates for testing the predictive limits of conventional continuum descriptions, although they do not establish that all effective continuum models must fail. 


\subsection{End-to-end complexity and speedup regime}
\label{sec:lattice_complexity}

Under the locality and efficient-access assumptions described in this section, the coherent quantum Hamiltonian-simulation cost is linear in the simulation time. Hence, for a fixed number of traversal times, the quantum complexity scales as
\begin{equation}
    \comp_Q=\tilde{\O}\left(N^{1/3}\right),
\end{equation}
which should be compared with the classical complexity $\O(N^{4/3})$ for that problem discussed in Sec.~\ref{sec:continuum_complexity}. Thus, one recovers a quartic speedup for coherent evolution. 

An end-to-end advantage, however, additionally requires that relevant classical observables can be extracted from the final quantum state without introducing a system-size-dependent overhead. This issue is particularly important in the microscopic setting: resolving the full lattice is computationally meaningful only when the observable of interest requires such lattice-scale simulation to be accurately estimated, since otherwise a coarser continuum model may suffice. In Sec.~\ref{sec:msbo} we characterized such microscopically sensitive resolved bulk observables that do not bring extra overhead cost related to their extraction, while in Sec.~\ref{sec:binary-random-msbo} we gave an explicit model of heterogeneous material with such an observable that can be efficiently simulated on a quantum computer. 

We have thus established a regime where a quartic advantage can be expected. More precisely, we identified this regime as the one in which the lattice model is necessary, the quantum readout remains efficient, and a quartic separation is not ruled out by an evident shortcut. The separation is lost if the readout cost grows polynomially with $N$, e.g., because the required normalized precision tightens with $N$, or if the output is obtained to the required accuracy from a coarser continuum model with polynomially fewer degrees of freedom. 


\section{Discussion and conclusions}
\label{sec:discussion}

Quantum algorithms for elastic dynamics offer the prospect of a quartic advantage over classical simulation, but realizing this separation requires more than an efficient quantum representation of the equations of motion. In this work, we identified how spatial regularity, physical scale, and observable sensitivity determine the attainable end-to-end advantage. By treating continuum elasticity and microscopic harmonic lattices within a common framework, we showed how these physical features constrain the costs of encoding the material, simulating its dynamics, and extracting the requested information.

For continuum elasticity, the lattice represents a numerical approximation of a system at fixed volume, and its size must therefore be chosen according to the required physical accuracy. Accounting for this dependence changes the complexity comparison obtained by considering the number of discretization points alone. For an attained spatial convergence order $m$, the quantum and classical costs scale as $\tilde{\O}((\bar t/\epsilon)^{1+1/m})$ and $O(\bar t(\bar t/\epsilon)^{4/m})$, respectively, with $\bar{t}$ denoting the simulation time and $\epsilon$ the accuracy of the simulation. Thus, the end-to-end speedups are of order $(m+4)/(m+1)$ in simulation time and $4/(m+1)$ in accuracy. Clearly, higher regularity, which enables higher-order classical discretizations, reduces the potential quantum advantage. For second-order convergence, the speedup is quadratic in simulation time, while the accuracy dependence improves from $\epsilon^{-2}$ to $\epsilon^{-3/2}$. Note that at sufficiently high convergence order, the quantum algorithm has a worse accuracy dependence than the classical benchmark, due to the quantum algorithmic cost of observable extraction. The quantum representation nevertheless retains an exponential memory advantage over explicitly storing the discretized elastic state, under the assumed efficient input access. 

For a fixed volume system that is physically discrete (e.g., a lattice or granular system), one can consider a range of computational problems. The regime where one is forced to use a Fourier bandwidth description $\Theta(1/a)$ would require us to use the maximal number of degrees of freedom to simulate the problem. For such systems, the number of degrees of freedom $N$ measures the size of the material rather than the refinement of a continuum approximation. For a three-dimensional local lattice evolved over a fixed number of traversal times, quantum simulation has cost $\tilde{\O}(N^{1/3})$, compared with $\O(N^{4/3})$ for standard explicit classical evolution. This quartic separation can extend to the complete computation at fixed observable accuracy, provided that material access, initial-state preparation, and readout introduce no additional polynomial dependence on system size. This in turn raises challenges for a quantum algorithm since highly complex lattice data itself might give rise to a data-loading bottleneck.

The choice of observable is essential to this regime. Microscopic heterogeneity alone does not imply that a microscopic simulation is necessary, since a continuum description may still reproduce the requested output. We therefore introduced microscopically resolved bulk observables: quantities that require lattice-scale simulation to be estimated while remaining efficiently extractable from the quantum state. This property makes discarding the microscopic degrees of freedom inadequate for the requested accuracy, while their normalized bulk signal avoids a readout cost that grows with system size. Estimating such observables is consequently a candidate problem for quartic end-to-end quantum advantage.

Pseudorandom heterogeneous materials provide a concrete setting in which these requirements can be brought together. Our binary-disorder model illustrates how a long-wavelength displacement can produce an order-one energy fraction stored in microscopic degrees of freedom, yielding a bulk observable sensitive to material structure beyond the homogenized continuum description. The numerical dynamical examples support this mechanism. Efficient `quantum cheese'~\cite{barthe2026make} encodings supply coherent access to complex material geometries through compact reversible generative rules, addressing the material-input cost without explicitly storing every local coefficient.

The resulting quartic advantage remains a comparison with standard classical lattice evolution, rather than a lower bound against all classical methods. Establishing the difficulty of estimating the selected dynamical observables with specialized classical algorithms, and quantifying fault-tolerant resource requirements, are important next steps. Our results nevertheless identify a physically motivated regime in which quartic advantage can be sustained: large microscopic elastic systems with succinct material descriptions and efficiently measurable observables that require lattice-scale information. More broadly, they show that the computational advantage of quantum elastic simulation is governed jointly by the regularity of the dynamics, the physical meaning of the simulated degrees of freedom, and the sensitivity of the information sought.

\bigskip

\textbf{Author contributions}: Authors are listed alphabetically. DJ proposed the project and developed theory for the microscopic elasticity regime. KK developed the discrete oscillator mapping and the quantum algorithm for continuum elasticity, contributed to the quantum algorithm for microscopic elasticity, and helped shape the discussion of the trade-off between regularity and quantum advantage. ML studied how spatial regularity affects quantum advantage in continuum elasticity and helped identify lattice simulation estimation problems with potential quartic advantage. TV developed the initial continuum-to-lattice formulation and quantum algorithm, including the Hamiltonian embedding and complexity analysis. All authors wrote the paper.

\bigskip

\textbf{Acknowledgments:} We thank all our colleagues at PsiQuantum for useful conversations and feedback, and in particular Paul Mannix and Alice Barthe. During the preparation of this manuscript, the authors used ChatGPT (GPT-6, OpenAI) to assist with drafting and revising text, organizing explanations, and checking clarity and consistency. The authors reviewed and edited the generated material and take full responsibility for the contents of the manuscript.

\bigskip

\appendix
\renewcommand{\theHequation}{\Alph{section}.\arabic{equation}}


\section{Elasticity theory of heterogeneous materials}
\label{app:elastic-theory}

In this appendix, we develop the theory of elasticity from a lattice perspective, taking discrete lattice dynamics as the starting point and obtaining continuum elasticity from its long-wavelength sector. We begin in Sec.~\ref{app:elastic_basic} with elementary oscillator examples that introduce the displacement variables, deformation map, constitutive map, and interaction space used throughout the paper. We then generalize these ideas in Sec.~\ref{app:elastic_matrix} to formulate lattice elasticity in terms of mass, deformation, constitutive, and stiffness operators. Next, in Sec.~\ref{app:elastic_fourier}, we use translation symmetry and Fourier analysis to distinguish the geometric structure of the lattice from spatial variations in its material properties. Subsequently, in Sec.~\ref{app:elastic_continuum}, we derive continuum strain, energy, stress, and equations of motion from the long-wavelength limit of the lattice model. Finally, in Sec.~\ref{app:elastic_isotropic}, we specialize the resulting continuum theory to isotropic heterogeneous materials, introduce the volumetric–deviatoric decomposition of strain, and derive the equation describing the continuum elasticity dynamics.


\subsection{Basic concepts of lattice elasticity}
\label{app:elastic_basic} 

Consider a finite lattice whose vertices are indexed by $ n\in\Lambda$. At each vertex, the displacement has three Cartesian components. We therefore introduce the vertex and configuration spaces
\begin{equation}
    \label{eq:hilbert_v_c}
    \H_V
    =
    \operatorname{span}\left\{| n\rangle: n\in\Lambda\right\},
    \qquad
    \H_3
    =
    \operatorname{span}\left\{|i\rangle:i\in\{1,2,3\}\right\},
\end{equation}
and represent the displacement field as a vector $u(t)\in
    \H_V\otimes\H_3$:
\begin{equation}
    u(t)
    =
    \sum_{ n\in\Lambda}
    \sum_{i=1}^{3}
    u_{ n,i}(t)
    | n\rangle\otimes|i\rangle.
\end{equation}
The component $u_{ n,i}(t)$ is the displacement of vertex $n$ in the Cartesian direction $i$. Physically, $u$ is real-valued, although it is convenient to work in the complexification of this space when using Fourier analysis.

Our aim is to give a general description for a heterogeneous elastic lattice system. Ultimately, this is a generalization of the elementary 1-d simple harmonic oscillator, which is fully defined by the potential energy formula
\begin{equation}
    U = \frac{1}{2} \kappa x^2,
\end{equation}
where $x$ is the displacement of the oscillator and $\kappa$ is the `spring constant' that sets the energy scale. The equation of motion for the oscillator is then obtained by considering variations $x \rightarrow x + dx$, so that 
\begin{equation}
    dU = \kappa x \, dx.
\end{equation}
Then, the restoring force is $-dU/dx=-\kappa x$, and Newton's Law for a particle of mass $m$ is $m\ddot{x} = -\kappa x$. In the above, we see that $x$ is the kinematical degree of freedom, $m$ is the inertial parameter, while $\kappa$ is a parameter that encodes a constitutive physical property of the oscillator system. 

A slightly less trivial example is given by a 3-oscillator system on 3 collinear lattice points spaced by a lattice spacing~$a$. We can denote the displacement of each oscillator from its respective lattice point by $u_1, u_2,$ and $u_3$. Suppose neighboring oscillators are coupled by a spring that stores energy according to the relative displacement of the oscillator pair. Thus, while the system has the three displacement variables $u = (u_1,u_2, u_3)$, the energetic degrees of freedom have a different dimension, and are only determined by the variables $(u_2-u_1, u_3-u_2)$. We can thus define \emph{strain} variables
\begin{align}
    \xi_1 &:= u_2-u_1, \qquad
    \xi_2 :=  u_3-u_2 
    \label{eq:3-in-a-line}.
\end{align}
The strain vector, $\xi =( \xi_1,\xi_2)$, is linearly related to the displacement $u$ by a \emph{strain-displacement matrix} $D$ via $\xi = D u$, with the matrix components of $D$ obtained from Eq.~\eqref{eq:3-in-a-line}. The strain $\xi$ is the pure kinematic data for the system. With spring constants $\kappa_1$ and $\kappa_2$, the total potential energy of the system is
\begin{equation}
    \label{eq:toy_potential}
    U = \frac{1}{2} \kappa_1 \xi_1^2 + \frac{1}{2} \kappa_2 \xi_2^2,
\end{equation}
which can also be expressed in terms of the displacement field $u$ as
\begin{equation}
    U(u) = \frac{1}{2} (D u)^\dagger C (D u).
\end{equation}
Here, $C = \mathrm{diag}(\kappa_1, \kappa_2)$ is a \emph{constitutive matrix} that encodes the physical properties of the two springs. While the strain vector $\xi$ is 2-dimensional, Newton's laws describe the three masses associated to the three sites, and the force on mass $m_j$ is obtained from $U$ via differentiating with respect to $u_j$:
\begin{equation}
    {M} \ddot{u} = - \nabla U =: -{K} u.
\end{equation}
In the above, ${M}=\mathrm{diag}(m_1,m_2,m_3)$ is a \emph{mass matrix}, $\nabla$ is the gradient with respect to the function's argument $u=(u_1,u_2,u_3)$, ${K}$ is a \emph{stiffness matrix} that can easily be computed from $U(u)$, and the resultant expression is the equation of motion for the full 3-body system. In what follows, we treat separately the main ingredients introduced in this subsection, in particular the inertial properties, kinematics, and constitutive response, and generalize them to the heterogeneous elastic lattice.


\subsection{Operator formulation of lattice elasticity}
\label{app:elastic_matrix} 

At the highest level, the elastic system is a 3-d lattice system that has Hamiltonian dynamics in the vicinity of some stable equilibrium point. It is therefore determined by the full potential energy $U$ of the system and its inertial properties. In what follows, we will leverage Hilbert space notation to blend more naturally with the quantum algorithm, and separate out a vertex register that encodes which vertex we are considering on the lattice. For isotropic point masses whose values may vary spatially, we define the mass matrix $ M$ acting on $\H_V\ot\H_3$ as
\begin{equation}
     M = \sum_{ n\in\Lambda} m_n \ketbra{n}{n} \otimes I_3
\end{equation}
Newton's law then takes the form
\begin{equation}
     M\ddot u(t) = f_{\mathrm{ext}}(t) + f_{\mathrm{int}}(t),
\end{equation}
where $f_{\mathrm{ext}}$ is the externally applied force and $f_{\mathrm{int}}$ is the force generated by the internal interactions of the lattice. For simplicity, in our work we assume $f_{\mathrm{ext}} =0$.

It is now important to distinguish displacement variables $u$ from those variables that are energetically relevant. Let $\H_I$ be an interaction space whose basis vectors label the local bonds, cells, or deformation channels of the lattice. We introduce a strain-displacement map
\begin{equation}
     D: \H_V\otimes\H_3 \longrightarrow \H_I,
\end{equation}
and define the generalized strain
\begin{equation}
    \xi :=  D u.
\end{equation}
The components of $\xi$ are precisely all energy-bearing degrees of freedom that are linearly related to~$u$. They may represent bond extensions, shear deformations, volumetric deformations, angle changes, or other local deformation measures, and so the specifics of the map $D$ depend on the problem at hand. Whether one includes any particular degree of freedom in the specification $\xi$ depends on whether that degree of freedom carries potential energy (for whatever reason). The kinematical variable $\xi$ should therefore be understood as the lattice generalization of strain data in the 3-site toy model from the previous section to a much broader class of systems, and provides the lattice variables that describe perturbations away from some stable equilibrium point. 

The strains are energetically counted via a generalization of the potential energy from Eq.~\eqref{eq:toy_potential}. This potential energy depends on the physics of the lattice, in terms of stiffness coefficients encoding the material properties (i.e.,~generalized `spring constants'). More precisely, the elastic potential energy $U$ is a quadratic form
\begin{equation}
    U(u):=\frac{1}{2} \xi^\dagger C \xi =\frac{1}{2}
    \langle  D u, C D u\rangle_{\H_I},
    \label{eq:elastic-energy}
\end{equation}
where $ C:\H_I\longrightarrow\H_I$ is a linear operator known as the constitutive map. It encodes the material properties and, in order for $U(u)$ to be a stable potential, must be self-adjoint and positive semidefinite. The constitutive map $C$ generalizes the spring constant $\kappa$ in the single-oscillator model, and maps the strain $\xi$ to stress $\sigma$:
\begin{equation}
    \sigma := C\xi= C D u.
\end{equation}

The energy functional $U(u)$ defines the entire elastic problem, and we can now deduce the equation of motion for the system from a variational approach. For a variation $u \rightarrow u+v$ of the displacement field, we have that the potential energy changes by
\begin{equation}
    \delta U(u;v)
    =
    \mathrm{Re}\langle  D v, C D u\rangle_{\H_I} + \O(\|v\|^2)
    =
    \mathrm{Re} \langle v, D^\dagger C D u\rangle_
    {\H_V\otimes\H_3}+ \O(\|v\|^2).
\end{equation}
This holds true for arbitrary perturbations, and by taking infinitesimal $v$, we see that the restoring force, $f_{\mathrm{int}} =-\nabla U$, for the elastic system is given by
\begin{equation}
    f_{\mathrm{int}}(u)=-D^\dagger C D u.
\end{equation}
Defining the stiffness operator
\begin{equation}
    K:=  D^\dagger C D,
\end{equation}
Newton's law in the absence of external driving becomes
\begin{equation}
    \label{eq:lattice_newton}
     M\ddot u(t) = - K u(t).
\end{equation}

\begin{figure}
    \centering
    \includegraphics[width=0.45\linewidth]{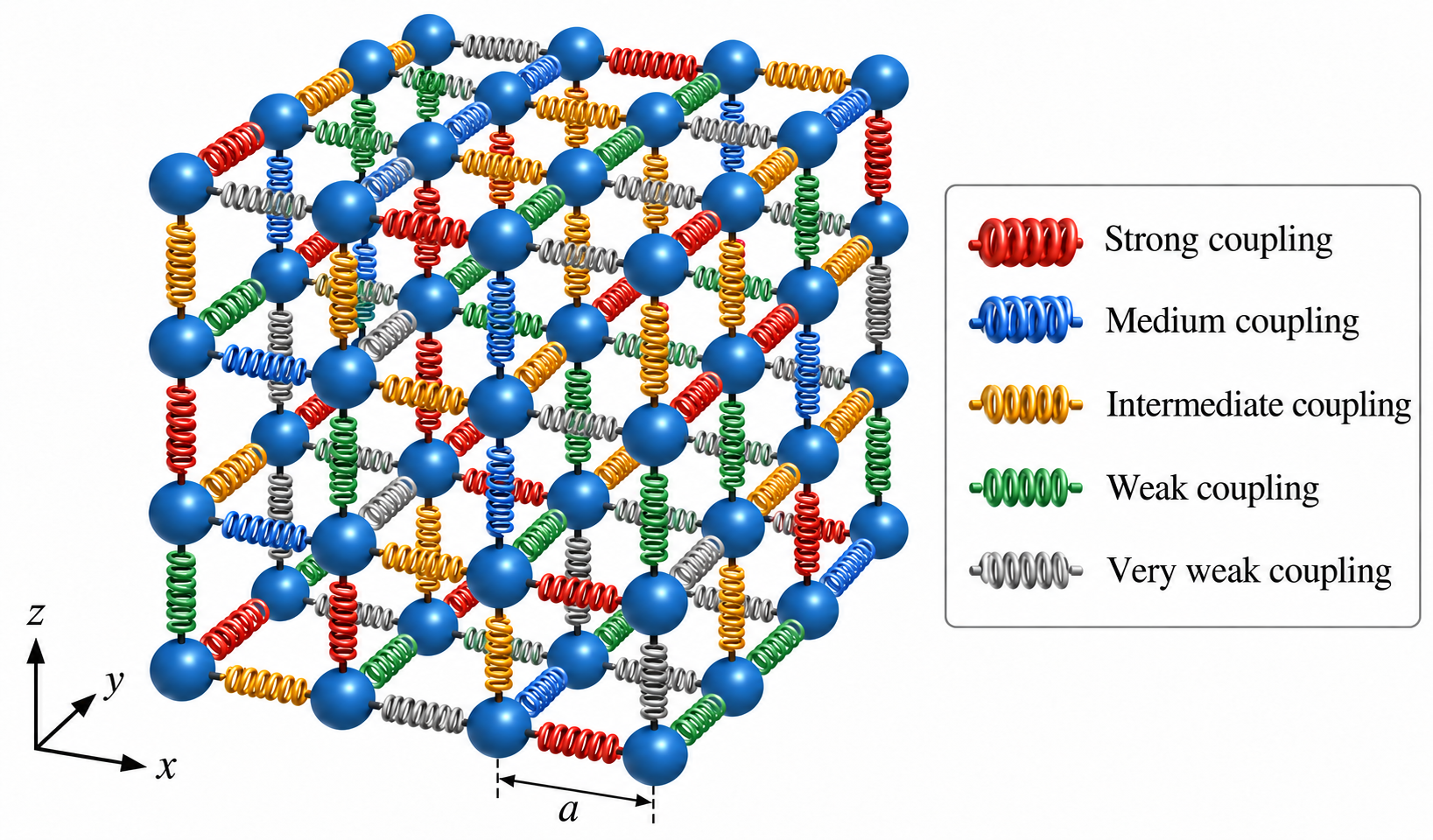}
    \caption{\textbf{Elastic dynamics of heterogeneous materials.} We consider a heterogeneous lattice of oscillators where the constitutive properties can vary across the lattice. The displacement $u_{{n},i}$ at any site ${n}$ and in any axis direction $i\in\{1,2,3\}$ corresponds to a local deformation of the lattice. These displacements are linearly related via the strain-displacement map $\D$ to the strain variable $\xi=\D u$ that encodes the kinematics of the non-trivial energy-carrying degrees of freedom of the lattice (e.g., local couplings represented here by nearest-neighbors springs). The actual material properties, in terms of varying coupling strengths, are encoded in the constitutive matrix $\C$ (shown here schematically as varying spring strengths).}
    \label{fig:hetero-lattice}
\end{figure}

The factorization of ${K}$ therefore has the stress-strain formulation
\begin{equation}
    u
    \xrightarrow{\  D\ }
    \xi
    \xrightarrow{\  C\ }
    \sigma
    \xrightarrow{\  D^\dagger\ }
     K u.
\end{equation}
The operator $D$ computes local strains from displacements, $C$ determines the material response to these strains, and $D^\dagger$ assembles the resulting local stresses into forces on the vertices. The above provides a simple way to understand general elastic systems, with Eq.~\eqref{eq:elastic-energy} being the fundamental equation providing the description of the dynamics of the system. The separation into these operators is particularly useful for describing heterogeneous lattice systems. Figure~\ref{fig:hetero-lattice} illustrates this distinction on a regular cubic lattice, where the local interaction geometry is identical at every site but the coupling strengths vary spatially. The common geometric structure is encoded in $D$, which can therefore remain \emph{covariant under lattice translations}, while the spatially varying constitutive properties are encoded in $C$, which generally breaks translation symmetry. Consequently, Fourier transformation is simple for $D$, whereas the Fourier representation of $C$ reveals the spatial spectrum of the inhomogeneity and its effect on the dynamics. We develop this representation in the next subsection.

\subsection{Translation symmetry and Fourier representation}
\label{app:elastic_fourier}

To analyze heterogeneous materials that break translation symmetry and exhibit variations across different length scales, we will employ a discrete Fourier representation of the elastic lattice system. One can construct this representation by examining how the discrete group of lattice translations acts on each of its components. For definiteness, consider a periodic cubic lattice
\begin{equation}
    \Lambda=\mathbb Z_{L_1}\times\mathbb Z_{L_2}\times\mathbb Z_{L_3},\qquad N=L_1L_2L_3, 
\end{equation}
with lattice spacing $a$. The vertex and configuration spaces from Eq.~\eqref{eq:hilbert_v_c} thus satisfy $\dim(\H_V\otimes\H_3)=3N$. We now assume that the local interaction variables can be labeled by a lattice label \( n\) and an internal strain or interaction channel $s$. Writing
\begin{equation}
    \H_S=\operatorname{span}\{|s\rangle:s=1,\ldots,r\},
\end{equation}
we may identify
\begin{equation}
    \H_I=\operatorname{span}\{| n,s\rangle_I\}\simeq\H_V\otimes\H_S,
\end{equation}
and so we can equally write $| n,s\rangle_I\longleftrightarrow| n\rangle_V\otimes|s\rangle_{S}$. The integer $r$ is the number of local strain channels per lattice cell, namely the number of energy-bearing degrees of freedom that depend on the displacement $u_{{n}}$. Consequently, the dimension of the interaction space is $\dim\H_I=Nr.$ 

A lattice translation \(\ell\in\Lambda\) acts by $
T_{\ell}^{(V)}| n\rangle =| n+\ell\rangle $. We assume a regular lattice with the same strain channels at each site. Since translations do not change Cartesian or strain-channel labels,
\begin{equation}
    T_{\ell}^{(VC)}=T_{\ell}^{(V)}\otimes I_{\H_3},
    \qquad
    T_{\ell}^{(I)}=T_{\ell}^{(V)}\otimes I_{\H_S}.
\end{equation}
Having the same strain channels across the lattice means we can take the strain-displacement map $D$ to be covariant, and to obey
\begin{equation}
    T_{\ell}^{(I)}D=D\bigl(T_{\ell}^{(V)}\otimes I_{\H_3}\bigr).
\end{equation}
Note that the pair $(D,C)$ is not uniquely defined, since only the form of $U(u)$, and thus $D^\dagger C D$, matters. Therefore one could artificially have that $D$ breaks covariance, but on a regular lattice this is not a natural thing to have, and it makes sense to have all of the translation-breaking aspects within $C$.

Let
\begin{equation}
    \mathcal K
    :=
    \left\{
        k:\;
        k_j=\frac{2\pi m_j}{L_j a},
        \quad
        m_j=-\lfloor L_j/2\rfloor,\ldots,\lceil L_j/2\rceil-1.
    \right\}
\end{equation}
be the reciprocal lattice, and define $\H_K$ as the corresponding Hilbert space. The discrete Fourier transform $F_V:\H_V\to\H_K$ is given by
\begin{equation}
    F_V
    =
    \frac{1}{\sqrt N}
    \sum_{k\in\mathcal K}\sum_{n\in\Lambda}
    e^{-\i k\cdot x_n}
    \ketbra{k}{n},
\end{equation}
where $x_n=an$. Equivalently, the corresponding Fourier modes in $\H_V$ are
\begin{equation}
    F_V^\dagger|k\rangle
    =
    \frac{1}{\sqrt N}
    \sum_{n\in\Lambda}
    e^{\i k\cdot x_n}|n\rangle.
\end{equation}
The transformation can be extended by acting trivially on the internal degrees of freedom:
\begin{equation}
    F_{VC}=F_V\otimes I_{\H_3},
    \qquad
    F_I=F_V\otimes I_{\H_S}.
\end{equation}
Hence $\H_V\otimes\H_3
\simeq \bigoplus_{{k}}\mathbb{C}^{3} \simeq \mathbb{C}^{3N}$ and $\H_I \simeq \bigoplus_{{k}}\mathbb{C}^{r}\simeq \mathbb{C}^{rN} $.
Given components $u_{n,i}
\quad\text{and}\quad
\xi_{ n,s}=( D u)_{ n,s}$, the Fourier space components are then given explicitly as
\begin{align}
    \label{eq:fourier_coeffs}
    \widehat u_i( k)&=\frac{1}{\sqrt N}\sum_{ n}e^{-\i k\cdot x_{ n}}u_{ n,i} \qquad
    \widehat \xi_s( k)=\frac{1}{\sqrt N}\sum_{ n}e^{-\i k\cdot x_{ n}}\xi_{ n,s}.
\end{align}
The covariance of $D$ implies that it does not mix modes~\cite{terras1999fourier}, and so is block diagonal in Fourier space: 
\begin{equation}
    \widehat{D}:= F_{I} D F_{VC}^\dagger
    =
    \bigoplus_{ k}\widehat{{D}}( k),
    \qquad
    \widehat{{D}}( k):\mathbb C^3\longrightarrow\mathbb C^r.
    \label{eq:block-diag-D}
\end{equation}

For local strains of the form
\begin{equation}
    (D u)_{n,s}
    =
    \sum_{\delta,i}
    d_{s,i}(\delta)
    u_{n+\delta,i},
\end{equation}
its Fourier representation is
\begin{equation}
    \widehat{\xi}_{s}(k)
    =
    \sum_{i}
    \widehat{ D}_{s,i}(k)
    \widehat u_{i}(k),
    \qquad
    \widehat{ D}_{s,i}(k)
    =
    \sum_{\delta}
    d_{s,i}(\delta)
    e^{\i k\cdot a\delta}.
\end{equation}
For example, a forward difference in direction $j$ results in $e^{\i k_j a}-1$. If the constitutive map $C$ is also translation invariant, then it is block diagonal in the same decomposition and
$
\widehat{K}( k)
=
\widehat{{D}}( k)^\dagger
\widehat{{C}}( k)
\widehat{{D}}( k).
$
In contrast, for spatially varying material parameters, \(\widehat{ D}\) remains diagonal in  $k$ when the lattice geometry is regular, but \(\widehat{C}\) generally couples different wavevectors:
\begin{equation}
    \widehat{ C}( k, k')
    =
    \frac{1}{N}
    \sum_{ n}
    e^{\i( k'- k)\cdot x_{ n}}
     C(x_n).
\end{equation}
Thus material inhomogeneity appears in Fourier space as wavevector mixing.


\subsection{Long-wavelength continuum limit}
\label{app:elastic_continuum}

We have described the core features within a lattice setting, however very often one considers a continuum description with the elastic dynamics given by smooth partial differential equations. These are related to lattice models as follows. Take the discrete lattice system as fundamental. Focus on a cubic lattice with spacing \(a\), and let \(u_{ n}(t)\in\mathbb R^3\) denote the displacement of the site \( n\) at position $x_n=an$, which evolves according to Eq.~\eqref{eq:lattice_newton}. No spatial homogeneity of the masses or constitutive parameters is assumed. We may nevertheless Fourier transform the lattice degrees of freedom to obtain the exact dynamics in the Fourier space given by
\begin{equation}
    \sum_{ k'}
    \left[
    \widehat{ M}( k, k')
    \frac{\mathrm d^2}{\mathrm dt^2}
    \widehat u( k',t)
    +
    \widehat{ K}( k, k')
    \widehat u( k',t)
    \right]
    =
    0,
\end{equation}
where $\widehat u( k,t)$ was defined in Eq.~\eqref{eq:fourier_coeffs} and
\begin{equation}
    \widehat{ K}( k, k')=\widehat{D}( k)^\dagger\widehat{C}( k, k')\widehat{D}( k'),
\end{equation}
due to translation-covariance of $\D$ (recall Eq.~\eqref{eq:block-diag-D}).

The continuum description is extracted from the long-wavelength expansion of the strain-displacement map $D$. A simple 1-d toy model suffices to explain how this occurs. Consider a ring of oscillators with lattice points $x_n = na$, $n\in \mathbb{Z}_N$ and $L=Na$. The nearest-neighbor bonds imply
\begin{equation}
    \xi_n = (D u)_n = u_{n+1} - u_n,
\end{equation}
with periodicity $u_{N+1}=u_1$. The discrete Fourier transform gives that
\begin{equation}
    \widehat{\xi}(k) = (e^{\i ka}-1)\widehat{u}(k),
\end{equation}
and so the strain-displacement map is purely diagonal with $\widehat{D}(k) = (e^{\i ka}-1)$. The continuum limit is obtained by taking $a\rightarrow 0$, with $L$ held fixed. We observe that expanding in powers of $ka$ gives
\begin{equation}
    \widehat{D}(k) = \i ka + \O(a^2k^2). 
\end{equation}
Now, noting that for a continuum field $\partial_x u(x)= \lim_{a \rightarrow 0} (u_{n+1}-u_n)/a$, we see that the order-$a$ term is precisely the spatial derivative and dominates in the $a \rightarrow 0$ limit. This pattern holds more generally, and for a 3-d model and \(| k|a\ll1\), we have that 
\begin{equation}
    \label{eq:D_long_wave}
    \widehat{D}( k) = a\,\widehat{\D}( k)+ \O\!\left(a^{2}| k|^{2}\right),
\end{equation}
where $\widehat{\D}( k)$ is the continuum limit displacement map given in Fourier space. Equivalently,
\begin{equation}
    \widehat{\mathcal D}( k)  =  \lim_{a\to0}  \frac{1}{a}\widehat{ D}( k).
\end{equation}

Therefore, the continuum strain $\varepsilon$ is obtained from the leading-order term in the long-wavelength expansion $\D$ of the discrete strain-displacement map $D$, while modes satisfying $|k|a=\O(1)$ retain genuinely microscopic lattice behavior. This kinematic statement does not require spatially homogeneous material parameters. If the inhomogeneity varies smoothly on length scales much larger than $a$, the long-wavelength sector produces a variable-coefficient continuum theory. If it varies on the scale of the lattice spacing, long- and short-wavelength modes are mixed, and obtaining a continuum description requires an additional homogenization or effective-medium argument. For a local strain-displacement map $D$ and a displacement field $u$ that varies smoothly on the lattice scale, a Taylor expansion shows that the leading long-wavelength contribution to the discrete strain $\xi=D u$ is proportional to $a\nabla u$. The zeroth-order term vanishes because $D$ must annihilate uniform translations, whereas the invariance under rigid-body rotations implies that the leading term can depend only on the symmetric part of $\nabla u$. After identifying the continuum strain channels with the independent components of a symmetric tensor, the resulting continuum strain field is therefore
\begin{equation}
    \varepsilon_{ij}(x_n)
    :=
    \frac{1}{2}
    \left(
        \partial_i u_{j}(x_n)+\partial_j u_{i}(x_n)
    \right),
    \label{eq:continuum-strain}
\end{equation}
where $u(x)$ is a continuum interpolating field with $u(x_n)=u_n$. Any fixed change of basis or redundancy in the discrete strain channels can be absorbed into the constitutive map and does not alter the continuum elastic theory. Thus, at leading order in the lattice spacing, the discrete and continuum strains are related by
\begin{equation}
    \xi_n
    =
    (D u)_n
    =
    a\,\varepsilon(x_n)
    +
    \O(a^2),
    \label{eq:discrete-continuum-strain}
\end{equation}
which is the real-space counterpart of Eq.~\eqref{eq:D_long_wave}.

The same long-wavelength scaling connects the discrete and continuum energies. In three dimensions, the discrete strains scale as $\O(a)$ and the local constitutive matrix weights as $\O(a)$. Consequently,
\begin{equation}
    U=\frac12
    \xi^{\dagger}
     C
    \xi
    =
    \frac12
    \sum_{ n}
    a^3\,
    \varepsilon^\dagger( x_{ n})
    \C( x_{ n})
    \varepsilon( x_{ n})
    +
    \mbox{(higher orders in $a$)}.
    \label{eq:continuum-elastic-energy_pre}
\end{equation}
Here, $a^3$ becomes the volume element in the continuum limit, whereas $\C$ is the continuum elasticity tensor that acts on the strain tensor $\varepsilon$ to produce the continuum stress tensor $\tau$:
\begin{equation}
    \tau(x)
    :=
    \C(x)\varepsilon(x).
\end{equation}
Thus, $\tau$ is a second-order stress tensor, whereas $\C$ is a fourth-order tensor with four spatial indices:
\begin{equation}
    \tau_{ij}(x)
    =
    \C_{ijkl}(x)\varepsilon_{kl}(x).
    \label{eq:continuum-stress-constitutive}
\end{equation}
Holding the physical volume fixed and taking $a\rightarrow0$ in Eq.~\eqref{eq:continuum-elastic-energy_pre} gives the expression for the continuum elastic energy
\begin{equation}
    U
    =
    \frac{1}{2}
    \int_V
    \varepsilon^\dagger(x)
    \C(x)
    \varepsilon(x)
    \,\mathrm d^3x.
    \label{eq:continuum-elastic-energy}
\end{equation}
Finally, introducing $\rho(x)$ as the spatial mass density, the variation of Eq.~\eqref{eq:continuum-elastic-energy} gives the continuum elasticity equation of motion
\begin{equation}
    \rho(x)\ddot u(x,t)
    =
    \nabla\cdot\tau(x,t),
    \label{eq:general-continuum-eom}
\end{equation}
which is the continuum counterpart of the lattice equation $ M\ddot u=-Ku$. 


\subsection{Isotropic continuum elasticity}
\label{app:elastic_isotropic}

We now specialize the continuum theory obtained above to locally isotropic materials. At each point, the strain tensor belongs to the six-dimensional space of symmetric tensors,
\begin{equation}
    \varepsilon(x)\in\operatorname{Sym}(3).
\end{equation}
This space has a natural orthogonal decomposition into shape-changing and volume-changing components. Define the scalar volumetric strain and the deviatoric strain by
\begin{subequations}
\begin{align}
    \varepsilon^{\vol}
    &:=
    \operatorname{tr}\varepsilon
    =
    \nabla\cdot u,
    \label{eq:continuum-volumetric-strain}\\
    \varepsilon^{\dev}
    &:=
    \varepsilon
    -
    \frac{1}{3}\varepsilon^{\vol}I,
    \label{eq:continuum-deviatoric-strain}
\end{align}
\end{subequations}
and note that $\varepsilon^{\vol}$ here is the same quantity as $\gamma$ in Eq.~\eqref{eq:gamma_evol}. The strain tensor can then be reconstructed as
\begin{equation}
    \varepsilon
    =
    \varepsilon^{\dev}
    +
    \frac{1}{3}\varepsilon^{\vol}I.
\end{equation}
The deviatoric component is trace-free and describes shear and differential stretching, whereas $\varepsilon^{\vol}$ measures the first-order relative change in volume. At the level of independent local strain channels, this gives
\begin{equation}
    \H_S
    \simeq
    \H_{\dev}\oplus\H_{\vol},
    \qquad
    \dim\H_{\dev}=5,
    \qquad
    \dim\H_{\vol}=1.
    \label{eq:strain-space-decomposition}
\end{equation}
These dimensions count independent physical strain components. A lattice encoding may retain a redundant set of Cartesianly labelled channels, as does the construction in Sec.~\ref{sec:discrete_mapping}, without changing the underlying five-plus-one decomposition.

The existence of the quadratic elastic energy, together with the symmetry of the strain and stress tensors, gives the standard major and minor symmetries of the elasticity tensor $\C$:
\begin{equation}
    \C_{ijkl}
    =
    \C_{klij},
    \qquad
    \C_{ijkl}
    =
    \C_{jikl},
    \qquad
    \C_{ijkl}
    =
    \C_{ijlk}.
    \label{eq:syms}
\end{equation}
Local isotropy, namely invariance of the elastic response under rotations, then restricts the elasticity tensor to
\begin{equation}
    \C_{ijkl}(x)
    =
    \lambda(x)\delta_{ij}\delta_{kl}
    +
    \mu(x)
    \left(
        \delta_{ik}\delta_{jl}
        +
        \delta_{il}\delta_{jk}
    \right),
    \label{eq:C_tensor}
\end{equation}
where $\lambda(x)$ and $\mu(x)$ are the Lam\'e parameters. Substituting the above into Eq.~\eqref{eq:continuum-stress-constitutive} and using the definitions from Eqs.~\eqref{eq:continuum-volumetric-strain}-\eqref{eq:continuum-deviatoric-strain} gives
\begin{align}
    \tau
    &=
    \left(\lambda+\frac{2}{3}\mu\right)\varepsilon^{\vol}I + 
    2\mu\varepsilon^{\dev}.
    \label{eq:isotropic-stress}
\end{align}
Thus, $\mu$ controls the response to shape-changing deformations, while $\lambda+2\mu/3$ (known as the bulk modulus) controls the response to changes in volume. The corresponding elastic energy is
\begin{equation}
    U=
    \int_V
    \left[
        \mu(x)
        \left\|
            \varepsilon^{\dev}(x)
        \right\|^2
        +
        \frac{\lambda(x)+\frac{2}{3}\mu(x)}{2}
        \left(
            \varepsilon^{\vol}(x)
        \right)^2
    \right]
    \mathrm d^3x.
    \label{eq:isotropic-continuum-energy}
\end{equation}
Nonnegative elastic energy requires $\mu\geq0$ and $\lambda+2\mu/3\geq0$, with strict inequalities for a strictly stable material.

Substituting Eq.~\eqref{eq:isotropic-stress} into the general elastic equation of motion, Eq.~\eqref{eq:general-continuum-eom}, gives
\begin{subequations}
\begin{align}
    \rho \Ddot{u}_i&= \sum_j\partial_j \tau_{ij}=\sum_{j,k,l}\partial_j(\C_{ijkl}\varepsilon_{kl})=\sum_{j,k,l}\partial_j(\C_{ijkl}\partial_k u_l),\label{eq:macro_3}\\
&=\sum_{j} \left [\lambda\partial_i \partial_j u_j + (\partial_j u_j)\partial_i \lambda + \mu \partial_i \partial_j u_j + (\partial_i u_j)\partial_j \mu+ (\partial_j \mu)\partial_j u_i + \mu \partial_j \partial_j u_i \right], \label{eq:macro_4}
\end{align}
\end{subequations}
or, in vector form,
\begin{align}
    \rho\frac{\partial^2u}{\partial t^2}
    ={}&
    (\lambda+\mu)\nabla(\nabla\cdot u)
    +
    (\nabla\cdot u)\nabla\lambda
    +
    \left[
        \nabla u+(\nabla u)^{\mathsf T}
    \right]\nabla\mu
    +
    \mu\nabla^2u,
    \label{eq:mat_wave_eq_vector}
\end{align}
which is precisely the equation of motion for a heterogeneous isotropic material from Eq.~\eqref{eq:wave_eq}.


\section{Sparse lattice model for continuum isotropic elasticity}
\label{app:micro}

Appendix~\ref{app:elastic-theory} established how continuum elasticity emerges from the long-wavelength limit of a general elastic lattice, but it did not select a unique lattice realization: many discrete models can have the same continuum limit. Here we address the complementary constructive problem required by the quantum algorithm by explicitly constructing a lattice model that combines controlled accuracy with a structure suitable for efficient quantum encoding. We introduce local discrete strains, together with spatially varying deviatoric and volumetric stiffness coefficients, and derive the resulting sparse lattice equations of motion. We then show that these equations reproduce the continuum elasticity equation with a controlled second-order spatial discretization error, derive its nondimensional form used in the main text, and finally extend the construction to arbitrary even-order finite-difference stencils. 


\subsection{Coupled oscillators model}
\label{app:oscillators}

We work in physical units throughout Appendix~\ref{app:oscillators} and \ref{app:consistency}, as this convention makes the continuum limit in Appendix~\ref{app:consistency} direct, without requiring any prior rescaling. The nondimensional variables used in the main text and in the quantum algorithm are introduced in Appendix~\ref{app:nondim}. Consider then a three-dimensional periodic Cartesian lattice with a constant lattice spacing $a$ and a total number of $N$ sites. A site at position $\bar{{x}}$ has a mass $\bar{\rho}(\bar{\v{x}})a^3$ and is displaced by $\bar{\v{u}}(\bar{\v{x}})$. The sites are coupled to each other, and the strength of interactions is determined by stiffness coefficients that are denoted with $\bar{\kappa}$, with a superscript specifying the type of interaction. We use two types of cell interactions, labeled by the interaction superscript $\{\dev,\vol\}$. The first label denotes deviatoric strains, which penalize changes in the local shape of a lattice cell while preserving its volume. The second label denotes volumetric strains, which penalize changes in the local volume of a lattice cell. To define these interactions, we introduce discrete strain variables $\bar{\xi}$ associated with the continuum strain tensor from Eq.~\eqref{eq:continuum-strain}:
\begin{equation}
    \label{eq:stran_tensor_discrete}
    \bar{\xi}_{ij}(\bar{\v{x}}) := \frac{1}{4}\left[\bar{u}_i(\bar{\v{x}}+a\v{e}_j)-\bar{u}_i(\bar{\v{x}}-a\v{e}_j)+\bar{u}_j(\bar{\v{x}}+a\v{e}_i)-\bar{u}_j(\bar{\v{x}}-a\v{e}_i)\right].
\end{equation}
We also employ its volumetric–deviatoric decomposition:
\begin{equation}
    \bar{\xi}_{ij}(\bar{\v{x}}) = \bar{\xi}_{ij}^\dev(\bar{\v{x}})+ \frac{\delta_{ij}}{3} \bar{\xi}^\vol(\bar{\v{x}})
\end{equation}
with
\begin{align}
    \bar{\xi}^\vol(\bar{\v{x}}) &= \sum_{i=1}^3 \bar{\xi}_{ii}(\bar{\v{x}}) = \frac{1}{2}\sum_{i=1}^3\left[\bar{u}_i(\bar{\v{x}}+a\v{e}_i)-\bar{u}_i(\bar{\v{x}}-a\v{e}_i)\right],
\end{align}
which mimics the continuum decomposition from Eqs.~\eqref{eq:continuum-volumetric-strain}-\eqref{eq:continuum-deviatoric-strain}. Then, the considered interactions are described by the following discrete potential:
\begin{align}
    \label{eq:potential_new}
    U[\bar{{u}}]
    =
    \frac{1}{2}\sum_{\bar{{x}}} \bar{\kappa}^{\dev}(\bar{\v{x}}) \sum_{i,j=1}^3 (\bar{\xi}^\dev_{ij}(\bar{{x}}))^2
    +
    \frac{1}{2}\sum_{\bar{{x}}} \bar{\kappa}^\vol(\bar{{x}})(\bar{\xi}^\vol(\bar{{x}}))^2,
\end{align}
which is the discrete counterpart of the continuum elastic energy from \eqref{eq:isotropic-continuum-energy}. The resulting force leads to the following equation of motion for the displacements $\bar{u}_i(\bar{\v{x}})$ along direction $i$ of discrete oscillators centered at the lattice sites $\bar{\v{x}}$: 
\begin{align}
\begin{split}
    \label{eq:oscillator_eq_of_motion}
    \bar{\rho}(\bar{{x}})\frac{\partial^2 \bar{u}_i(\bar{{x}})}{\partial \bar{t}^2}
    =
    &\frac{1}{2a^3}\sum_{j=1}^3
    \left[
    \bar{\kappa}^{\dev}(\bar{{x}}+a{e}_j)
    \bar{\xi}^\dev_{ij}(\bar{{x}}+a{e}_j)
    -
    \bar{\kappa}^{\dev}(\bar{{x}}-a{e}_j)
    \bar{\xi}^\dev_{ij}(\bar{{x}}-a{e}_j)
    \right] \\
    &+
    \frac{1}{2a^3}    \left[
    \bar{\kappa}^{\vol}(\bar{{x}}+a{e}_i)
    \bar{\xi}^\vol(\bar{{x}}+a{e}_i)
    -
    \bar{\kappa}^{\vol}(\bar{{x}}-a{e}_i)
    \bar{\xi}^\vol(\bar{{x}}-a{e}_i)
    \right].
\end{split}
\end{align}


\subsection{Recovering continuum elasticity}
\label{app:consistency}

By Taylor expanding Eq.~\eqref{eq:oscillator_eq_of_motion} with respect to $a$, one obtains (we defer the details of this expansion to Appendix~\ref{app:consistency_details}):
\begin{align}
    \label{eq:discrete_tensor_eq_of_motion}
    \bar{\rho}(\bar{{x}}) \frac{\partial^2\bar{u}_i(\bar{x})}{\partial \bar{t}^2} = 
    \partial_j\left[\frac{1}{a}\left(\bar{\kappa}^\dev(\bar{{x}})A^\dev_{ijkl}
    +\,\bar{\kappa}^\vol(\bar{{x}})A^\vol_{ijkl}\right) \partial_k \bar{u}_l(\bar{{x}})\right]+\O(a^2),
\end{align}
where the effect of the two considered interactions may be packaged into the tensors $A^\dev$ and $A^\vol$ with
\begin{equation}
    \label{eq:DV_tensors}
    A^\dev_{iiii} = \frac{2}{3},\qquad
    A^\dev_{iijj}=-\frac{1}{3},\qquad A^\dev_{ijij}= A^\dev_{ijji}=\frac{1}{2},\qquad
    A^\vol_{iiii}=A^\vol_{iijj} = 1,
\end{equation}
where it is assumed $i \neq j$. Now, if we compare Eq.~\eqref{eq:discrete_tensor_eq_of_motion} with Eq.~\eqref{eq:macro_3}, we clearly see that the coupled oscillator model recovers the continuum elasticity equation, up to the discretization error, with the elasticity tensor $\C$ expressed by the oscillator model parameters via
\begin{equation}
    \C(\bar{\v{x}}) = \frac{1}{a} \left(\bar{\kappa}^\dev(\bar{{x}})A^\dev
    +\,\bar{\kappa}^\vol(\bar{{x}})A^\vol\right). 
\end{equation}
Writing explicitly the nonzero entries of the elasticity tensor from Eq.~\eqref{eq:C_tensor}:
\begin{equation}
    \C_{iiii} = {\lambda}+2{\mu},\qquad
    \C_{iijj}={\lambda},\qquad
    \C_{ijij}= C_{ijji}={\mu},
\end{equation}
where again it is assumed that $i \neq j$, we see that the oscillator system from Eq.~\eqref{eq:oscillator_eq_of_motion} with stiffness coefficients chosen according to:
\begin{equation}   
    \label{eq:kappa_choice}
    \bar{\kappa}^\dev(\bar{{x}})=2a{\mu}(\bar{{x}}),\qquad
    \bar{\kappa}^\vol(\bar{{x}})= a\left[{\lambda}(\bar{{x}})+\frac{2}{3}{\mu}(\bar{{x}})\right],
\end{equation}
allows one to simulate the the continuum elasticity dynamics from Eq.~\eqref{eq:mat_wave_eq_vector} for a material with Lam\'e constants ${\lambda}$ and ${\mu}$. Note that the constraint that all stiffness coefficients must be positive (to keep the potential convex) translates into the standard constraint on Lam\'e parameters: ${\mu}\geq 0$ and ${\lambda}\geq -(2/3){\mu}$. Hence, materials with all physical Lam\'e constants can be simulated this way.


\subsection{Non-dimensionalization}
\label{app:nondim}

We now explain how to rescale the variables in Eq.~\eqref{eq:oscillator_eq_of_motion} to arrive at Eq.~\eqref{eq:discrete_eom} presented in the main text, i.e., such that the equation to be solved by the quantum algorithm is nondimensional and of order unity. Let us denote the non-dimensional quantities without an overbar, the scaling factors with a $0$ subscript (where applicable), and the original dimensional quantities with an overbar. First, we choose
\begin{align}
    \label{eq:new_variables}
    \bar{\rho}(\bar{{x}}) = \rho_0 \rho(\bar{{x}}),\qquad \bar{\kappa}^\dev(\bar{\bar{{x}}})= a\lambda_0 {\kappa}^\dev(\bar{{x}}),\qquad \bar{\kappa}^\vol(\bar{{x}})= a\lambda_0 \kappa^\vol(\bar{{x}}),
\end{align}
where $\rho_0$ and $\lambda_0$ are such that non-dimensional ${\rho}$, ${\kappa}^\dev$ and ${\kappa}^\vol$ are $\O(1)$, and that ${\kappa}^\dev$ and ${\kappa}^\vol$ are smaller than 1, while ${\rho}$ is greater than 1. E.g., one can choose $\rho_0=\min_{\bar{\v{x}}}\bar{\rho}(\bar{\v{x}})$, and analogously (just with maximization) for stiffness coefficients. We also rescale:
\begin{align}
    \label{eq:nondim_m_kappa}
    \bar{t}= t_0 {t}, \qquad \bar{{u}}(\bar{{x}},\bar{t})=u_0 {{u}}(\bar{{x}},{t}),
\end{align}
where we are free to choose $t_0$ and $u_0$. Using the new dimensionless variables, Eq.~\eqref{eq:oscillator_eq_of_motion} becomes:
\begin{align}
\begin{split}
    \label{eq:oscillator_eq_of_motion_rescaled_1}
    {\rho}(\bar{{x}})\frac{\partial^2 {u}_i(\bar{{x}})}{\partial {t}^2}
    =
    & \frac{t_0^2\lambda_0}{a^2 \rho_0}\Bigg(\frac{1}{2}\sum_{j=1}^3
    \left[
    {\kappa}^{\dev}(\bar{{x}}+a{e}_j)
    {\xi}^\dev_{ij}(\bar{{x}}+a{e}_j)
    -
    {\kappa}^{\dev}(\bar{{x}}-a{e}_j)
    {\xi}^\dev_{ij}(\bar{{x}}-a{e}_j)
    \right] \\
    &+
    \frac{1}{2}    \left[
    {\kappa}^{\vol}(\bar{{x}}+a{e}_i)
    {\xi}^\vol(\bar{{x}}+a{e}_i)
    -
    {\kappa}^{\vol}(\bar{{x}}-a{e}_i)
    {\xi}^\vol(\bar{{x}}-a{e}_i)
    \right]\Bigg).
\end{split}
\end{align}

Clearly, in order to eliminate the scaling factors, we must choose
\begin{align}
    \label{eq:t_0}
    t_0= a\sqrt{\frac{\rho_0}{\lambda_0}}.
\end{align}
Moreover, it is enough to choose $u_0=\O(1)$ to guarantee
\begin{align}
    \norm{{{u}}(\bar{\v{x}})}=\O(1),\quad \norm{\frac{d{{u}}(\bar{\v{x}})}{d{t}}}=\O(a).
\end{align}
Now, in order to recover Eq.~\eqref{eq:discrete_eom}, we need to replace the position $\bar{\v{x}}$ in physical units with a multi-integer index $\v{n}=(n_1,n_2,n_3)$ with $n_k\in\{1,\dots,N^{1/3}\}$ via:
\begin{equation}
    \bar{\v{x}} =  a \v{n}.
\end{equation}
This means that, e.g., instead of ${\rho}(\bar{\v{x}})$ we will write ${\rho}_{\v{n}}$ and instead of ${\xi}^\dev_{ij}(\bar{\v{x}}+a\v{e}_j)$ we will write ${\xi}^\dev_{\v{n}+\v{e}_j,ij}$. Then, Eq.~\eqref{eq:oscillator_eq_of_motion_rescaled_1} becomes:
\begin{align}
    \label{eq:oscillator_eq_of_motion_rescaled_2}
    {\rho}_{\v{n}}\frac{\partial^2 {u}_{\v{n},i}}{\partial {t}^2}
    =
    & \frac{1}{2}\sum_{j=1}^3
    \left[
    {\kappa}^{\dev}_{\v{n}+\v{e}_j}
    {\xi}^\dev_{{\v{n}+\v{e}_j},ij}
    -
    {\kappa}^{\dev}_{\v{n}-\v{e}_j}
    {\xi}^\dev_{{\v{n}-\v{e}_j},ij}
    \right] 
    +
    \frac{1}{2}    \left[
    {\kappa}^{\vol}_{\v{n}+\v{e}_i}
    {\xi}^\vol_{\v{n}+\v{e}_i}
    -
    {\kappa}^{\vol}_{\v{n}-\v{e}_i}
    {\xi}^\vol_{\v{n}-\v{e}_i}
    \right],
\end{align}
which recovers Eq.~\eqref{eq:discrete_eom} from the main text. Note that due to Eqs.~\eqref{eq:kappa_choice}~and~\eqref{eq:new_variables}, we have:
\begin{equation}
    {\kappa}^\dev_{\v{n}} = \frac{2{\mu}(a\v{n})}{\lambda_0},\qquad {\kappa}^\vol_{\v{n}} = \frac{{\lambda}(a\v{n})+\frac{2}{3}{\mu}(a\v{n})}{\lambda_0}.
\end{equation}

Finally, recall that the dimensional Eq.~\eqref{eq:oscillator_eq_of_motion} recovered the continuum elasticity equation up to discretization error scaling as $\O(a^2)$. Now, note that to get from Eq.~\eqref{eq:oscillator_eq_of_motion} to the dimensionless Eq.~\eqref{eq:oscillator_eq_of_motion_rescaled_2}, we essentially multiplied both sides of the former equation by $t_0^2/(\rho_0 u_0)$. Given the choice of $t_0$ from Eq.~\eqref{eq:t_0}, this means both sides of Eq.~\eqref{eq:oscillator_eq_of_motion} were multiplied by $a^2/(u_0 \lambda_0)$. Since $u_0=\O(1)$ and $\lambda_0=\O(1)$, this means that the error term in the dimensionless equation scales as $\O(a^4)$.


\subsection{Details of the Taylor expansion}
\label{app:consistency_details}

The expansion of the discretized strain tensor from Eq.~\eqref{eq:stran_tensor_discrete} can be obtained by Taylor expanding each displacement $\bar{\v{u}}$ separately:
\begin{equation}
    \bar{\xi}_{ij}(\bar{\v{x}}) = \frac{1}{4} \left[2a \partial_j \bar{u}_i(\bar{\v{x}})+2a \partial_i \bar{u}_j(\bar{\v{x}})\right] +\O(a^3) = a\varepsilon_{ij}(\bar{\v{x}}) + \O(a^3).
\end{equation}
Thus, $\bar{\xi}_{ij}/a$ recovers the strain tensor $\varepsilon_{ij}$ from Eq.~\eqref{eq:continuum-strain} with second-order accuracy. Using the above, we can also get the expansion for $\bar{\xi}^\vol(\bar{\v{x}})$:
\begin{equation}
    \label{eq:vol_expansion}
    \bar{\xi}^\vol(\bar{\v{x}}) = a \sum_{k=1}^{3} \varepsilon_{kk}(\bar{\v{x}}) + \O(a^3),
\end{equation}
and so:
\begin{equation}
    \label{eq:dev_expansion}
    \bar{\xi}^\dev_{ij}(\bar{\v{x}}) = a\left(\varepsilon_{ij}(\bar{\v{x}})-\frac{\delta_{ij}}{3}\sum_{k=1}^{3} \varepsilon_{kk}(\bar{\v{x}})\right) + \O(a^3).
\end{equation}

Let us then expand:
\begin{equation}
    \bar{\kappa}^\dev(\bar{\v{x}}\pm a\v{e}_j)\bar{\xi}^\dev_{ij}(\bar{\v{x}}\pm a\v{e}_j) = \bar{\kappa}^\dev(\bar{\v{x}})\bar{\xi}^\dev_{ij}(\bar{\v{x}}) \pm a\partial_j (\bar{\kappa}^\dev(\bar{\v{x}})\bar{\xi}^\dev_{ij}(\bar{\v{x}})) +\frac{a^2}{2} \partial_j^2 (\bar{\kappa}^\dev(\bar{\v{x}})\bar{\xi}^\dev_{ij}(\bar{\v{x}})) + \O(a^5),
\end{equation}
and analogously:
\begin{equation}
    \bar{\kappa}^\vol(\bar{\v{x}}\pm a\v{e}_i)\bar{\xi}^\vol(\bar{\v{x}}\pm a\v{e}_i) = \bar{\kappa}^\vol(\bar{\v{x}})\bar{\xi}^\vol(\bar{\v{x}}) \pm a\partial_i (\bar{\kappa}^\vol(\bar{\v{x}})\bar{\xi}^\vol(\bar{\v{x}})) +\frac{a^2}{2} \partial_i^2 (\bar{\kappa}^\vol(\bar{\v{x}})\bar{\xi}^\vol(\bar{\v{x}})) + \O(a^5),
\end{equation}
where we note that $\O(a^5)$ scaling is due to $\O(a)$ dependence of $\bar{\kappa}$ and $\bar{\xi}$. Using the above, Eq.~\eqref{eq:oscillator_eq_of_motion} can be expanded as:
\begin{align}
    \bar{\rho}(\bar{{x}})\frac{\partial^2 \bar{u}_i(\bar{{x}})}{\partial \bar{t}^2} &=
    \frac{1}{a^2}\sum_{j=1}^3 \partial_j (\bar{\kappa}^\dev(\bar{\v{x}})\bar{\xi}^\dev_{ij}(\bar{\v{x}})) + \frac{1}{a^2}\partial_i (\bar{\kappa}^\vol(\bar{\v{x}})\bar{\xi}^\vol(\bar{\v{x}})) + \O(a^2)\\
    & = \frac{1}{a}\sum_{j,k,l=1}^3\partial_j \left[\left(A^\dev_{ijkl}\bar{\kappa}^\dev(\bar{\v{x}}) +  A^\vol_{ijkl}\bar{\kappa}^\vol(\bar{\v{x}})\right)\partial_k \bar{u}_l(\bar{\v{x}})\right]+\O(a^2),
\end{align}
where:
\begin{align}
    A^\dev_{ijkl} = \frac{1}{2}(\delta_{ik}\delta_{jl} + \delta_{jk}\delta_{il}) - \frac{1}{3}\delta_{ij}\delta_{kl},\qquad
    A^\vol_{ijkl}  = \delta_{ij}\delta_{kl},
\end{align}
which yields the non-vanishing elements as in Eq.~\eqref{eq:DV_tensors}.


\subsection{Generalization to the \texorpdfstring{$m$-th}{m-th}-order spatial discretization}
\label{app:kordermodel}

We consider a family of microscopic models with an arbitrary even spatial order:
\begin{align}
    m=2p,
    \qquad
    p\in\mathbb{N}.
\end{align}
For fixed $m$, the interactions extend over at most $2p$ lattice sites in each coordinate direction. We introduce the centered finite-difference coefficients~\cite{khan1999closed}
\begin{align}
    \label{eq:order_k_coefficients_main}
    c_0^{(m)}
    &=0,
    &
    c_{-r}^{(m)}
    &=-c_r^{(m)},
    &
    c_r^{(m)}
    &=
    \frac{(-1)^{r+1}}{r}
    \frac{(p!)^2}{(p-r)!(p+r)!},
    \qquad
    r=1,\ldots,p.
\end{align}
 We consider the evolution of the system of oscillators, where the displacement $u_{\v{n},i}$ in direction $i$ of the oscillator at site $\v{n}$ evolves according to
\begin{align}
\begin{split}
    \rho_{\v{n}}
    \frac{\partial^2u_{\v{n},i}}{\partial t^2}
    =
    &\sum_{j=1}^{3}
    \sum_{r=-p}^{p}
    c_r^{(m)}
    \kappa_{\v{n}+r\v{e}_j}^{\dev}
    \xi_{\v{n}+r\v{e}_j,ij}^{\dev,(m)}
    +
    \sum_{r=-p}^{p}
    c_r^{(m)}
    \kappa_{\v{n}+r\v{e}_i}^{\vol}
    \xi_{\v{n}+r\v{e}_i}^{\vol,(m)},
\end{split}
\end{align}
where the discrete volumetric and deviatoric strains are defined, respectively, as:
\begin{align}
    \label{eq:order_k_volumetric_strain_main}
    \xi_{\v{n}}^{\vol,(m)} &=
    \sum_{\ell=1}^{3}
    \sum_{r=-p}^{p}
    c_r^{(m)}
    u_{\v{n}+r\v{e}_\ell,\ell}, \\
\label{eq:order_k_deviatoric_strain_main}
    \xi_{\v{n},ij}^{\dev,(m)}
    &=
    \frac{1}{2}
    \sum_{r=-p}^{p}
    c_r^{(m)}
    \left[
        u_{\v{n}+r\v{e}_j,i}
        +
        u_{\v{n}+r\v{e}_i,j}
    \right]
    -
    \frac{\delta_{ij}}{3}
    \sum_{\ell=1}^{3}
    \sum_{r=-p}^{p}
    c_r^{(m)}
    u_{\v{n}+r\v{e}_\ell,\ell}.
\end{align}

This dimensionless microscopic model is related to the macroscopic elasticity model discussed in the main text in exactly the same way as the $m=2$ model. The quantum encoding is also essentially unchanged, with the order $m$ only entering linearly the sparsity (and hence the complexity) of the final algorithm. 


\section{Alternative initial state preparation}
\label{app:priorstateprep}

Following Lemma~10 of Ref.~\cite{babbush2023exponential}, we assume access to the state-preparation oracles for the displacement and velocity fields,
\begin{equation}
O_u:
|0\rangle
\mapsto
|u(0)\rangle,\qquad 
O_{\dot{u}}:
|0\rangle
\mapsto
|\dot{u}(0)\rangle,
\label{eq:initial_data_oracles}
\end{equation}
and assume that the norms $\|u(0)\|$ and $\|\dot{u}(0)\|$ are known. We will also employ the unitary block-encodings of $\sqrt{M}$ and~$\tilde{B}$. We have already constructed the latter and showed that its block-encoding prefactor $\alpha_{\tilde{B}}=\O(1)$. As for the former, although we have chosen the units so that $\sqrt{M^{-1}}$ can be block-encoded with a unit prefactor, we still get $\alpha_{\sqrt{M}}=\O(1)$ under the assumption that the ratio between the largest density $\rho_{\v{n}}$ and the smallest one is a constant (i.e., independent of spatial discretization $a$).

We start by preparing the first ancillary qubit in a superposition, followed by applications of $O_u$ and $O_{\dot u}$ controlled on the ancilla: 
\begin{equation}
    \ket{0}\ot\ket{0} \mapsto (\beta\ket{0}-\i\sqrt{1-\beta^2}\ket{1}) \ot \ket{0} \mapsto \beta \ket{0}\ot\ket{\dot{u}(0)} - \i\sqrt{1-\beta^2} \ket{1}\ot \ket{u(0)},
\end{equation}
where the real parameter $\beta$ is yet to be specified. Next, we apply the unitary block-encoding of $\sqrt{M}$ to the main system, followed by the unitary block-encoding of $\tilde{B}$ controlled on the ancilla to be in the one state:
\begin{equation}
    \beta \ket{0}\ot\ket{\dot{u}(0)} - \i\sqrt{1-\beta^2} \ket{1}\ot \ket{u(0)} \mapsto \frac{\beta}{\alpha_{\sqrt{M}}} \ket{0}\ot\sqrt{M}\ket{\dot{u}(0)} - \frac{\i\sqrt{1-\beta^2}}{\alpha_{\sqrt{M}}\alpha_{\tilde{B}}} \ket{1}\ot \tilde{B}\sqrt{M}\ket{u(0)},
\end{equation}
with the final state being subnormalized due to block-encoding prefactors for $\sqrt{M}$ and $\tilde{B}$. Now, the choice:
\begin{equation}
    \beta = \frac{\|\dot u(0)\|}{\sqrt{\|\dot u(0)\|^2 + \alpha^2_{\tilde{B}}\|u(0)\|^2}}
\end{equation}
results in obtaining $\chi\ket{\psi(0)}$ with a subnormalization factor
\begin{equation}
    \chi = \sqrt{\frac{E}{E_{\rm max}}},\qquad E_{\mathrm{max}}
    :=
    \frac{\alpha_{\sqrt M}^{2}}{2}
    \left(
        \|\dot u(0)\|^{2}
        +
        \alpha_{\tilde B}^{2}\|u(0)\|^{2}
    \right).
\end{equation}

The normalized target state can be obtained from the above subnormalized state using fixed-point amplitude amplification. Preparing $|\psi(0)\rangle$ to error $\epsilon_{\mathrm{SP}}$ therefore requires
\begin{equation}
    \label{eq:Q_SP_old}
    Q_{\mathrm{SP}}
    =
    \O\left(
        \frac{1}{\chi}
        \log\frac{1}{\epsilon_{\mathrm{SP}}}
    \right)
\end{equation}
applications of the preparation circuit and its inverse. For smooth $u(0)$ nonzero on a finite measure portion of the domain, $E$ scales as $\O(Na^2)=\O(1/a)$ and $E_{\mathrm{max}} = \Theta(N)$, which gives $Q_{\mathrm{SP}} = \tilde\O(N^{1/3})$.


\section{Observable extraction accuracy}
\label{app:info}


\subsection{Approximating continuum energies with Riemann sums}

Let us denote by $u_{\v{n},i}^{\mathrm{cont}}(t)$ the dimensionless solution of the continuum elasticity equation, Eq.~\eqref{eq:wave_eq}, sampled at the
lattice sites:
\begin{equation}
    u_{\v{n},i}^{\mathrm{cont}}(t)
    :=
    \frac{
        \bar{u}_i(a\v{n},\bar{t})
    }{u_0}.
\end{equation}
The corresponding lattice approximations to the kinetic and potential energies of the continuum elastic system in region $\Rc$ are
\begin{equation}
    T_{\Rc}^{\mathrm{cont}}(t)
    :=
    \frac{1}{2}
    \norm{
        \Pi_{\Rc}^{V}\sqrt M\,
        \dot u^{\mathrm{cont}}(t)
    }^2,\qquad U_{\Rc}^{\mathrm{cont}}(t)
    :=
    \frac{1}{2}
    \norm{
        \Pi_{\Rc}^{I}B
        u^{\mathrm{cont}}(t)
    }^2.
\end{equation}

To see this explicitly, define the common energy-conversion factor
\begin{equation}
    C_E:=a\lambda_0u_0^2.
\end{equation}
From Eq.~\eqref{eq:non_dim}, we have
\begin{equation}
    \rho_{\v{n}}
    =
    \frac{\bar\rho(a\v{n})}{\rho_0},
    \qquad
    \dot u_{\v{n}}^{\mathrm{cont}}(t)
    =
    \frac{t_0}{u_0}
    \frac{\partial\bar u(a\v{n},\bar t)}
         {\partial\bar t},
    \qquad
    t_0^2
    =
    \frac{a^2\rho_0}{\lambda_0}.
\end{equation}
Consequently,
\begin{align}
    C_E T_{\Rc}^{\mathrm{cont}}(t)
    &=
    \frac{a\lambda_0u_0^2}{2}
    \sum_{\v{n}\in\Rc}
    \rho_{\v{n}}
    \left|
        \dot u_{\v{n}}^{\mathrm{cont}}(t)
    \right|^2 =
    \frac{a^3}{2}
    \sum_{\v{n}\in\Rc}
    \bar\rho(a\v{n})
    \left|
        \frac{\partial\bar u(a\v{n},\bar t)}
             {\partial\bar t}
    \right|^2.
\label{eq:continuum-kinetic-riemann-sum}
\end{align}
The last expression is the lattice Riemann sum for the physical continuum kinetic energy:
\begin{equation}
    \bar T_{\Rc}^{\mathrm{wave}}(\bar t)
    :=
    \frac{1}{2}
    \int_{\Rc}
    \bar\rho(\bar{\v{x}})
    \left|
        \frac{\partial\bar u(\bar{\v{x}},\bar t)}
             {\partial\bar t}
    \right|^2
    \mathrm d^3\bar{\v{x}}.
\end{equation}

For the potential energy, we first note that the discrete strains are linear in the displacement. Since $u^{\mathrm{cont}}=\bar u/u_0$, the strain expansions derived in Appendix~\ref{app:micro} (recall Eqs.~\eqref{eq:vol_expansion}-\eqref{eq:dev_expansion}) give
\begin{align}
    \xi_{\v{n}}^{\mathrm{vol}}
    [u^{\mathrm{cont}}]
    &=
    \frac{a}{u_0}
    \sum_{k=1}^{3}
    \varepsilon_{kk}(a\v{n},\bar t)
    +
    \O\!\left(\frac{a^3}{u_0}\right),
    \\
    \xi_{\v{n},ij}^{\mathrm{dev}}
    [u^{\mathrm{cont}}]
    &=
    \frac{a}{u_0}
    \left[
        \varepsilon_{ij}(a\v{n},\bar t)
        -
        \frac{\delta_{ij}}{3}
        \sum_{k=1}^{3}
        \varepsilon_{kk}(a\v{n},\bar t)
    \right]
    +
    \O\!\left(\frac{a^3}{u_0}\right).
\end{align}
Using
\begin{equation}
    \kappa_{\v{n}}^{\mathrm{dev}}
    =
    \frac{2\mu(a\v{n})}{\lambda_0},
    \qquad
    \kappa_{\v{n}}^{\mathrm{vol}}
    =
    \frac{
        \lambda(a\v{n})
        +\frac{2}{3}\mu(a\v{n})
    }{\lambda_0},
\end{equation}
together with \(C_E=a\lambda_0u_0^2\), we obtain
\begin{align}
    C_EU_{\Rc}^{\mathrm{cont}}(t)
    &=
    \frac{a^3}{2}
    \sum_{\v{n}\in\Rc}
    \left\{
        2\mu(a\v{n})
        \sum_{i,j=1}^{3}
        \left[
            \varepsilon_{ij}(a\v{n},\bar t)
            -
            \frac{\delta_{ij}}{3}
            \sum_{k=1}^{3}
            \varepsilon_{kk}(a\v{n},\bar t)
        \right]^2
    \right.
    \nonumber\\
    &\hspace{7em}\left.
        +
        \left[
            \lambda(a\v{n})
            +\frac{2}{3}\mu(a\v{n})
        \right]
        \left[
            \sum_{k=1}^{3}
            \varepsilon_{kk}(a\v{n},\bar t)
        \right]^2
    \right\}
    +
    \O(a^2).
\label{eq:continuum-potential-riemann-sum}
\end{align}
The leading term is a Riemann-sum approximation to the physical continuum potential energy
\begin{align}
    \bar U_{\Rc}^{\mathrm{wave}}(\bar t)
    &:=
    \frac{1}{2}
    \int_{\bar\Rc}
    \left\{
        2\mu(\bar{\v{x}})
        \sum_{i,j=1}^{3}
        \left[
            \varepsilon_{ij}(\bar{\v{x}},\bar t)
            -
            \frac{\delta_{ij}}{3}
            \sum_{k=1}^{3}
            \varepsilon_{kk}(\bar{\v{x}},\bar t)
        \right]^2
    \right.
    \nonumber\\
    &\hspace{7em}\left.
        +
        \left[
            \lambda(\bar{\v{x}})
            +\frac{2}{3}\mu(\bar{\v{x}})
        \right]
        \left[
            \sum_{k=1}^{3}
            \varepsilon_{kk}(\bar{\v{x}},\bar t)
        \right]^2
    \right\}
    \,\mathrm d^3\bar{\v{x}}.
\end{align}

In conclusion, assuming smooth fields and a second-order spatial quadrature, we have
\begin{equation}
    \label{eq:wave_energies}
    C_E T_{\Rc}^{\mathrm{cont}}(t)
    =
    \bar T_{\Rc}^{\mathrm{wave}}(\bar t)
    +
    \O(a^2),
    \qquad
    C_EU_{\Rc}^{\mathrm{cont}}(t)
    =
    \bar U_{\Rc}^{\mathrm{wave}}(\bar t)
    +
    \O(a^2).
\end{equation}


\subsection{Approximating the Riemann sums with discrete model energies}

We assume that the total energies of the dimensionless discrete and continuum models initially coincide exactly, $E^{\mathrm{cont}}=E$, and since both evolutions conserve their respective total energies, they remain aligned at all times. Thus, the energy fractions are given by
\begin{subequations}
\begin{align}
    \label{eq:continuum-kinetic-fraction}  
    \frac{T_{\Rc}^{\mathrm{cont}}(t)}{E} &=
    \frac{T_{\Rc}(t)}{E}
    +
    \frac{
        T_{\Rc}^{\mathrm{cont}}(t)-T_{\Rc}(t)
    }{E}
    =
    \langle\psi(t)|
    P_{\Rc}^{T}
    |\psi(t)\rangle
    +
    \Delta_{\Rc}^{T}(t),\\
    \label{eq:continuum-potential-fraction}  
    \frac{U_{\Rc}^{\mathrm{cont}}(t)}{E} &=
    \frac{U_{\Rc}(t)}{E}
    +
    \frac{
        U_{\Rc}^{\mathrm{cont}}(t)-U_{\Rc}(t)
    }{E}
    =
    \langle\psi(t)|
    P_{\Rc}^{U}
    |\psi(t)\rangle
    +
    \Delta_{\Rc}^{U}(t),
\end{align}
\end{subequations}
where
\begin{equation}
    \label{eq:potential-fraction-error}
    \Delta_{\Rc}^{T}(t)
    :=
    \frac{
        T_{\Rc}^{\mathrm{cont}}(t)-T_{\Rc}(t)
    }{E},\qquad \Delta_{\Rc}^{U}(t)
    :=
    \frac{
        U_{\Rc}^{\mathrm{cont}}(t)-U_{\Rc}(t)
    }{E},
\end{equation}
are the errors in the energy fractions caused by the discrete oscillator mapping.

In order to bound these errors, let us first define
\begin{equation}
    e(t):=u(t)-u^{\mathrm{cont}}(t),
\end{equation}
and note the following inequality arising from the reverse triangle inequality:
\begin{equation}
        \left|\norm{x}^2-\norm{y}^2\right|
        =
        (\norm{x}+\norm{y})\left|\norm{x}-\norm{y}\right|
        \leq
        \left(\norm{x}+\norm{y}\right)\norm{x-y}.
\end{equation}
We then have the following:
\begin{align}
    \left|\Delta_{\Rc}^{T}(t)\right|
    &=
    \frac{1}{2E}
    \left|
        \norm{
           ( \Pi_{\Rc}\ot I_3)\sqrt M\,
        \dot u^{\mathrm{cont}}(t)
        }^2
        -
        \norm{
            (\Pi_{\Rc}\ot I_3)\sqrt M\,
        \dot u(t)
        }^2
    \right|
    \nonumber\\
    &\leq
    \frac{
        \sqrt{2T_{\Rc}^{\mathrm{cont}}(t)}+\sqrt{2T_{\Rc}(t)}
    }{2E}
    \norm{
        (\Pi_{\Rc}\ot I_3)\sqrt M\,
        \dot e(t)
    }\nonumber\\
    &\leq
    \sqrt{\frac{2}{E}}
    \norm{
        \sqrt M\,
        \dot e(t)
    },\label{eq:error_kinetic}
\end{align}
where the last inequality comes from the fact that the kinetic energy in region $\Rc$ is bounded by the total energy $E$ in the full volume, and that $\Pi_{\Rc}$ is a projector. Analogously, one gets
\begin{align}
    \left|\Delta_{\Rc}^{U}(t)\right|
    &\leq
    \sqrt{\frac{2}{E}}
    \norm{
        B e(t)
    }.\label{eq:error_potential}
\end{align}

Next, we recall that in Appendix~\ref{app:micro} we proved that the dimensionless continuum solution $u^{\mathrm{cont}}(t)$ does not exactly satisfy the discrete equation, Eq.~\eqref{eq:eom_for_u}, and the mismatch,
\begin{equation}
    \label{eq:consistency-residual}
    r(t)
    :=
    M\ddot u^{\mathrm{cont}}(t)
    +
    B^\dagger B u^{\mathrm{cont}}(t),
\end{equation}
is of order $\O(a^4)$ component-wise. Subtracting Eq.~\eqref{eq:consistency-residual} from Eq.~\eqref{eq:eom_for_u} gives
\begin{equation}
    \label{eq:displacement-error-equation}
    M\ddot e(t)+B^\dagger Be(t)=-r(t),
\end{equation}
We assume that the discrete initial conditions are obtained by sampling the continuum initial conditions, so that
\begin{equation}
    e(0)=\dot e(0)=0.
\end{equation}
Now, multiplying Eq.~\eqref{eq:displacement-error-equation} from the left by
\(\dot e^{\mathsf T}(t)\) yields
\begin{equation}
    \frac{\mathrm d}{\mathrm dt}
    \left[
        \frac{1}{2}\norm{\sqrt M\,\dot e(t)}^2
        +
        \frac{1}{2}\norm{B e(t)}^2
    \right]
    =
    -\dot e^{\mathsf T}(t)r(t).
\end{equation}
We can bound the right-hand side using the Cauchy--Schwarz inequality as follows:
\begin{align}
    \left|
        \dot e^{\mathsf T}(t)r(t)
    \right|
    &=
    \left|
        \bigl(\sqrt M\,\dot e(t)\bigr)^{\mathsf T}
        \bigl(M^{-1/2}r(t)\bigr)
    \right|
    \nonumber\\
    &\leq
    \norm{\sqrt M\,\dot e(t)}
    \norm{M^{-1/2}r(t)}
    \nonumber\\
    &\leq
    \left[
        \norm{\sqrt M\,\dot e(t)}^2
        +
        \norm{B e(t)}^2
    \right]^{1/2}
    \norm{M^{-1/2}r(t)}.
\label{eq:weighted-cauchy-schwarz}
\end{align}
On the other hand, we have
\begin{align}
    &\frac{\mathrm d}{\mathrm dt}
    \left[
        \frac{1}{2}\norm{\sqrt M\,\dot e(t)}^2
        +
        \frac{1}{2}\norm{B e(t)}^2
    \right]=
    \left[
        \norm{\sqrt M\,\dot e(t)}^2
        +
        \norm{B e(t)}^2
    \right]^{1/2}
    \frac{\mathrm d}{\mathrm dt}
    \left[
        \norm{\sqrt M\,\dot e(t)}^2
        +
        \norm{B e(t)}^2
    \right]^{1/2}.
\end{align}
Combining the above equations and canceling the common square-root factor yields
\begin{equation}
    \frac{\mathrm d}{\mathrm dt}
    \left[
        \norm{\sqrt M\,\dot e(t)}^2
        +
        \norm{B e(t)}^2
    \right]^{1/2}
    \leq
    \norm{M^{-1/2}r(t)}.
\end{equation}
Then, using $e(0)=\dot e(0)=0$ and integrating from 0 to $t$, we obtain
\begin{equation}
    \left[
        \norm{\sqrt M\,\dot e(t)}^2
        +
        \norm{B e(t)}^2
    \right]^{1/2}
    \leq
    \int_0^t
    \norm{M^{-1/2}r(s)}
    \,\mathrm ds.
\label{eq:energy-norm-error-bound}
\end{equation}
Finally, since $r(s)$ has $3N$ components and the dimensionless masses are bounded away from zero, we also have
\begin{equation}
    \label{eq:consistency-residual-norm}
    \norm{M^{-1/2}r(s)}
    =
    \O\!\left(\sqrt{N}\,a^4\right).
\end{equation}
Together, this gives
\begin{equation}
    \left[
        \norm{\sqrt M\,\dot e(t)}^2
        +
        \norm{B e(t)}^2
    \right]^{1/2}
    =
    \O\left(t\sqrt N\,a^4\right).
\end{equation}

Using the above in Eq.~\eqref{eq:error_kinetic} and \eqref{eq:error_potential}, we conclude that
\begin{equation}
    |\Delta_\Rc^U(t)|\leq \O\left(t\sqrt{N/E}\,a^4\right),\qquad |\Delta_\Rc^T(t)|\leq \O\left(t\sqrt{N/E}\,a^4\right).
\end{equation}
Recall that Eq.~\eqref{eq:non_dim} yields $t=\Theta(\bar t/a)$, and note that for a nonvanishing smooth elastic field, the dimensionless velocities and strains are $\O(a)$, so that the total dimensionless energy satisfies $E=\Theta(Na^2)$. Consequently,
\begin{equation}
    |\Delta_\Rc^U(t)|\leq \epsilon_{\rm DM},\qquad |\Delta_\Rc^T(t)|\leq \epsilon_{\rm DM}, 
\end{equation}
with
\begin{equation}
    \epsilon_{\rm DM} = \O\left(\bar{t} a^2\right).
\end{equation}

Finally, combining Eq.~\eqref{eq:wave_energies} with Eqs.~\eqref{eq:continuum-kinetic-fraction}-\eqref{eq:continuum-potential-fraction} and using the above bounds, we end up with the following expressions for the continuum energy fractions of the elastic system:
\begin{align}
    \frac{\bar T_{\Rc}^{\mathrm{wave}}(\bar t)}{\bar{E}^{\rm wave}} &= \langle\psi(t)|
    P_{\Rc}^{T}
    |\psi(t)\rangle  +\O(\bar{t}a^2),\\
    \frac{\bar U_{\Rc}^{\mathrm{wave}}(\bar t)}{\bar{E}^{\rm wave}} &= \langle\psi(t)|
    P_{\Rc}^{U}
    |\psi(t)\rangle  +\O(\bar{t}a^2).
\end{align}


\section{Continuum models for a one-dimensional nearest-neighbor lattice}
\label{app:1d_lattice}


\subsection{Homogeneous case}

If $\kappa_n=\kappa$, Taylor expansion of a smooth interpolating displacement $u(x,t)$ gives
\begin{equation}
     u(x+a,t)-2u(x,t)+u(x-a,t)
     =a^2\partial_x^2u(x,t)
      +\frac{a^4}{12}\partial_x^4u(x,t)+\cdots.
\end{equation}
Keeping the leading term in Eq.~\eqref{eq:lattice} gives
\begin{equation}
     \rho\,\partial_t^2u=\mathcal C\,\partial_x^2u,
     \qquad
     \rho:=\frac ma,
     \qquad
     \mathcal C:=a\kappa.
     \label{eq:homogeneous-continuum}
\end{equation}
Here $\rho$ is mass per length and $\mathcal C$ is the one-dimensional elastic coefficient. Equivalently, the lattice energy is approximated by 
\begin{equation}
     U[u]\approx\frac12\int\mathcal C(\partial_xu)^2\,\mathrm dx.
\end{equation}
The wave speed $c^2=\mathcal C/\rho=\kappa a^2/m$ agrees with the long-wavelength lattice dispersion,
\begin{equation}
     \omega_{\mathrm{lat}}^2(k)
     =\frac{4\kappa}{m}\sin^2\!\left(\frac{ka}{2}\right)
     =c^2k^2+O(c^2a^2k^4),
\end{equation}
while for $|k|=O(\lambda^{-1})$ the relative correction to the continuum dispersion is $O((a/\lambda)^2)$. 


\subsection{Homogenization gives harmonic average}

We identify a continuum elastic coefficient by matching the relaxed energy under any prescribed mean strain $e$. In an interval $\mathcal{I}_x$ of length $\ell=N_\ell a$, with bond extensions $\xi_n$, define the average strain
\begin{equation}
e(x) := \frac{1}{\ell} \sum_{n\in \mathcal{I}_x}\xi_n.
\end{equation}
 The minimum energy is
\begin{equation}
     U_{\min}(x,e)
     :=\min_{\{\xi_n\}:\,\sum_{n\in \mathcal{I}_x}\xi_n=\ell e}
           \frac12\sum_{n\in \mathcal{I}_x}\kappa_n\xi_n^2.
     \label{eq:minimization}
\end{equation}
Introducing a Lagrange multiplier $f$, stationarity gives
\begin{equation}
     \frac{\partial}{\partial\xi_n}
     \left[
      \frac12\sum_{j\in \mathcal{I}_x}\kappa_j\xi_j^2
      -f\left(\sum_{j\in \mathcal{I}_x}\xi_j-\ell e\right)
     \right]=0,
     \qquad\text{hence}\qquad
     \kappa_n\xi_n=f.
\end{equation}
Thus, every bond carries the same force $f$. Substituting $\xi_n = f/\kappa_n$ yields
\begin{equation}
     f=\frac{\ell e}{\sum_{n\in \mathcal{I}_x}\kappa_n^{-1}},
     \qquad
     U_{\min}(x,e)
     =\frac12 f^2\sum_{n\in \mathcal{I}_x}\kappa_n^{-1}
     =\frac{(\ell e)^2}{2\sum_{n\in \mathcal{I}_x}\kappa_n^{-1}}.
\end{equation}
Matching this energy to that of a continuum interval under uniform strain, $U_{\min}(x,e)=\tfrac12\ell\mathcal C_\ell(x)e^2$, defines
\begin{equation}
      \mathcal C_\ell(x)
      =a\left(
        \frac1{N_\ell}\sum_{n\in \mathcal{I}_x}\frac1{\kappa_n}
       \right)^{-1}.
     \label{eq:local-coefficient}
\end{equation}

For a fixed partition into representative intervals, assign each interval its coefficient from Eq.~\eqref{eq:local-coefficient}. This gives the continuum model at  displacement scale $\lambda\gg\ell$
\begin{equation}
     U_{\mathrm{cont}}[u]
     =\frac12\int\mathcal C_\ell(x)(\partial_xu)^2\,\mathrm dx,
     \qquad
     \rho\,\partial_t^2u
     =\partial_x\!\left[\mathcal C_\ell(x)\partial_xu\right],
     \qquad \rho=\frac ma.
     \label{eq:disordered-continuum}
\end{equation}
At coefficient discontinuities, the equation is understood in its weak form. Alternatively, one could introduce a smooth interpolatation. 

If $p_\ell(x)$ is the actual fraction of type-1 bonds in $\mathcal{I}_x$, then
\begin{equation}
     \mathcal C_\ell(x)
     =a\left[
      \frac{1-p_\ell(x)}{\kappa^{(0)}}
      +\frac{p_\ell(x)}{\kappa^{(1)}}
     \right]^{-1}.
\end{equation}
The local coefficients fluctuate because the sample is finite. As the number of independent bonds in a representative interval grows, $p_\ell(x)$ converges to $p$ and the coefficient approaches
\begin{equation}
     \mathcal C_{\mathrm{eff}}
     =a\left[
      \frac{1-p}{\kappa^{(0)}}
      +\frac{p}{\kappa^{(1)}}
     \right]^{-1}.
     \label{eq:effective-coefficient}
\end{equation}
Thus a statistically uniform material with microscopic disorder has a constant leading homogenized coefficient. Keeping the interval-dependent coefficients retains finite-sample composition fluctuations. Using many
bonds per interval and many intervals per wavelength simultaneously requires the scale separation in Eq.~\eqref{eq:separation}.


\subsection{Fourier projection gives arithmetic average}

Writing $(Du)_n=u_{n+1}-u_n$ and
$C=\operatorname{diag}(\kappa_n)$, the lattice dynamics is defined by
\begin{equation}
    m\ddot u=-Ku,
    \qquad
    K=D^\dagger C D.
\end{equation}
For the normalized Fourier vectors $\ket{k_r}$, we have
\begin{equation}
    \langle n|k_r\rangle
    =\frac1{\sqrt N}e^{\i k_r x_n},
    \qquad
    k_r=\frac{2\pi r}{Na},
    \qquad
    x_n = n a.
\end{equation}
The difference operator acts diagonally on Fourier modes:
\begin{equation}
    D|k_r\rangle=d_r|k_r\rangle,
    \qquad
    d_r=e^{\mathrm{i}k_r a}-1.
\end{equation}
Indeed,
\begin{equation}
    \bra{n}D|k_r\rangle=\sum_{m=0}^{N-1} \bra{n}D\ket{m}\bra{m}k_r\rangle = \braket{n+1}{k_r}-\braket{n}{k_r} = (e^{\i k_r a}-1)\braket{n}{k_r}.
\end{equation}
Hence:
\begin{align}
    \widehat K_{rs}
    =\langle k_r|D^\dagger C D|k_s\rangle
    =(D|k_r\rangle)^\dagger C(D|k_s\rangle)
    =d_r^*d_s\,\langle k_r|C|k_s\rangle.
\end{align}
Since $C=\sum_n\kappa_n\ketbra{n}{n}$,
\begin{align}
    \langle k_r|C|k_s\rangle
    =\sum_{n=0}^{N-1}\kappa_n
      \langle k_r|n\rangle\langle n|k_s\rangle
    =\frac1N\sum_{n=0}^{N-1}\kappa_n
      e^{-\i k_{r-s} x_n}
      =\widehat\kappa_{r-s}.
\end{align}
Therefore,
\begin{equation}
    \widehat K_{rs}
    =d_r^*\,\widehat\kappa_{r-s}\,d_s,
    \qquad
    \widehat\kappa_q
    :=\frac1N\sum_{n=0}^{N-1}
        \kappa_n e^{-\i k_q x_n}.
    \label{eq:projected-chain-stiffness}
\end{equation}

Now, we want to retain a fixed set of macroscopic Fourier modes:
\begin{equation}
    P_R=\sum_{r=-R}^{R} \ketbra{k_r}{k_r},
    \qquad
    K_{\mathrm{LL}}=P_RKP_R,
\end{equation}
where $R$ is independent of $N$, and we shall take $N>2R$. The projected equations are
\begin{equation}
    m \frac{\mathrm d^2 \hat{u}_r}{\mathrm dt^2}
    =-\sum_{s=-R}^{R}
        d_r^*\widehat\kappa_{r-s}d_s\,\widehat u_s,
    \qquad |r|\leq R.
\end{equation}

Each spring constant has the same expectation,
\begin{equation}
    \mathbb E[\kappa_n]
    =(1-p)\kappa^{(0)}+p\kappa^{(1)}
    =:\kappa_{\mathrm A}.
\end{equation}
Therefore, by linearity of expectation,
\begin{align}
    \mathbb E[\widehat\kappa_{r-s}]
    =\frac1N\sum_{n=0}^{N-1}
      \mathbb E[\kappa_n]\,
      e^{-\i k_{r-s} x_n}
    =\frac{\kappa_{\mathrm A}}N
      \sum_{n=0}^{N-1}e^{-\i k_{r-s} x_n}.
\end{align}
If $r=s$, every summand equals one, giving $\mathbb E[\widehat\kappa_0]=\kappa_{\mathrm A}$. If $r\neq s$, then $0<|r-s|\leq2R<N$, so $z:=e^{-\i k_{r-s} x_n}\neq1$. The geometric sum consequently vanishes:
\begin{equation}
    \sum_{n=0}^{N-1}e^{-\i k_{r-s} x_n}=\sum_{n=0}^{N-1}e^{-2\pi\i(r-s)n/N}
    =\frac{1-z^N}{1-z}=0.
\end{equation}
Hence
\begin{equation}
    \mathbb E[\widehat\kappa_{r-s}]
    =\kappa_{\mathrm A}\delta_{rs}.
\label{eq:macroscopiclimit}
\end{equation}

Moreover, the variance reads
\begin{align}
    \mathbb E\!\left[
        \left|
        \widehat\kappa_{r-s}
        -\kappa_{\mathrm A}\delta_{rs}
        \right|^2
    \right]
    &=
    \frac1{N^2}
    \sum_{n,n'}
    \mathbb E[
        (\kappa_n-\kappa_{\mathrm A})
        (\kappa_{n'}-\kappa_{\mathrm A})
    ]
    e^{-2\pi\mathrm{i}(r-s)(n-n')/N}
    \nonumber\\
   &=  \frac1{N^2}
    \sum_{n}
    \mathbb E[
        (\kappa_n-\kappa_{\mathrm A})^2
    ] 
    =\frac{N\sigma_\kappa^2}{N^2}=\frac{\sigma_\kappa^2}{N},
\end{align}
where $\sigma_\kappa^2$ is the variance of the spring constants. Thus, for the set of retained modes,
\begin{equation}
    \widehat\kappa_{r-s}
    \xrightarrow[N\to\infty]{\mathbb P}
    \kappa_{\mathrm A}\delta_{rs}.
\end{equation}

Let us identify the dynamics on the macroscopic propagation timescale for the retained wave numbers. Since
\begin{equation}
    d_r=\frac{2\pi\i r}{N}+O(N^{-2}),
\end{equation}
Eqs.~\eqref{eq:projected-chain-stiffness} and \eqref{eq:macroscopiclimit} imply
\begin{equation}
    N^2\widehat K_{\mathrm{LL},rs}
    \xrightarrow[N\to\infty]{\mathbb P}
    \kappa_{\mathrm A}(2\pi r)^2\delta_{rs}.
\end{equation}
Introducing the macroscopic time $\tau=t/N$, the limiting
retained-mode equations are therefore
\begin{equation}
    m\frac{\mathrm d^2 \hat{u}_r}{\mathrm d\tau^2}
    =-\kappa_{\mathrm A}(2\pi r)^2 u_r, \qquad |r|\leq R.
\end{equation}
The above are simply the Fourier expansion of the arithmetic-mean wave equation
\begin{equation}
    m\,\frac{\partial^2 u}{d\tau^2 }
    =\kappa_{\mathrm A}\partial_y^2 u
\end{equation}
for a band-limited displacement $u = \sum_{r=-R}^R \hat{u}_r e^{2\pi i r y}$, with $y= x/(Na) \in [0,1)$. Thus the limiting projected dynamics agree with the arithmetic-mean
wave equation restricted to the retained Fourier modes.


\section{Details for the microscopically resolved bulk observable}
\label{app:MSBO-details}

The initial fraction of the total elastic energy stored in microscopic Fourier components of the energy-weighted strain is given by
\begin{equation}
    U_{\mathrm{mic}}(0)
    =
    \langle\psi(0)|
    P_{\mathrm{mic}}^U
    |\psi(0)\rangle
    =
    \frac{
        \left\|
            (\Pi_{\mathrm{mic}}\ot I_S) s(0)
        \right\|_2^2
    }{
        2E
    },
\end{equation}
To evaluate it, we introduce a binary field $\zeta$ on the lattice $\Lambda$ together with its discrete Fourier transform:
\begin{equation}
    \zeta_{ n}:=\eta_{ n}-p,
    \qquad
    \widehat\zeta(k)
    :=
    \frac{1}{\sqrt N}
    \sum_{ n\in\Lambda}
    \zeta_{ n}
    e^{-\i k\cdot x_{ n}}.
    \label{eq:centered-binary-fourier-transform}
\end{equation}
These satisfy
\begin{equation}
    \mathbb E[\zeta_{ n}]
    =
    0,
    \qquad
    \mathbb E[
        \zeta_{ n}\zeta_{ m}
    ]
    =
    p(1-p)\delta_{ n, m},
\end{equation}
resulting in
\begin{align}
    \mathbb E
    \left[
        \widehat\zeta( q)
        \widehat\zeta( q')^*
    \right]
    &=
    \frac{1}{N}
    \sum_{ n, m}
    \mathbb E[
        \zeta_{ n}\zeta_{ m}
    ]
    e^{-\i q\cdot x_{ n}}
    e^{\i q'\cdot x_{ m}}=
    p(1-p)
    \delta_{ q, q'},
    \label{eq:binary-fourier-covariance}
\end{align}
so that the expected disorder power is constant across the Brillouin zone.

We then define the mean constitutive coefficient in the selected strain channel and its square-root contrast by
\begin{equation}
    \overline c
    :=
    (1-p)c^{(0)}+pc^{(1)},
    \qquad
    \Delta_s
    :=
    \sqrt{c^{(1)}}-\sqrt{c^{(0)}},
    \label{eq:binary-channel-parameters}
\end{equation}
as well as
\begin{equation}
    m_s
    :=
    (1-p)\sqrt{c^{(0)}}
    +
    p\sqrt{c^{(1)}}.
    \label{eq:mean-square-root-stiffness}
\end{equation}
Now, it can be seen by inspection that for $\eta_{{n}} \in \{0,1\}$, we have the identity
\begin{equation}
    \sqrt{c_{ n}}=m_s+\Delta_s\zeta_{ n}.
\end{equation}
Putting this into Eq.~\eqref{eq:binary-energy-amplitude}, we obtain the initial energy-weighted strain
\begin{equation}
    s_{ n}(0)=e^{\i k_0\cdot x_{ n}}
    \left(
        m_s+\Delta_s\zeta_{ n}
    \right) v.
    \label{eq:centered-binary-energy-amplitude}
\end{equation}

Taking the discrete Fourier transform gives
\begin{equation}
    \widehat s( k,0) = \left[\sqrt N\,m_s\delta_{ k, k_0}+\Delta_s\widehat\zeta( k- k_0)\right] v.
    \label{eq:binary-energy-amplitude-fourier-transform}
\end{equation}
The first term in the brackets is the long-wavelength component, while the second term consists of Fourier `sidebands' generated by the constitutive disorder. Introducing
\begin{equation}
    \mathcal K_{\mathrm{mic}}^{(N)}
    :=
    \left\{
         k:
        \| k a\|_\infty
        \geq
        \gamma_{\mathrm{mic}}
    \right\},
\end{equation}
and noting that $\|k_0a\|_\infty<\gamma_{\mathrm{mic}}$, we get that the coherent Fourier component at $ k_0$ is removed by the microscopic projector, and we therefore obtain
\begin{align}
    \left\|
        (\Pi_{\mathrm{mic}}\ot I_S) s(0)
    \right\|_2^2
    &=
    \sum_{ k\in\mathcal K_{\mathrm{mic}}^{(N)}}
    \left\|
        \widehat s( k,0)
    \right\|_2^2
    =
    \|v\|_2^2
    \Delta_s^2
    \sum_{ k\in\mathcal K_{\mathrm{mic}}^{(N)}}
    \left|
        \widehat\zeta( k- k_0)
    \right|^2.
    \label{eq:projected-microscopic-power}
\end{align}

We will now produce a high-probability estimate of the above. Let
\begin{equation}
    \phi_{\mathrm{mic}}
    :=
    \frac{\left|
        \mathcal K_{\mathrm{mic}}^{(N)}
    \right|}{N}
    \label{eq:microscopic-mode-fraction-definition}
\end{equation}
be the fraction of lattice wavevectors contained in the microscopic sector. Then, define the shifted set
\begin{equation}
    \mathcal Q_N
    :=
    \left\{
         k- k_0:
         k\in
        \mathcal K_{\mathrm{mic}}^{(N)}
    \right\},
\end{equation}
where wavevector subtraction is understood modulo the reciprocal lattice. Since this shift is a permutation of the discrete Fourier grid,
\begin{equation}
    |\mathcal Q_N|
    =
    N \phi_{\mathrm{mic}}.
    \label{eq:QN_size}
\end{equation}
Introducing
\begin{equation}
    X_N
    :=
    \sum_{ q\in\mathcal Q_N}
    |\widehat\zeta( q)|^2,
\end{equation}
and employing Eq.~\eqref{eq:binary-fourier-covariance} together with Eq.~\eqref{eq:QN_size}, we get
\begin{equation}
    \mathbb E[X_N]
     =
    N\phi_{\mathrm{mic}}p(1-p).
    \label{eq:expected-microscopic-fourier-power}
\end{equation}

To control fluctuations, define
\begin{equation}
    A_N
    :=
    F_V^\dagger
    \left(
        \sum_{ q\in\mathcal Q_N}
        \ketbra{q}{q}
    \right)
    F_V,
\end{equation}
so that
\begin{equation}
    X_N
    =
    \zeta^\dagger A_N\zeta.
\end{equation}
The matrix $A_N$ is an orthogonal projector of rank $N \phi_{\mathrm{mic}}$, and hence
\begin{equation}
    \operatorname{Tr}(A_N)
    =
    N \phi_{\mathrm{mic}},
    \qquad
    \|A_N\|
    =
    1,
    \qquad
    \|A_N\|_{\mathrm F}^2
    =
    N \phi_{\mathrm{mic}}.
\end{equation}
Since the entries of $\zeta$ are independent, centered, and bounded, a direct quadratic-form variance estimate gives
\begin{equation}
    \operatorname{Var}(X_N)
    \leq
    C_p
    \|A_N\|_{\mathrm F}^2
    =
    C_p N \phi_{\mathrm{mic}}
    =
    \O(N),
    \label{eq:microscopic-power-variance}
\end{equation}
where $C_p$ depends only on $p$. Thus, for every fixed $0<\delta<1/2$, Chebyshev's inequality gives
\begin{equation}
    \mathbb P
    \left(
        \left|
            X_N-N \phi_{\mathrm{mic}} p(1-p)
        \right|
        \geq
        N^{1/2+\delta}
    \right)
    =
    \O(N^{-2\delta}).
    \label{eq:microscopic-power-concentration}
\end{equation}
Since for a large cubic lattice we have
\begin{equation}
    \phi_{\mathrm{mic}}
    =
    1-
    \left(
        \frac{\gamma_{\mathrm{mic}}}{\pi}
    \right)^3
    +
    o(1),
    \label{eq:microscopic-mode-fraction}
\end{equation}
with probability tending to one, we get
\begin{equation}
    X_N
    =
    N
    \phi_{\mathrm{mic}}
    p(1-p)
    +
    o(N).
\end{equation}
Substituting this result into Eq.~\eqref{eq:projected-microscopic-power} yields Eq.~\eqref{eq:microscopic-energy-numerator2} claimed in the main text:
\begin{equation}
    \left\|
        (\Pi_{\mathrm{mic}}\ot I_S) s(0)
    \right\|_2^2
    =
    N\|v\|_2^2
    \phi_{\mathrm{mic}}
    p(1-p)
    \Delta_s^2
    +
    o_{\mathbb P}(N).
\end{equation} 

\bibliography{ref}    

\end{document}